\documentclass[aps,prb,twocolumn,floats]{revtex4-2}
\usepackage{amssymb}
\usepackage{amsbsy}
\usepackage{amsmath}
\usepackage{color}
\usepackage{float}
\usepackage{graphicx} 
\usepackage[colorlinks,linktocpage,bookmarks=false,citecolor=blue,linkcolor=red,urlcolor=blue]{hyperref}

\newcommand{\beq}{\begin{equation}}
\newcommand{\eeq}{\end{equation}}
\newcommand{\bea}{\begin{eqnarray}}
\newcommand{\eea}{\end{eqnarray}}

\newcommand{\si}{\sigma}

\newcommand{\non}{\nonumber}

\newcommand{\red}{\textcolor{red}}

\begin{document}

\title{Heating suppression via two-rate random and quasiperiodic drive protocols}

\author{Krishanu Ghosh$^1$, Sayan Choudhury$^2$, Diptiman Sen$^3$, and K. Sengupta$^1$}

\affiliation{$^{(1)}$School of Physical Sciences, Indian
Association for the Cultivation of Science, 2A and 2B Raja S. C.
Mullick Road, Jadavpur 700032, India \\
$^{(2)}$Harish Chandra Research Institute, A CI of Homi Bhabha National Institute,
Chhatnag Road, Jhunsi, Prayagraj, Uttar Pradesh 211019, India \\
$^{(3)}$Center for High Energy Physics, Indian Institute of Science, Bengaluru 560012, India}

\date{\today}

\begin{abstract}
We study a random and quasiperiodically driven one-dimensional non-integrable PXP spin chain in a magnetic field for two distinct drive protocols. Each of these protocols involves square pulses with two driving frequencies which are integer multiples of each other. For the first class of protocols, the duration of the pulse is changed randomly by an amplitude $dT$ while for the second class we use a random/quasiperiodic dipolar drive, where the quasiperiodicity is implemented using the Thue-Morse (TM) or Fibonacci sequences. For both protocols, we identify parameter regimes for which the thermalization of the driven chain is drastically slowed down due to proximity to a two-rate drive induced exact dynamical freezing. We also study the properties of these driven system moving slightly away from the freezing limit. For the first type of protocols, we 
show the existence of special value of $dT$ for which the thermalization rate remains small and provide an analytic explanation for such slow thermalization. For the second class of protocols, in contrast to random/quasiperiodic drives involving a single frequency studied earlier, we find  that the TM quasiperiodic drive leads to a distinctly slower thermalization than that for drive protocols which are either periodic or follow a random or quasiperiodic Fibonacci sequence. We provide a qualitative semi-analytic understanding of these phenomena either using an exact calculation for small system sizes or carrying out a perturbative analysis in the large drive-amplitude limit. Our analysis brings out the central role of such two-frequency protocols in the reduction of heating in driven quantum systems. We discuss experiments which can test our theory.

\end{abstract}

\maketitle

\section{Introduction}
\label{intro}

The study of non-equilibrium dynamics of closed quantum systems has been a subject of intense research activity \cite{rev1,rev2,rev3,rev4,rev5,rev6,rev7,rev8,rev9,rev10,rev11,rev12,rev13,rev14,rev15,rev16,rev17}. The initials studies concentrated
on quench and ramp protocols; one of the main focus of such studies was to understand the universal aspects of 
excitations production when closed quantum systems are quenched or ramped through quantum critical points \cite{rev1,rev2,rev3,rev4,subir1,dsen, anatoly0_part1,anatoly0_part2,sdas0_part1,sdas0_part2,sdas1,sdas2_part1,sdas2_part2,rajdeep1_part1,rajdeep1_part2,rajdeep2_part1,rajdeep2_part2, trip1_part1,trip1_part2, kibble1,zurek1,anatoly1,anatoly2,dsen1,dsen2_part1,polkovnikov_barankov_part2,sdas3_part1,sdas3_part2,sdas4,adutta1_part1,adutta1_part2,rajdeep3}.  
The later studies focused mostly on periodic drives \cite{rev5,rev6,rev7,rev8,rev9,rev10,rev11,rev12,rev13,rev14,rev15,rev16,rev17}. Such a shift in focus stems mostly from the presence of several phenomena found in periodically driven systems; these do not have any analogs either in equilibrium or for aperiodically driven systems. These include dynamical freezing \cite{adas1,adas2,adas3,pekker1,deb1,uma1,camilo1,apal1,koch1,adasnew,tb1,tb2_part1,Turkeshi_Schiro_part2,cd1}, presence of Floquet scars \cite{pretko1,bm1,mituza1_part1,Sug_Kun_Saito_part2,bm2_part1,lukinsc1,papic1,liu1}, signatures of prethermal Hilbert-space fragmentation  \cite{sg1,sg2,xu1,zhang1}, dynamical localization \cite{dynloc1,dynloc2,dynloc3,dynloc4,dynloc5,tanay1,fava1_part1,holthus_part1,rg1,galit1,martinez1,tun1,guoexp}, dynamical phase transitions \cite{heyl1,heyl2,ks1,ks2,amit1,ks3}, generation of drive-induced topological Floquet states \cite{topo1,topo2,topo3,topo4,topo5,topo6,topo7,topo8,topo9}, and realization of prethermal time crystalline phases\cite{tcrev1,tcrev2,tcrev3,tcrev4,tcrev5,tcpap1_part1,tcpap1_part2,tcpap2_part1,tcpap2_part2,tcpap3}.

The properties of aperiodically driven quantum systems have also been studied extensively in recent years \cite{q1,q2,q3,q4,q5,q6,q7,q8,q9,q10,q11,q12,q13,q14,q15,q16,q17,q18,q19,q20,q21,q22,q23,q24}. Typically, in such studies aperiodicity is introduced in two distinct ways. The first method involves a random modification of the time period of an otherwise periodically driven system with a square-pulse protocol with unitaries having time period $T(\eta)$ \cite{q3,q5,q11}, where $T(\eta) = T+\eta dT$, $\eta \in [-1,1]$ is chosen randomly, and $dT/T \le 1$. The time evolution via such unitaries leads to realization of a random drive. The second protocol involves taking combination of multiples of periodic unitaries $U_{a} = \exp[-i H_{a} T/\hbar]$ of time period $T$ with distinct Hamiltonians $H_{a}$ with $a=1,2$; combining $n$ such unitaries in different sequences leads to random or quasi-periodic multipolar drives \cite{q12,q13,q14}. For example a dipolar drive would corresponds to repeated application of two unitaries $U_{12}= U_1 U_2$ and $U_{21}= U_2 U_1$ following a random or quasiperiodic sequence. It was shown in Ref.\ \onlinecite{q13} that for small $T$ such multipolar drives leads to fastest (slowest) thermalization for random (periodic) drive protocols; the quasiperiodic drive protocols always lead to a thermalization rate between random and periodic drives. In this regime, using an argument which relies on the validity of Magnus expansion at small $T$, Ref.\ \onlinecite{q13} established a bound on the heating rate for such protocols. 

More recently, a two-rate periodic (two-tone) protocol, which allows for realization of exact Floquet flat bands for a large class of periodically driven non-integrable Hamiltonians, has been put forth \cite{flat1}. Such a protocol typically constitutes two drive frequencies which are integer multiples of each other; for a square-pulse protocol, it was shown that the flat band is realized for any integers. The presence of such flat bands leads to violation of the eigenstate thermalization hypothesis (ETH)~\cite{eth1,eth2a,eth2b,eth3,eth4} and results in complete dynamical freezing and, consequently, suppression of heating in such driven systems. Interestingly, the presence of such a flat band leads to a perturbative regime with suppressed thermalization; it was noted this perturbative regime depends only on the proximity to the flat band and does not require the presence of large drive amplitude or frequencies. The implication of such a two-rate drive for quasiperiodic and random drive protocols has not been studied so far.

In this work, we study an experimentally realizable 
non-integrable spin chains driven quasiperiodically or randomly by two frequencies using a square pulse protocol. The model we consider is the well-known PXP chain \cite{subir2,subir3,abanin1,abanin2} which describes the physics of ultracold Rydberg atoms. The Hamiltonian of such a chain is given by
\begin{eqnarray} 
H_1 &=& \sum_j [w(t) \tilde \sigma_j^x -\lambda(t) \sigma_j^z], \quad \tilde \sigma_j^x = P_{j-1} \sigma_j^x P_{j+1}, \label{ham2} 
\end{eqnarray} 
where $\si_j^x, ~\si_j^z$ denote Pauli matrices at a 
site labeled $j$, and $P_j= (1-\sigma_j^z)/2$ is an operator which projects out the spin-up
state at site $j$. The effect of the $P_j$'s in
Eq.~\ref{ham2} is to project out all states with two-neighboring up-spins from the Hilbert space. This constraint makes the model non-integrable and also reduces the dimension of the Hilbert space of the spin model which facilitates numerical analysis; for a chain of size $L \gg 1$, the Hilbert space dimension of the model is a constant times $\varphi^L$, where $\varphi=(1+\sqrt{5})/2$ is the golden ratio \cite{abanin1,abanin2}.  Our choice for the study of this spin chain is motivated by the fact that it describes the physics of experimentally realizable Rydberg atom arrays \cite{exp1,exp2,exp3,exp4}; this allows for a possibility of experimental verification of some of our results. 

In what follows, we shall consider square pulse drive protocols where the ratio of the drive frequencies is fixed: $\omega_2/\omega_1=2$ and $\omega_1= 2\pi/T$, where $T$ denote the time period of the drive in the absence of randomness or quasiperiodicity. We note that the qualitative features of our results hold for any $\omega_2/\omega_1=n_0$ where $n_0\in Z$; in this work, we set $n_0=2$ for all numerical results. For $n_0=2$, during each of these square pulses, the PXP Hamiltonian can be written as 
\begin{eqnarray} 
H_1(a,b) &=& \sum_j [(a w +\delta_w) \tilde \sigma_j^x -(b \lambda +\delta_{\lambda})\sigma_j^z], \label{prot1}
\end{eqnarray} 
where $a,b =\pm 1$ and $\delta_w$ and $\delta_{\lambda}$ are the static components of $w(t)$ and $\lambda(t)$ respectively. In the absence of quasiperiodicity or randomness, these pulses implement a two-rate protocol which leads to exact Floquet flat bands for $\delta_w=\delta_{\lambda}=0$; the width of the Floquet spectrum close to this limit ($\delta_w,\delta_{\lambda} \ll w,\lambda$) is controlled by $\delta_w$ and/or $\delta_{\lambda}$ \cite{flat1}. 

To implement randomness and quasiperiodicity, we modify the periodic drive protocol in two distinct ways. First, we modify the time period of the square pulses randomly by an amount $dT \ll T/4$; each of these pulses is given a duration of $T_i= T/4+ \eta_i dT$ where $\eta_i \in [-1,1]$ are random  numbers. Second, we study the effect of random or quasiperiodic dipolar drives on these spin chains. To this end, we consider unitaries $U_{\pm}$ for the PXP Hamiltonian given by
\begin{eqnarray} 
U_{\pm} &=& e^{-i H_1(\pm 1,-1) T/(4 \hbar)} e^{-i H_1(\pm 1,1) T/(4\hbar)}, \label{diunit} 
\end{eqnarray} 
and construct dipole unitaries $U_1= U_+ U_-$ and $U_2= U_- U_+$ using them. The dipolar drive is then realized by random or quasiperiodic sequences of $U_1$ and $U_2$ \cite{q13,q14} acting on the initial state; a periodic drive with time period $T$ corresponds to a sequence of either only $U_1$'s or $U_2$'s. We note that such a dipolar drive leads to exact dynamic freezing for $\delta_w=\delta_{\lambda}=0$ for any quasiperiodic or random drive sequence; this results in the divergence of the thermalization time as $\delta_w,\delta_{\lambda} \to 0$. This points to the possibility of achieving large thermalization times and hence reduction of heating for small $\delta_{w}$ and/or $\delta_{\lambda}$.

The central results that we obtain are as follows. First, we study the PXP model driven by the two-rate square pulse protocol where the time period is changed randomly by $dT$. We show that for $\lambda \gg w$ and $\delta_w=\delta_{\lambda}=0$, the evolution under such a random drive leads to slow thermalization for $\lambda dT/\hbar= p \pi/2$, where $p\in Z$ for $\delta_{\lambda}=\delta_w=0$. Interestingly, in this limit, the time scale of evolution and hence the thermalization time is controlled only by $dT$ for any $\lambda$, $w$; it is completely independent of $T$ for the protocol chosen. 

Second, we also study the driven PXP model away from the point $\delta_w=\delta_{\lambda}=0$ by tuning $\delta_w/w$ (keeping $\delta_{\lambda}=0$) for a fixed $\lambda dT/\hbar= \pi/2$. For $\delta_w \ne 0$, the thermalization time depends on $T$; our analysis shows the existence of special drive frequencies, characterized by an integer $p$ and given by $\omega_p^{\ast}=2 \pi/T_p^{\ast}$, at which the thermalization becomes slow. These special frequencies depend on the details of the protocol chosen; we  find that for $\lambda T/(4 \pi \hbar) \le 6$, they occur close to $\lambda T_p^{\ast}/(4 \pi \hbar) \simeq p/2$ provided that
$w(t)$ is driven with the shorter time period (around $T/2$ with some randomness) and $\lambda(t)$ has the longer time period (around $T$ with some randomness). In contrast, we find them to occur at  $\lambda T_p^{\ast}/(4 \pi \hbar) \simeq (p+1/2)$ if $\lambda(t)$ has the shorter drive period.  We provide an analytic understanding of this phenomenon using a Floquet perturbation theory which provides analytic expressions of these special drive periods and brings out the reason for slow thermalization at these drive frequencies. Our analysis shows the role of the flat band limit behind this phenomena. We also find that for both protocols, the condition for the slow thermalization gradually changes as $T$ is increased and approaches $\lambda T_p^{\ast}/(4 \pi \hbar) \simeq (p+1/4)$ for $\lambda  T/(4 \pi \hbar) \ge 6$; this deviation is not captured by the first-order Floquet perturbation theory. 

\begin{figure}[h]
\includegraphics[width =0.98 \linewidth]{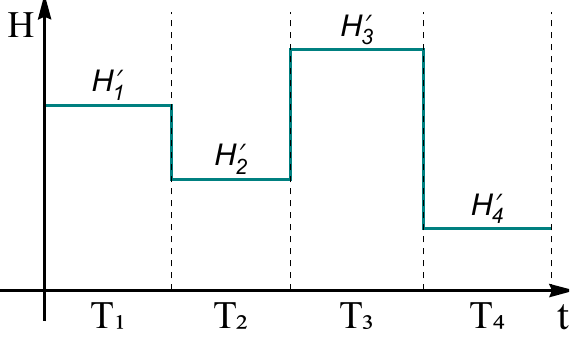}
\caption{ Schematic representation of the unitaries given by $U_i= \exp[-i H'_i T_i/\hbar]$ for the random drive protocol with $T_i= T/4 + \eta_i dT$. The Hamiltonians $H'_i$ depend on the precise choice of protocols. For 
$U_3$ (Eq.\ \ref{prot2}) $H'_1= H_2(-1-1)$, $H'_2= H_2(1,1)$, $H'_3= H_2(1,-1)$ and $H'_4 = H_2(-1,1)$. For $U_4$ (Eq.\ \ref{prot4}) where $w$ is varied with shorter time period,  $H'_1= H_3(1,1)$, $H'_2= H_3(-1,1)$, $H'_3= H_3(1,-1)$ 
and $H'_4= H_3(-1,-1)$ while for $U_5$ (Eq.\ \ref{prot5}) where $w$ is varied with the longer time period, $H'_1= H_3(1,1)$, $H'_2= H_3(1,-1)$, $H'_3= H_3(-1,1)$ 
and $H'_4= H_3(-1,-1)$. See text for details. \label{figpr1}}
\end{figure}

Third, for the dipolar drive protocol implemented by the unitaries $U_1$ and $U_2$ with finite $\delta_{\lambda}$, we show that the thermalization near the dynamic freezing limit, where $\delta_{\lambda} \ll w,\lambda$ and $\delta_w=0$, is fastest for periodic drives, slower for the random and quasiperiodic Fibonacci drives, and slowest for a Thue-Morse (TM) drive. 
This result is in contrast to that found in Refs.\ \onlinecite{q13,q14} for single-rate drive protocols. Moreover, the fidelity of the driven state approaches unity for the TM drive protocol at high drive-frequencies, whereas they remain finite, but substantially less than unity, for other protocols. For all protocols and for $\delta_{\lambda} \ll w,\lambda$, the fidelity of the driven state remain finite for a much longer time compared to their counterparts for a single-rate drive protocol showing a lack of heating in this regime. We provide an analytic understanding of these phenomena and demonstrate the role of the flat-band limit of the two-rate drive protocol behind it. We discuss our main results and provide a list of possible experiments using ultracold Rydberg atoms which may test our theory.  

The organization of the rest of the paper is as follows. In Sec.\ \ref{rd1}, we study protocols where the period of the square pulse involving two drive frequencies is modified randomly. Next, in Sec.\ \ref{dp1}, we study random or quasiperiodic dipolar drives involving two drive frequencies. Finally, we discuss our main results, provide a concrete experimental proposal to test our theory, and conclude in Sec.\ \ref{diss1}. Some details of our analysis 
is presented in the appendices.

\section{Random drive period} 
\label{rd1} 

In this section, we study the PXP model (Eq.\ \ref{ham2}) in the presence of a square-pulse protocol. The time periods of these pulse
are given by $T_i= T/4 + \eta_i dT$, where $\eta_i \in[-1,1]$ are chosen from a random distribution, and $T =2\pi/\omega_1$
where $\omega_1$ is the drive frequency. In what follows, we shall choose $dT \ll T/4$. The corresponding evolution operators for any 
four consecutive unitaries are chosen to be 
\begin{widetext}
\begin{eqnarray} 
U_3(T,dT;\{\eta_i\}) &=&  e^{-i H_2(-1,1)\,(T/4+\eta_4 dT)/\hbar} \,e^{-i H_2 (1,-1)\,(T/4+\eta_3 dT)/\hbar} e^{-iH_2 (1,1)\,(T/4+\eta_2 dT)/\hbar} \,e^{-iH_2 (-1,-1)\,(T/4+\eta_1 dT)/\hbar}, \nonumber\\ \label{prot2}
\end{eqnarray} 
\end{widetext} 
where $H_2(a,b)= H_1(\delta_{\lambda}= \delta_w=0)$ is given by Eq.\ \ref{prot1}. This protocol is represented schematically in Fig.\ \ref{figpr1}. The time-evolution of any initial state $|\psi(0)\rangle$ up to a time 
\begin{eqnarray} 
T_0 &=& \sum_{i=1}^{4m} T_i =  mT +\sum_{i=1}^{4 m} \eta_i dT  \label{time1} 
\end{eqnarray} 
proceeds via repeated application of the unitary $U_3$ and is given by
\begin{eqnarray} 
|\psi(T_0)\rangle &=& U_3(T,dt;\{\eta_i\}_m) U_3(T,dT; \{\eta_i\}_{m-1}) \cdots \nonumber\\
&& \cdots \times  U_3(T,dT; \{\eta_i\}_1) |\psi(0)\rangle.
\end{eqnarray}  
We note that in the absence of $\eta_i$, this corresponds to a periodic drive with period $T$. Moreover, for our chosen protocol, $|\psi(mT)\rangle = |\psi(0)\rangle$ for all $m$ and any $T$ for $dT=0$; this behavior occurs due to the presence of exact Floquet flat bands for such protocols which follows from the condition $H(a,b)= -H(-a,-b)$ yielding $U_3(dT=0)=I$ \cite{flat1}. We denote this limit as the exact dynamic freezing limit of two-rate drive protocols. 

Next, we consider such drives away from the exact freezing limit. This is done by introducing a static component of $w(t)$ \cite{flat1}. More specifically, we choose 
\begin{eqnarray} 
H_3(a,b) &=&  \sum_j \left[ (a w + \delta_w) \tilde \sigma_j^x -b \lambda \sigma_j^z \right], \label{ham3} 
\end{eqnarray} 
where the static component $\delta_w \ll w,\lambda$ is a measure of the proximity of the system to the freezing limit at $\delta_w=dT=0$.  Here we consider two separate classes of unitaries. For the first protocol, $w$ is changed after the shorter timescale $T/4 + \eta_i dT$ while for the second, $\lambda$ changes over this shorter timescale. These unitaries are denoted by $U_4$ (for $w$ with shorter timescale) and $U_5$ (for $\lambda$ with shorter timescale) and are given by

\begin{widetext} 

\begin{figure}[h]
\begin{tabular}{ccc}
\includegraphics[width=0.33\textwidth]{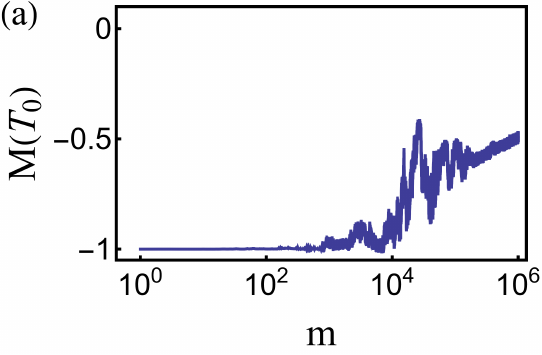}&
\includegraphics[width=0.33\textwidth]{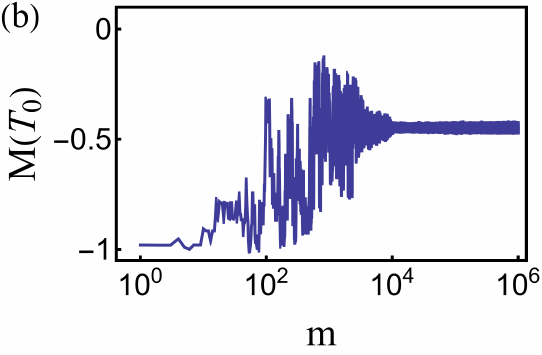}&
\includegraphics[width=0.33\textwidth]{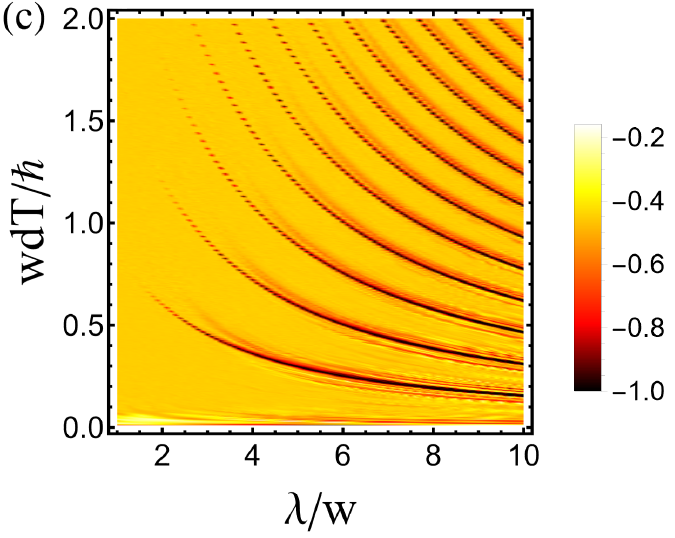}\\
\end{tabular} 
\caption{(a) Plot of $M(T_0)$  as a function of $m$ for $\lambda dT/\hbar =\pi/2$ with $w=1$ and $\lambda/w=10$. (b) Similar plot for $\lambda dT= \pi/(8\hbar)$. (c) Plot of average magnetization $\bar M$, averaged over 
$100$ cycles around $m=1000$  as a function of $w dT/\hbar$ and $\lambda/w$ showing slow dynamics of $\bar M$ for $\lambda dT/\hbar =p \pi/2$ with $p \in Z$.}
\label{fig1} 
\end{figure}

\end{widetext}

\begin{widetext}
\begin{eqnarray} 
U_4(T,dT;\{\eta_i\}) &=&  e^{-i H_3(-1,-1)\,(T/4+\eta_4 dT)/\hbar} \,e^{-i H_3 (1,-1)\,(T/4+\eta_3 dT)/\hbar} e^{-iH_3 (-1,1)\,(T/4+\eta_2 dT)/\hbar} \,e^{-iH_3 (1,1)\,(T/4+\eta_1 dT)/\hbar}, \nonumber\\ \label{prot4} \\
U_5(T,dT;\{\eta_i\}) &=&  e^{-i H_3(-1,-1)\,(T/4+\eta_4 dT)/\hbar} \,e^{-i H_3 (-1,1)\,(T/4+\eta_3 dT)/\hbar} e^{-iH_3 (1,-1)\,(T/4+\eta_2 dT)/\hbar} \,e^{-iH_3 (1,1)\,(T/4+\eta_1 dT)/\hbar}. \nonumber\\ \label{prot5}
\end{eqnarray} 
\end{widetext}

These protocols are represented schematically in Fig.\ \ref{figpr1}. The time evolution of the system proceeds via repeated application of these unitaries for $t\le T_0$ and lead to distinct features for $\delta_w \ne 0$; we shall discuss these differences in detail in Secs.\ \ref{numres} and \ref{analyt}. For $\delta_w=0$, $U_4$ and $U_5$ provide qualitatively similar dynamics to that given by $U_3$.

\subsection{Numerical Results} 
\label{numres} 

In this section, we carry out a numerical study of the magnetization of the PXP model defined as
\begin{eqnarray}
M(T_0) &=& \frac{1}{L} \langle \psi(T_0) | \sum_j \sigma_j^z |\psi(T_0)\rangle, \label{mag1} 
\end{eqnarray} 
where $L$ is the length of the spin chain, and we choose $L$ to be an even integer (we will set the lattice spacing equal to unity).

In what follows, we use exact diagonalization (ED) to numerically obtain the eigenvalues and eigenstates of $H_2(a,b)$ (for $\delta_w=0$) and $H_3(a,b)$ (for $\delta_w \ne 0$). Using these, we construct the matrix $\exp[-i H_{2,3}(a,b)\,(T/4+\eta_i dT)/\hbar]$ for any $\eta_i$, $dT$, and $T$, where the $\eta_i=\pm 1$ are chosen randomly. A repeated operation of such matrices on the state vector $|\psi(0)\rangle$ according to the protocol outlined in Eq.\ \ref{prot2} allows one obtain $|\psi(T_0)\rangle$. For the rest of the work, we chose $|\psi (0)\rangle = |\downarrow, \downarrow \cdots \downarrow \rangle$ leading to $M(0)=-1$.

\begin{figure}[h]
\includegraphics[width =0.48 \linewidth]{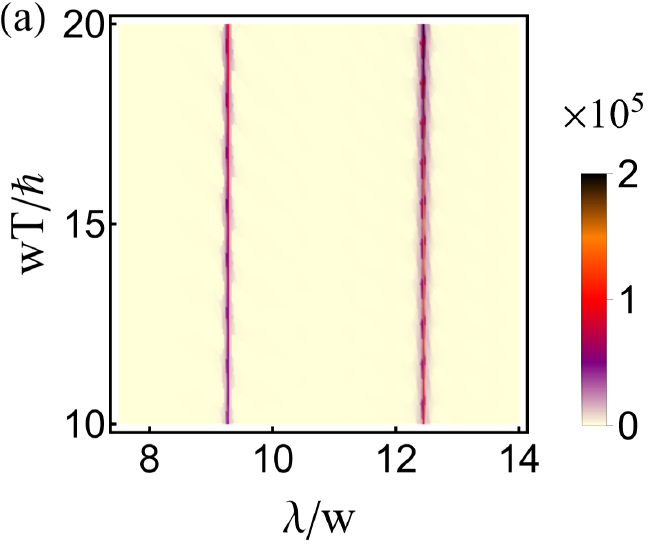}
\includegraphics[width =0.48 \linewidth]{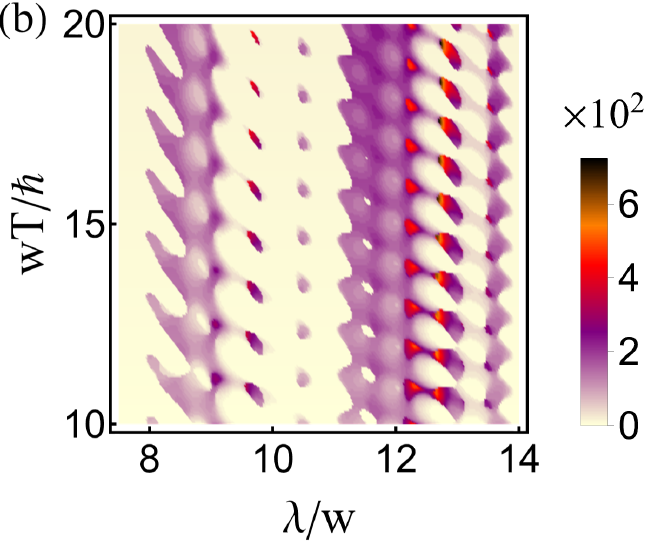}\\
\includegraphics[width =0.48 \linewidth]{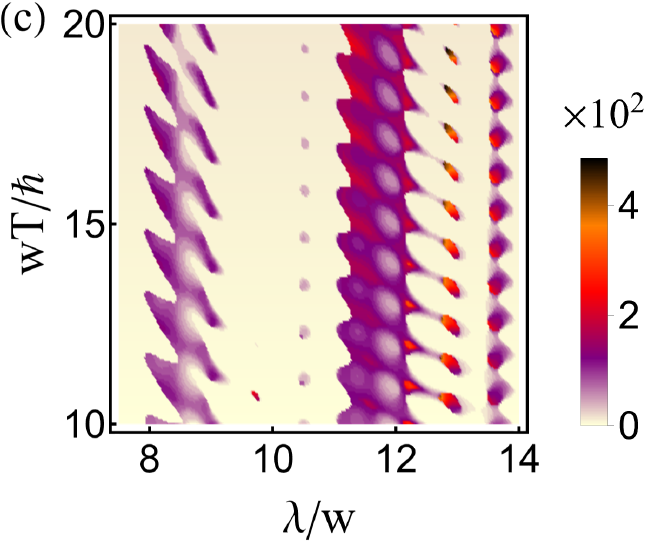}
\includegraphics[width =0.48 \linewidth]{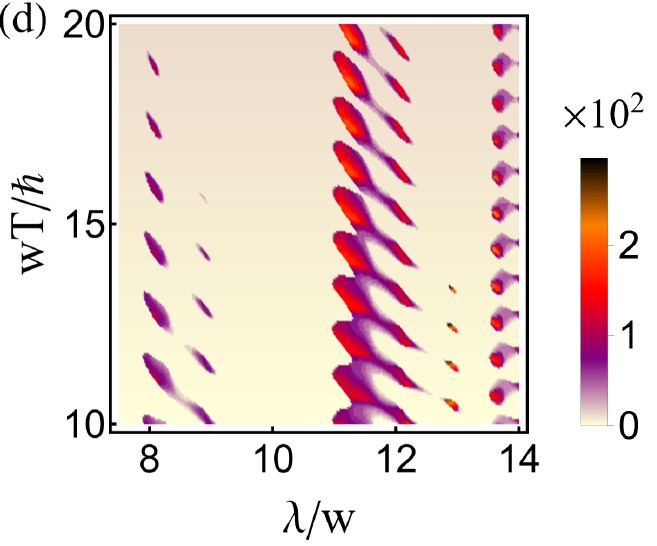}\\
\caption{ Plots of $T_{\rm th}/T=m_0$ as a function of $\lambda/w$ and $wT/\hbar$ for several representative values, (a) $\delta_w/w=0$ (b) $0.05$, (c) $0.1$ and (d) $0.2$. For all plots $w=1$, $\delta_{\lambda}=0$, and $w dT/\hbar =1/2$. \label{fig2}  }
\end{figure}

The results obtained from such numerics is shown in Fig.\ \ref{fig1} where we plot the magnetization $M(T_0)$ as a function of $m$ during its evolution obtained by repeatedly applying the unitaries given in Eq.\ \ref{prot2}. We find from Figs.\ \ref{fig1}(a) and (b) that for the protocol outlined above, $M$ shows a much slower growth for $\lambda dT/\hbar= \pi/2$ than that at $\lambda dT/\hbar= \pi/8$. A plot of the average magnetization (Fig.\ \ref{fig2}(c)) defined as
\begin{eqnarray}
\overline M &=& \sum_{m=950}^{1050} M(T_0)/100, \label{mag2}
\end{eqnarray}
shows that the change in the magnetization from its initial value is drastically reduced for special drive amplitudes 
at which $\lambda dT/\hbar = p \pi/2$, where $p \in Z$. 

Next, we study the behavior of the magnetization of the driven system when one moves away from the flat band limit by introducing a finite $\delta_w$. To study the nature of thermalization, we also introduce the notion of a thermalization time $T_{\rm th} = m_0 T$, where $m_0$ is the minimum number of cycles after which 
\begin{eqnarray}
\Delta M &=&  |M(T_{\rm th})/M(0)-1| \sim \epsilon M(0).
\end{eqnarray}  
In this definition $\epsilon$ is arbitrary; for concreteness, we choose $\epsilon=0.1$ for all 
our numerical results. 

The behavior of the thermalization time $T_{\rm th}$ is shown in Fig.\ \ref{fig2} as a function of $w T/\hbar$ and $\lambda/w$ with fixed $w dT/\hbar=1/4$ and for representative values of $\delta_w$. The protocol chosen is given by Eq.\ \ref{prot4}. We find, from Fig.\ \ref{fig2}(a), that in the flat band limit ($\delta_w=0$), $m_0$ becomes independent of $T$ and is maximal around lines given by $\lambda dT/\hbar = p \pi/2$ where $p \in Z$. For finite $\delta w$, $m_0$ decreases in magnitude. Its value still remain the largest around these vertical lines; however, there exist special directions in the $\lambda-T$ plane for which it remains large away from these lines. This indicates the presence of a perturbative regime controlled by the parameter $\delta_w/w$ in which thermalization can be drastically slowed down by specific choice of parameters.

\begin{figure}[H]
\centering
\includegraphics[width=0.49 \linewidth]{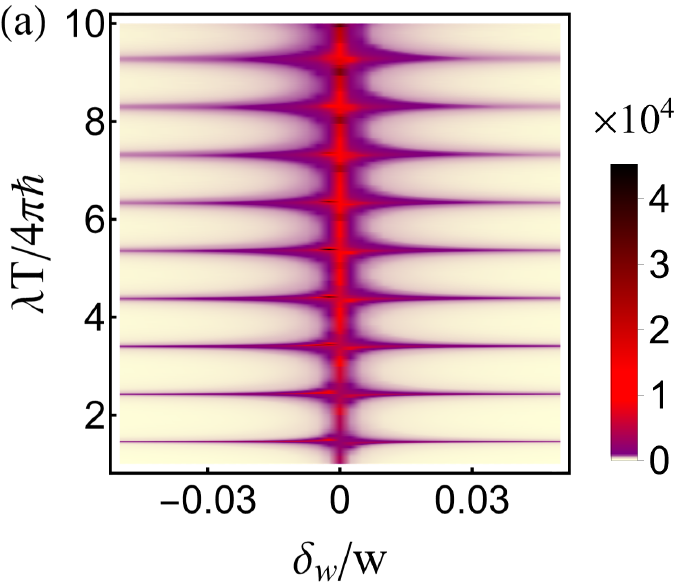}\hfill
\includegraphics[width=0.49 \linewidth]{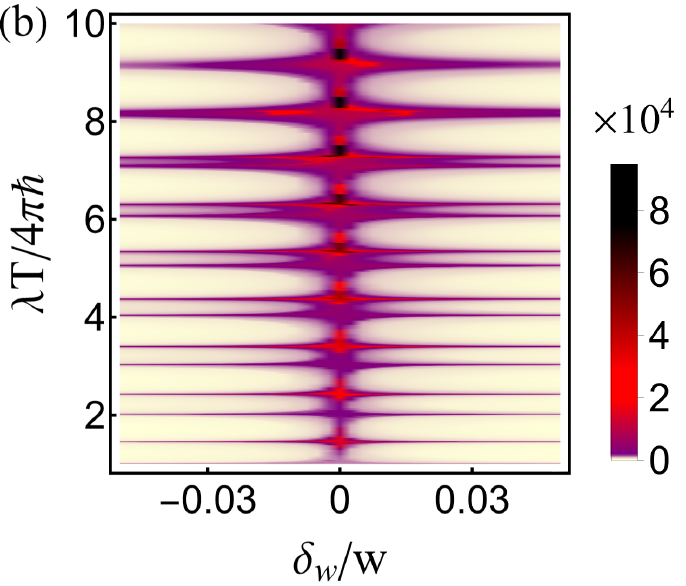}\hfill
\caption{Plots of $T_{th}/T=m_0$ as a function of $\lambda T/(4\pi \hbar)$ and $\delta_w/w$ for $dT=\pi/(2\lambda \hbar)$ and $\lambda/w=4 \pi$ for (a) the protocol given by Eq.\ \ref{prot5} where $\lambda$ has the shorter time period, and (b) the protocol given by Eq.\ \ref{prot4} for which $\lambda$ is varied with the longer time period. } \label{fig3}
\end{figure}

To explore this further, we study the behavior of the thermalization time as a function of $\delta_w/w$ for the protocols given by Eqs.\ \ref{prot4} and \ref{prot5} for $\lambda dT/\hbar=\pi/2$. The results of such a study is shown in Fig.\ \ref{fig3}. The left (right) panel of Fig.\ \ref{fig3} shows the behavior of $T_{\rm th}$ when the PXP chain is driven according to protocol given by Eq.\ \ref{prot5} (Eq.\ \ref{prot4}). For both cases, we find a vertical line showing slowest thermalization for $\delta_w=0$. Away from this limit, for finite $\delta_w/w$ and for both protocols, there exist special values of $T=T_p^{\ast}$ for which the thermalization remains slow. These correspond to the horizontal lines in the figure which are independent of $\delta_w$ for small $\delta_w/w \le 0.05$. 

We note that the structure of such lines of slow thermalization in the $\lambda T-\delta_w/w$ plane differs in the two cases. For the drive protocol given by Eq.\ \ref{prot5} where $\lambda$ has the shorter time period of
about $T/2$ and $w$ has the longer time period of 
about $T$, both with some randomness, the thermalization remains minimum along single horizontal lines for all $T$; these lines occur for $\lambda T_p^{\ast}/(4 \pi \hbar) \simeq (p+1/2)$ for small $T$. Their position gradually shift with increasing $T$ and reach $\lambda T_p^{\ast}/(4 \pi \hbar) \simeq (p+1/2)$ for $p\le 6$ as can be seen in Fig.\ \ref{fig3}(a). 

In contrast, for the drive protocol given by Eq.\ \ref{prot4} where $\lambda$ has the longer time period of
about $T$ and $w$ has the shorter time period of about $T/2$ with some randomness,
these horizontal lines split for small $T$ ($\lambda T/(4\pi \hbar) \le 6$) leading to two distinct closely spaced values of $T_p^{\ast}$ occurring close to $\lambda T_p^{\ast}/(4 \pi \hbar) \simeq p/2$. For $\lambda T/(4 \pi) > 6$, these two split horizontal lines merge and yield a single line close to $\lambda T_p^{\ast}/(4 \pi \hbar) \simeq (p+1/4)$ as can be seen in Fig.\ \ref{fig3}(b). This feature brings out the distinction between the two types of protocols described by Eqs.\ \ref{prot4} and \ref{prot5} for finite $\delta_w/w$. In the next section, we shall try to understand some of these numerical features using several semi-analytic methods. 

\subsection{Analytical Results}
\label{analyt} 

In this section, we aim to provide a semi-analytic understanding of certain features of the behaviors of the magnetization $M(T_0)$ and the thermalization time $T_{\rm th}$ for the randomly driven PXP chain. 
We first consider the dynamics for $\delta_w=0$. For the protocol described by Eq.\ \ref{prot2}, we find that in this limit $U_3$ is given by 
\begin{eqnarray} 
U_3(T,dT;\{\eta_i\}) &=&  e^{i H_2(-1,1)(\eta_3 -\eta_4)dT/\hbar} \nonumber\\
&& \times e^{i H_2 (1,1)\,(\eta_1 -\eta_2)dT/\hbar} \label{ueq1}
\end{eqnarray} 
This shows that the evolution of $|\psi(T_0)\rangle$ by repeated application of $U_3$ is completely independent of $T$; thus, $M(T_0)$ and hence $T_{\rm th}$ can depend only on two dimensionless quantities $w dT/\hbar$ and $\lambda/w$. This allows for the possibility of slow thermalization for any $T$; this feature is absent for the single-frequency random or quasiperiodic drive protocols. 

To understand the slow evolution of the magnetization at the special values of $dT$, $\lambda dT= p \pi/2$, we construct an effective unitary from $U_3$ in the limit of large $\lambda/w$. To this end, we first consider the unitary $U_{3a}= \exp[i H_2(1,1)(\eta_1-\eta_2) dT]$. For $\lambda/w\gg 1$, the leading term of this unitary is 
\begin{eqnarray} 
U_{3a}^{(0)} &=& e^{-i \lambda dT (\eta_1-\eta_2) \sum_j \sigma_j^z/\hbar}.
\end{eqnarray}
Since $\eta_2-\eta_1$ can take values $\pm 2,0$ and $\langle \sum_j \sigma_j^z \rangle$ is even for any Fock state for a chain with an even number of sites, we find that $U_{3a}^{(0)}=I$ for $\lambda dT/\hbar= p \pi/2$, where $p \in Z$. An exactly similar analysis holds for the other unitary $U_{3b} =\exp[i H(-1,1)(\eta_3-\eta_4) dT/\hbar]$. Thus the leading term of both the unitaries reduce to $I$ for $\lambda dT/\hbar= p \pi/2$; this naturally leads to a slow evolution of $M$ and hence a larger $T_{\rm th}$. 

The first subleading term for both these unitaries also vanishes at these points. To see this, we consider a perturbative expansion of $U_{3a}^{(0)}$ 
given by  $U_{3a}= U_{3a}^{(0)} + U_{3a}^{(1)} + \cdots$, where 
\begin{eqnarray} 
U_{3a}^{(1)} &=& - ~\frac{i}{\hbar}~ \int_0^{dT} U_{3a}^{(0) \dagger} H_1 U_{3a}^{(0)},  \nonumber\\
H_1 &=& w (\eta_1-\eta_2) \sum_j (\tilde \sigma_j^+ +{\rm H.c.}),  \label{pert1} 
\end{eqnarray} 
where $\tilde \sigma_j^x = \tilde \sigma_j^+ +\tilde \sigma_j^-$. Using the fact that for any Fock state $|m\rangle$ which has $m$ additional up-spins 
$\tilde \sigma_j^{\pm} |m\rangle = |m \pm 1\rangle$, one can evaluate the integral in Eq.\ \ref{pert1} to obtain \cite{rev12}
\begin{eqnarray} 
U_{3a}^{(1)} &=& - ~\frac{i w(\eta_1-\eta_2)}{\hbar}~ \sum_j (\tilde \sigma_j^+  \nonumber\\
&& \times \int_0^{dT} dt e^{-2i(\eta_1-\eta_2) \lambda t/\hbar}  +{\rm h.c.}).  \label{pert2} 
\end{eqnarray} 
The integral in Eq.\ \ref{pert2} vanishes for $\lambda dT/\hbar= p \pi/2$. A similar analysis holds for $U_{3b}$. Therefore we find that at these special values of $\lambda$, the evolution of $M$ proceeds via a
$O(w^2/\lambda^2)$ term. The details of the derivation of this term are given in Appendix \ \ref{appa}, and we find its expression to be 
\begin{eqnarray} 
H_{\rm eff} &=& \frac{w^2(\eta_2-\eta_1+\eta_3-\eta_4)}{2 \lambda} \hat C, \nonumber\\
\hat C &=& \sum_j [\tilde \sigma_j^z + P_{j-1} ( \sigma_j^+ \sigma_{j+1}^- + \sigma_j^- \sigma_{j+1}^+) P_{j+2} ], \label{seceqh} 
\end{eqnarray} 
where $\tilde \sigma_j^z= P_{j-1} \sigma_j^z P_{j+1}$. This explains the slow evolution of $M$ and consequent slow thermalization of the driven chain 
at special values given by $\lambda dT/\hbar= p \pi/2$. We note that this slow thermalization depends on the exact freezing limit 
which makes $U_3$ independent of $T$; hence the slow thermalization  occurs for any $T$. This feature has no analog for single-rate drive protocols.

Next, we consider the time evolution in the presence of a finite $\delta_w$. To this end, we first consider the regime $\lambda/w, \lambda T/\hbar \gg 1$ and the case where $\lambda$ is driven with the longer pulse duration
as in Eq.~\ref{prot4}. 
In this regime, one can compute $U_4(T,dT;\{\eta_i\})$ using perturbation theory. The leading term in such a perturbative expansion leads to 
\begin{eqnarray}
U_{4}^{(0)} &=& e^{-i \lambda \sum_j \sigma_j^z t/\hbar}, \quad t\le T_{01}\label{u4zeroth} \\ 
&=&  e^{-i \lambda \sum_j \sigma_j^z (2T_{01}-t)/\hbar}, \quad T_{01} < t\le T_{02}, \nonumber
\end{eqnarray} 
where $T_{01}= T/2+\sum_{i=1,2} \eta_i dT$ and $T_{02}= T+ \sum_{i=1}^4 \eta_i dT$. The leading correction to the evolution operator can be estimated perturbatively. The contribution to the first-order term from the part of $H_3$ in Eq.\ \ref{ham3} proportional to $w$ vanishes while that proportional to $\delta_w$ yields 
\begin{eqnarray} 
U_4^{(1)} &=& -~ \frac{i\delta_w}{\hbar} ~\int_0^{T_{02}} dt U_4^{(0) \dagger}  \tilde \sigma_j^x  U_4^{(0)}.  
\end{eqnarray} 
The evaluation of $U_4^{(1)}$ has been detailed in App.\ \ref{appa}. We find that for $\lambda dT/\hbar=\pi/2$ it vanishes for any choice of the $\{\eta_i\}$ at $T=T_p^{\ast}$ given by  
\begin{eqnarray}
\lambda T_p^{\ast}/(2 \hbar) = p \pi. \label{condp1} 
\end{eqnarray} 
This in turn leads to the condition $\lambda T_p^{\ast}/(4 \pi \hbar) = p/2$ and reproduces the horizontal lines in Fig.\ \ref{fig3}(b) for $\lambda T/(4 \pi \hbar) < 6$. We note that such a slow thermalization is a consequence of small $\delta_w/\lambda$; the evolution of magnetization proceeds via ${\rm O}(\delta_w^2/\lambda^2)$ terms and is therefore slow. However, the merging of these two lines for $\lambda T/(4 \pi \hbar)>6$ is not reproduced by this perturbative analysis.

The other protocol given by Eq.\ \ref{prot5} can also be analyzed in a similar manner. For this protocol, $\lambda$ has the shorter pulse duration. Consequently, for large $\lambda/w$, the leading order term in the perturbation theory is given by
\begin{eqnarray} 
U_{5}^{(0)} &=& e^{-i \lambda \sum_j \sigma_j^z t/\hbar}, \quad t\le T_{1}\label{u5zeroth} \\ 
&=&  e^{-i \lambda \sum_j \sigma_j^z (2T_{1}-t)/\hbar}, \quad T_{1} < t \le T_{2} \nonumber\\
&=& e^{-i \lambda \sum_j \sigma_j^z (t-(2T_2-T_1))/\hbar}, \quad T_{2} < t \le T_{3}\nonumber \\ 
&=&  e^{-i \lambda \sum_j \sigma_j^z (2(T_3+T_1)-T_2-t)/\hbar}, \quad T_{3} < t\le T_{4}, \nonumber
\end{eqnarray}
where $T_j= \sum_{i=1}^j (T/4+ \eta_i dT)$. Once again, the leading order perturbative correction to the evolution operator is given by   
\begin{eqnarray} 
U_5^{(1)} &=& -~ \frac{i\delta_w}{\hbar} \int_0^{T_4} dt U_5^{(0)\dagger} \tilde \sigma_j^x  U_5^{(0)}. \label{u51st} 
\end{eqnarray} 
The evaluation of $U_5^{(1)}$ is carried out in Appendix  \ref{appa}. We find that the first-order contribution to $U_5$ vanishes for any $\{\eta_i\}$ 
for $T=T_p^{\ast}$, where 
\begin{eqnarray}
\lambda T_p^{\ast}/(4\hbar) = (p+1/2) \pi \label{condp2}
\end{eqnarray} 
which yields the condition $\lambda T_p^{\ast}/(4\pi \hbar)= (p+1/2)$ for any $\delta_w$. This reproduces the horizontal lines of slow
thermalization in Fig.\ \ref{fig3}(a) for low $\lambda T/(4 \pi \hbar)<6$ ; however, the shift of these lines to $\lambda T_p^{\ast}/(4 \pi \hbar) \simeq (p+1/4)$ for $\lambda T/(4\pi \hbar) \ge 6$ is not reproduced by this perturbative analysis. 


\section{Dipolar drives}
\label{dp1} 

In this section, we consider quasiperiodic and random dipolar drive protocols. The protocol is detailed in Sec.\ \ref{dp1a}. This is followed by Sec.\ \ref{dp1b}, where we present our numerical results. Finally, we provide semi-analytic explanations of some of our numerical results in Sec.\ \ref{dp1c}.

\begin{figure}[h]
\includegraphics[width =0.48 \linewidth]{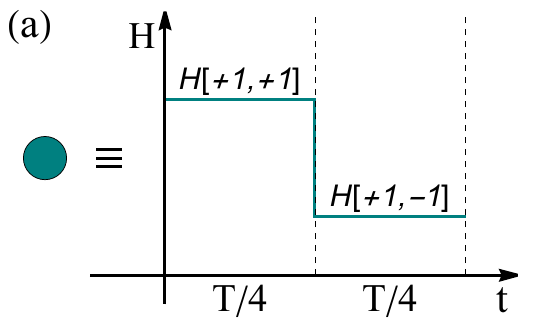}
\includegraphics[width =0.48 \linewidth]{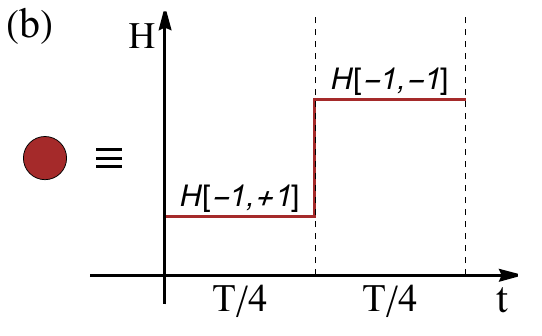}\\
\includegraphics[width =0.88 \linewidth]{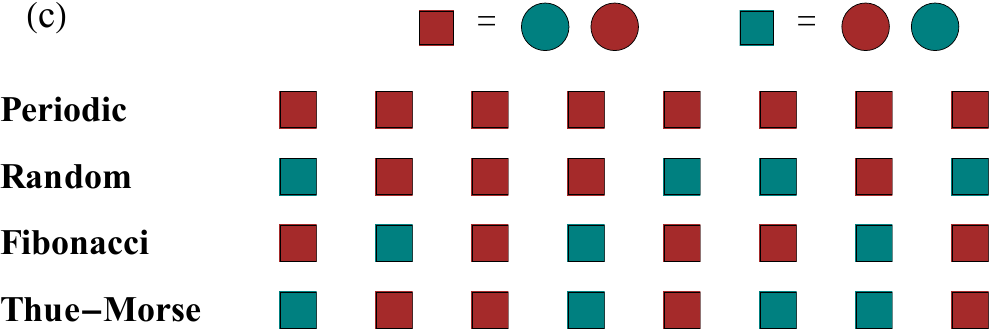}
\caption{ Schematic representation of the unitaries (a) $U_+$ denoted by green circles and (b) $U_-$ denoted by red circles. Each of these have a period of $T/2$. These are the building blocks for $T$-periodic unitaries $U_1= U_+U_-$ (red square) and $U_2= U_-U_+$ (green squares). $U_1$ and $U_2$ are used in the text to construct several dipolar drive protocols as shown in (c). See text for details. \label{figpr2}}
\end{figure}

\subsection{Protocols}
\label{dp1a}

In this section, we first chart out the protocol for random and quasiperiodic dipolar drives that we use 
to obtain the numerical results presented here. To this end, we follow Refs.\ \onlinecite{q12,q13} 
and define unitaries $U_{\pm} \equiv U_{\pm}(w,\lambda,\delta_{\lambda},\delta_{w};T)$ given by 
Eq.\ \ref{diunit}. These are schematically represented in Fig.\ \ref{figpr2}(a) and (b). We note that Eq.\ \ref{diunit} represents a protocol 
where $w$ is driven according to a longer period and that $\delta_w=\delta_{\lambda}=0$ corresponds 
to $U_{+} U_{-}= I= U_- U_+$, where the dynamics freezes for any initial state. Using these unitaries, we construct the dipoles 
\begin{eqnarray} 
U_1 &=& U_{+} U_-, \quad U_2=U_- U_+  \label{dipeq1} 
\end{eqnarray} 
The random and quasiperiodic dipolar drives correspond to the application of random and quasiperiodic sequences of these 
dipolar unitaries on any initial state; in contrast, a periodic drive, with periodic $T$, corresponds to the repeated application
of only $U_1$ or only $U_2$ on the state. These are schematically shown in Fig.\ \ref{figpr2}(c). The final state after the application of $m$ such unitaries is given by 
\begin{eqnarray} 
|\psi_m\rangle &=& U_{s_m} U_{s_{m-1}} \cdots U_{s_1} |\psi_0\rangle, \label{dipeq2}
\end{eqnarray} 
where $|\psi_0\rangle$ is the initial state (taken to be
the state with all spins down), and $s_i \in\{1,2\}$. The periodic protocol corresponds to 
$s_i=1$ or $s_i=2$ for all $i$, while the random protocol constitutes a random sequence of $s_i=1$ and $2$.

\begin{figure}[h]
\includegraphics[width=0.48 \linewidth]{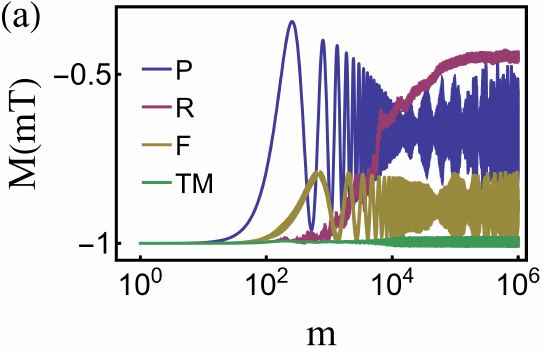}
\includegraphics[width=0.48 \linewidth]{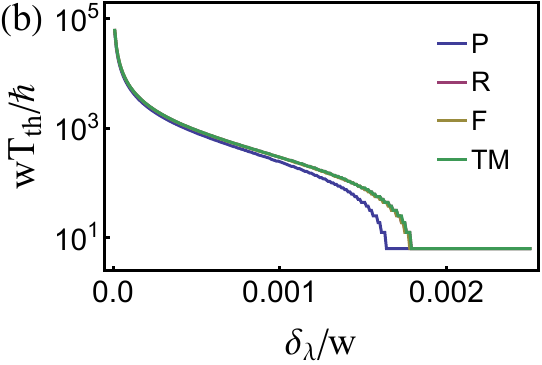}\\
\includegraphics[width=0.48 \linewidth]{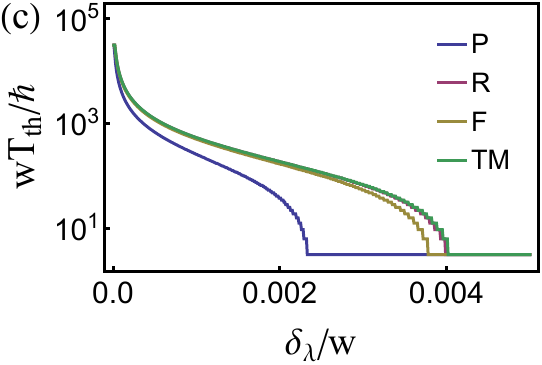}
\includegraphics[width=0.48 \linewidth]{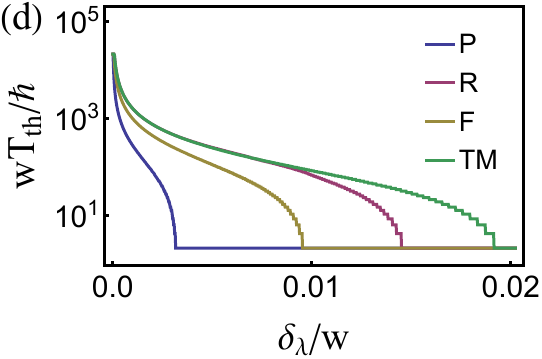}
\caption{ (a) Plot of the magnetization $M(mT)$ as a function of $m$ for several protocols for $wT/\hbar=\pi/2$. Plots of $w T_{\rm th}/\hbar$ as a function of $\delta_{\lambda}$ for $\lambda=w$, $\delta_w=0$, and (b) $\hbar \omega_D/w=1$ (c) $\hbar \omega_D/w=2$, and (d) $\hbar \omega_D/w=3$. For all plots $\lambda=w=1$. The symbols
P, R, F and TM denote periodic, random, Fibonacci and TM dipolar drives respectively.} \label{fig4} \end{figure}

The structure of $s_i$ for quasiperiodic drive depends on the sequence chosen; in this work, we shall study 
two such sequences, namely, TM and Fibonacci. For the TM sequence $s_i$ satisfies 
$s_1 =1$ and $s_{j+1}= {\bar s}_j s_j$, where ${\bar s}_j= 2(1)$ for $s_j= 1(2)$. Thus the unitaries for the first few $m$ read as
\begin{eqnarray}
U^1 &=& U_1, \quad U^2= U_2 U_1, \quad U^3= U_1 U_2 U_2 U_1, \non \\
U^4 &=& U_2 U_1 U_1 U_2 U_1 U_2 U_2 U_1, \label{tms}
\end{eqnarray}
and so on. For the Fibonacci sequence, at any level $j$, $s_j$ is constructed from 
elements of levels $j-1$ and $j-2$ with $s_1=1$ and $s_2=2$ chosen as $U^1= U_1$ and $U^2= U_1 U_2$. 
For $j \ge 3$, we define $s_j = s_{j-2} s_{j-1}$. Thus 
we have
\begin{eqnarray} 
U^1 &=& U_1, ~~U^2 = U_2 U_1, ~~U^{3} = U_1 U_2 U_1, 
\non \\
U^{4} &=& U_2 U_1 U_1 U_2 U_1, ~~U^5 = U_1 U_2 U_1 U_2 U_1 U_1 U_2 U_1,
\label{fs} 
\end{eqnarray}
and so on. In what follows, we shall compute the expectation value of $M$ (Eq.\ \ref{mag1}) under such a drive,
and we will define the thermalization time $T_{\rm th}= m_0 T$ to be the {\it least number of cycles $m_0$} after which 
$\Delta M = |M(T_{\rm th})/M(0)-1| \sim 0.05 M(0)$.

\subsection{Results} 
\label{dp1b} 

The results obtained from our numerics are shown in Fig.\ \ref{fig4}. Fig.\ \ref{fig4}(a) shows the behavior of magnetization as a function of number of drive cycles $m$ for a fixed $wT/\hbar=\pi/2$, $\lambda=w=1$, and $\delta_{\lambda}/w=0.005$. 
This plot illustrates that the magnetization changes much
more slowly for the TM protocol compared to the periodic protocol, while the
results for the random and Fibonacci protocols lie in between.
To illustrate this further, plots of the thermalization time $T_{\rm th}/T$ is shown in Figs.\ \ref{fig4}(b), (c) and (d) as a function of $\delta_{\lambda}/w$ for several representative values of $\hbar \omega_D/w$ and for $\lambda/w=1$ and $\delta_w=0$. The thermalization time $T_{\rm th}$ diverges as $\delta_{\lambda}/w \to 0$; this is a signature of the special point at which $U_1=U_2=I$ leading to a complete freezing of dynamics. Near this point, the thermalization time $T_{\rm th}/T$
can be quite long; it takes several hundred drive cycles or more for the magnetization to deviate by $5\%$ from its initial value. This behavior is present only for the
two-rate drive where exact dynamic freezing occurs for $\delta_{\lambda}=0$; it has no analog for single-rate drive protocols studied in the literature earlier \cite{q1,q2,q3,q4,q5,q6,q7,q8,q9,q10,q11,q12,q13,q14,q15,q16,q17,q18,q19,q20}. We note that this amount to an effective suppression of heating in such driven systems. 

\begin{figure}[H]
\centering
\includegraphics[width=0.49 \linewidth]{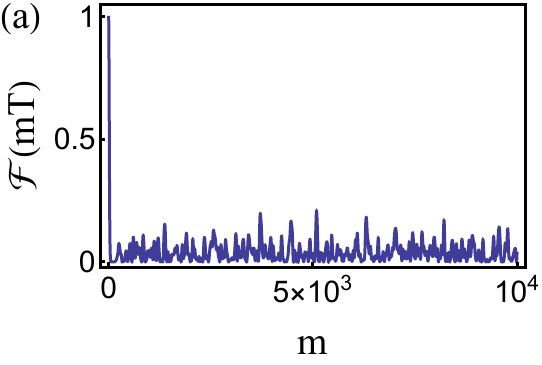}\hfill
\includegraphics[width=0.49 \linewidth]{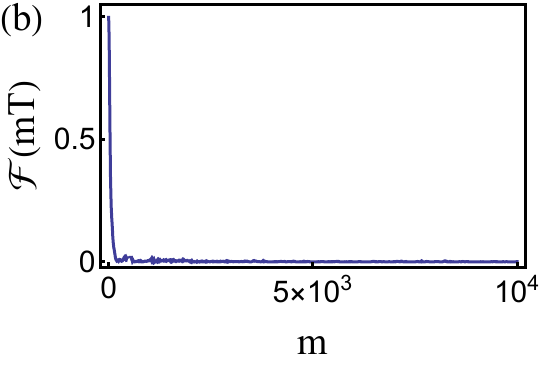}\hfill
\includegraphics[width=0.49 \linewidth]{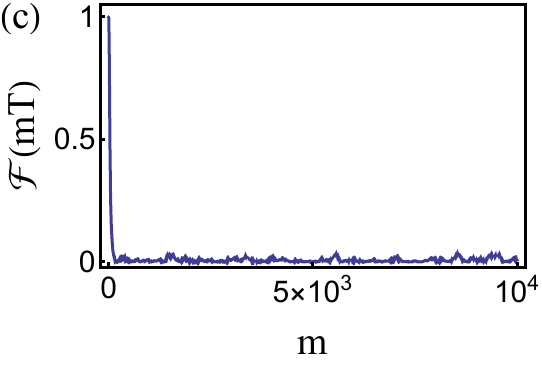}\hfill
\includegraphics[width=0.49 \linewidth]{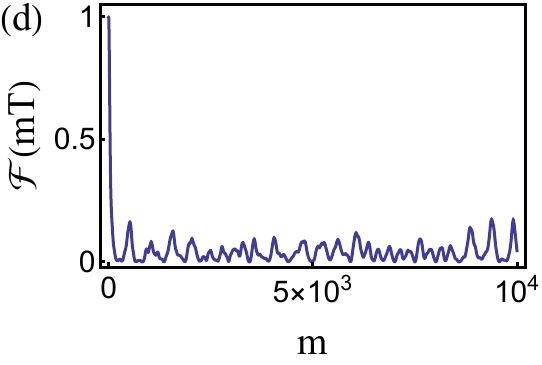}\hfill
\caption{Plots of ${\mathcal F}(mT)$ as a function of number of drive cycles $m$ for $\delta_{\lambda}/w=0.01$, $\delta_w=0$, $\lambda/w=1$, and $wT/\hbar=\pi$ for 
(a) periodic protocol with only $U_1$ acting on the state, (b) random protocol, (c) quasiperiodic protocol using a Fibonacci sequence and (d) quasiperiodic protocol using a TM sequence. For all plots $\lambda=w$. \label{fig5}}
\end{figure}

More interestingly, as confirmed by the plots shown in Fig.\ \ref{fig4}, the thermalization of such driven systems seems to be fastest for the periodic drive, slower for the random
and Fibonacci drives, and slowest for the TM protocol. This behavior is in complete contrast to the one found in earlier literature where quasiperiodic dynamics with single drive frequency has been studied; for example, it was found in Refs.\ \onlinecite{q12,q13} that periodic (random) drive protocols leads to slowest (fastest) heating. The contrast between the different protocols become more significant as the drive frequency $\omega_D$ is increased. This behavior is also most prominent in an intermediate range of $0.005 \le \delta_{\lambda}/w \le 0.02$. For smaller $\delta_{\lambda}/w\le 0.002$, all protocols lead to a divergent thermalization time due to the exact freezing of dynamics; in contrast, for larger $\delta_{\lambda} /w>0.02$, the system rapidly thermalizes for any drive protocol within the first few drive cycles.

\begin{figure}[htb]
\centering
\includegraphics[width=0.49 \linewidth]{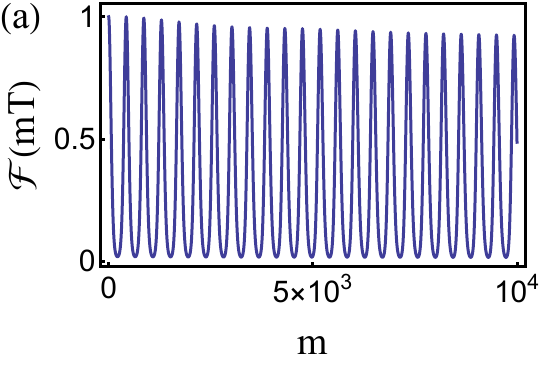}\hfill
\includegraphics[width=0.49 \linewidth]{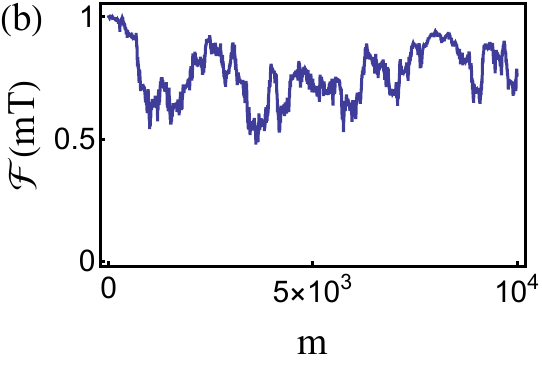}\hfill
\includegraphics[width=0.49 \linewidth]{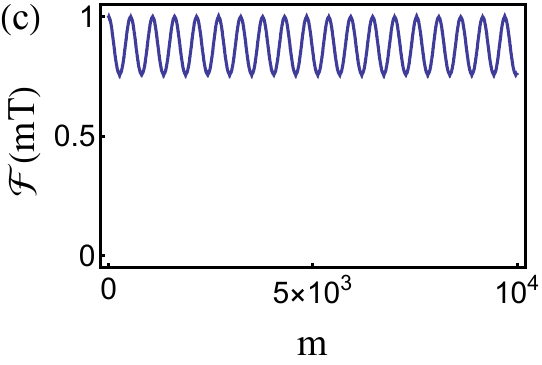}\hfill
\includegraphics[width=0.49 \linewidth]{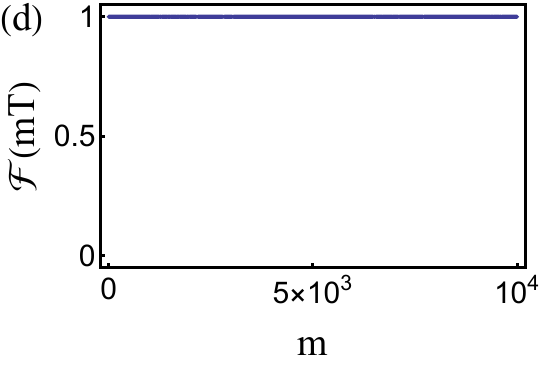}\hfill
\caption{Plots of ${\mathcal F}(mT)$ as a function of number of drive cycles $m$ for $wT/\hbar=\pi/4$ for (a) periodic protocol with only $U_1$ acting on the state, (b) random protocol, (c) quasiperiodic protocol using Fibonacci sequence and (d) quasiperiodic protocol using TM sequence. All other parameters are same as in Fig.\ \ref{fig5}.} \label{fig6}
\end{figure}

\begin{figure}[htb]
\centering
\includegraphics[width=0.49 \linewidth]{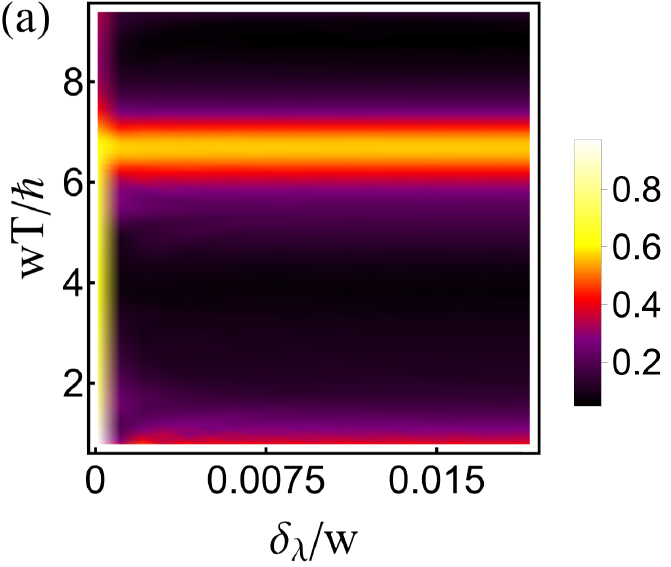}\hfill
\includegraphics[width=0.49 \linewidth]{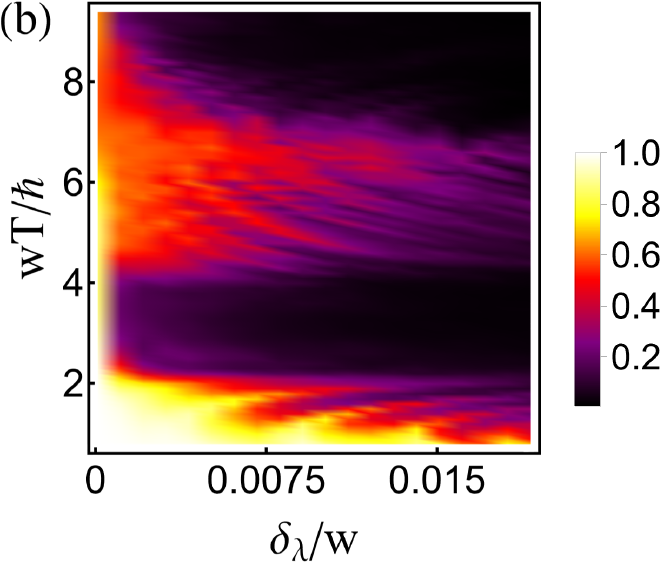}\hfill
\includegraphics[width=0.49 \linewidth]{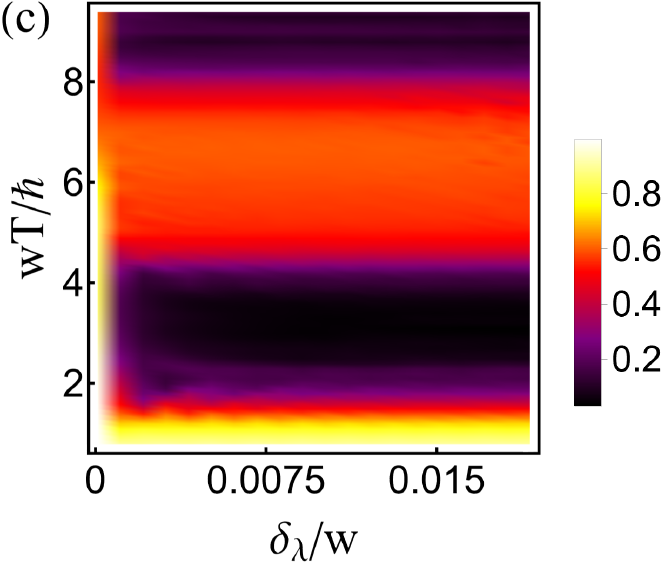}\hfill
\includegraphics[width=0.49 \linewidth]{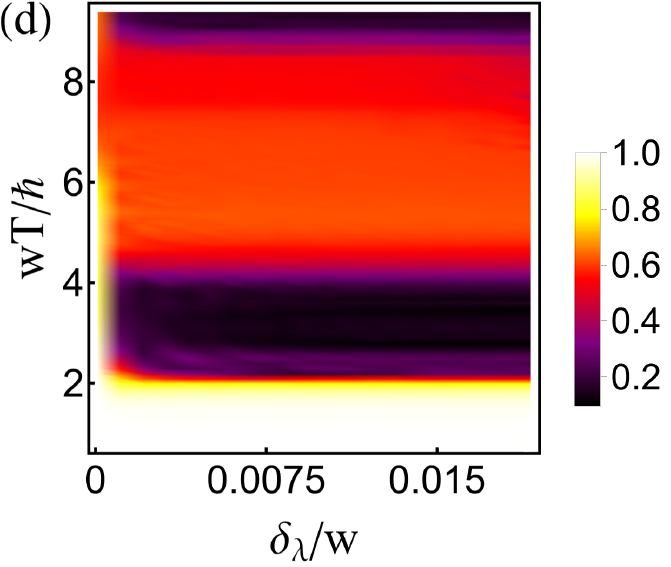}\hfill
\caption{Plots of ${\mathcal F}_{\rm av}$ as a function of $\delta_{\lambda}/w$ and $wT/\hbar$ for $\delta_w=0$ and $\lambda/w=1$ for 
(a) periodic protocol with only $U_1$ acting on the state, (b) random protocol, (c) quasiperiodic protocol using Fibonacci sequence, and (d) quasiperiodic protocol using TM sequence. For all plots $\lambda=w$. \label{fig7}}
\end{figure}

Next, we study the fidelity of the driven state defined as 
\begin{eqnarray} 
{\mathcal F}(mT) &=&  |\langle \psi(mT)|\psi(0)\rangle|^2 \label{fid1} 
\end{eqnarray} 
A representative plot of this shown for different drive protocols is shown in Fig.\ \ref{fig5} for $\hbar \omega_D/w=2$ and $\delta_{\lambda}=0.01$. We find that for the periodic, random, Fibonacci and TM protocols the driven state shows fast thermalization. This behavior is indicative of expected rapid thermalization for any protocol in a non-integrable driven model. In contrast, at a higher drive frequency $\hbar \omega_D/w=8$, such a thermalization seems to be completely absent or at least 
severely suppressed for the Fibonacci and TM protocols as can be seen in Fig. \ref{fig6}. This feature, which has no analogue for single-rate protocols, is particularly striking for the TM sequence where ${\mathcal F}\sim 1$ for $m_0\le 10000$ (Fig.\ \ref{fig6}(d)). 

A more detailed study of the time-averaged fidelity defined as 
\begin{eqnarray} 
{\mathcal F}_{\rm av} &=& \sum_{m_0=1}^{2500} {\mathcal F}(mT)/2500 \label{faveq} 
\end{eqnarray} 
also displays this suppression. This is shown in Fig.\ \ref{fig7} where $F_{\rm av}$ is plotted as a function of $\delta_{\lambda}//w$ and $\hbar \omega_D/w$. We find that the thermalization is generically suppressed for all protocols due to proximity to the dynamic freezing point at $\delta_{\lambda}=0$. This is evident from the fact that ${\mathcal F}_{\rm av}$ remain finite after a large number of drive cycles; this anomalous behavior is a signature of the exact freezing at $\delta_{\lambda}=\delta_w=0$ where $U_1=U_2=I$ and has no analog for the single-rate protocol. For all protocols, we find an enhanced averaged fidelity around $wT/\hbar=2 p \pi$ ($p \in Z$); the $p=1$ regime is shown in Fig.\ \ref{fig7}. The extension of such enhanced fidelity is broadest (in $wT/\hbar$) for the TM quasiperiodic protocol and narrowest for the periodic protocol. For all periodic and quasiperiodic protocols, this regime is almost independent of $\delta_w/\lambda$; in contrast, for the random drive protocol, this enhancement only occurs at low $\delta/\lambda$. Moreover, the averaged fidelity stays close to unity at low $wT/\hbar$ for the TM sequence (Fig.\ \ref{fig7}(d)) for all $\delta_w/\lambda$; this feature is not found for other protocols in any parameter regime. We shall explain these features in more detail in the next section.

\subsection{Perturbative analysis}
\label{dp1c}

In this section, we provide a semi-analytic understanding of two of our main results, namely a slower rate of thermalization for the random and quasiperiodic sequences drives compared to the periodic one and the near-unity value of ${\mathcal F}$ for the quasiperiodic TM protocol. We show that both of these properties can be understood by analyzing the structure of $U_1$ and $U_2$. We note from Eq.\ \ref{diunit} 
that for $\delta_w=0$, 
\begin{eqnarray} 
&& U_+(w,\delta_{\lambda};T) \equiv U_+(w,\lambda,\delta_{\lambda},0;T) \nonumber\\
&& = U_-(-w,\lambda,\delta_{\lambda},0;T) \equiv U_-(-w,\delta_{\lambda};T)
\end{eqnarray} 
for all $\lambda$. Thus, as can be seen from Eqs.\ \ref{diunit} and \ref{dipeq1}, for the protocol where $w$ is driven with the longer period we have
$U_1(w,\delta_{\lambda};T)= U_2(-w,\delta_{\lambda};T)$. Moreover, we note that $\delta_{\lambda}=0$, $U_1(w,0;T)= U_2(w,0;T)= I$ for any $w$ and $T$ since this corresponds to exact dynamic freezing. Using these properties, we find that a consecutive application of two unitaries $U_1$ and $U_2$ leads to an evolution operator $U_{12}(w,\delta_{\lambda};T)$, for a time period of $2T$, given by 
\begin{eqnarray}
U_{12}(w,\delta_{\lambda};T) &=& U_1(w,\delta_{\lambda};T) U_2(w,\delta_{\lambda};T) \nonumber\\
&=& U_1(w,\delta_{\lambda};T) U_1(-w,\delta_{\lambda};T) \label{uprop} 
\end{eqnarray} 
Defining $H_{\rm eff}$ via the relation $U_{12}(w,\lambda; T) = \exp[- i H_{\rm eff}(w,\lambda; T) 2T/\hbar]$, we first note that 
since at the exact freezing point $H_{\rm eff}$ vanishes, it admits a perturbative expansion in $\delta \lambda$ for any $w$ and $T$. 
Moreover, from Eq.\ \ref{uprop}, we find that the linear term of $H_{\rm eff}$ in such an expansion 
must be an even function of $w$. Thus for $\delta_{\lambda}/w, w T \ll 1$, one can write
\begin{eqnarray} 
H_{\rm eff} &=& \delta_{\lambda} \sum_j \sigma_j^z  + \delta_{\lambda} w^2 T^2 \sum_{j_1 j_2 j_3} c_{j_1 j_2 j_3} \tilde \sigma_{j_1}^x   \tilde \sigma_{j_2}^x \sigma_{j_3}^z + \cdots,  \nonumber\\ \label{heffpert}
\end{eqnarray}
where the ellipsis indicates higher order powers of $\delta_{\lambda}/w$ and $w T$, and the coefficients $c_{j_1 j_2 j_3}$ tend to a constant or zero in the high drive frequency limit as shown explicitly below for special cases. Importantly there can be no term in $H_{\rm eff}$ which is proportional to $\delta_{\lambda} w T$. A similar
argument works for $U_{31}(w,\delta_{\lambda};T) = U_2(w,\delta_{\lambda};T) U_1(w,\delta_{\lambda};T)$.

The first term in $H_{\rm eff}$ in Eq. \ref{heffpert}, acting on the Fock state $|\psi_0\rangle$, merely provides a phase since $|\psi_0\rangle$ is an eigenstate of $\sum_j \sigma_j^z$. Thus an application of either $U_1 U_2$ 
or $U_2 U_1$ leads to suppression of thermalization by at least a factor of $wT$ compared to the periodic protocol where only $U_1$ acts on the state allowing for a ${\rm O}(w \delta_{\lambda} T)$ term in $H_{\rm eff}$. Also, combinations of $U_1 U_2$ and $U_2 U_1$ appear most frequently in a quasiperiodic protocol with a TM sequence; this naturally leads to the slowest thermalization for such a protocol. This effect becomes stronger with increasing drive frequency and leads to a near unity fidelity of the driven state for the TM quasiperiodic drive where only opposite pairs like $U_1 U_2$ and $U_2 U_1$ appear. The effect is less drastic for Fibonacci and random sequences and is absent for periodic drives where evolution also occurs due to the actions of single $U_1$'s and $U_2$'s.

A more detailed argument is presented in Appendix \ref{appb}
showing that the effective lengths of sequences which
do {\it not} contain either $U_1 U_2$ or $U_2 U_1$ 
is the longest for the periodic protocol, shorter for
the random and Fibonacci protocols, and the shortest for
the TM sequence. Since the sequences $U_1 U_2$ and $U_2 U_1$
contribute the least to thermalization due to the smallness
of their off-diagonal elements, this explains
qualitatively why the rate of thermalization is largest
for the periodic protocol, smaller for the random and
Fibonacci protocols, and smallest for the TM protocol.

As an explicit demonstration of the slower thermalization for the quasiperiodic protocol, we carry out an exact analytical calculation of the evolution operator $U_1 U_2$ for small system of size $L=3$ using periodic boundary condition. 
For such a chain, the Hilbert space, in the total momentum $K=0$ sector, consists of two states given by \cite{bm2_part1}
\begin{eqnarray}
|\psi_1\rangle &=& |\downarrow, \downarrow, \downarrow\rangle, \label{states1} \\
|\psi_2\rangle &=& \frac{1}{\sqrt{3}} \left ( |\uparrow, \downarrow, \downarrow\rangle+ |\downarrow, \uparrow, \downarrow\rangle + |\downarrow, \downarrow, \uparrow\rangle \right). \nonumber
\end{eqnarray} 
In the space of these two states, the effective Hamiltonian is given by Eq.\ \ref{states1} and can be written, for $\delta_{\lambda}\ne 0$, as 
\begin{eqnarray} 
H_r(a,b) &=& \sqrt{3}a w \tau^x + \left(2 I + \tau^z \right) (\lambda b +\delta_{\lambda}), \label{heff1} 
\end{eqnarray} 
where $I$ denotes the $2 \times 2$ identity matrix and $\tau^x$ and $\tau^z$ are Pauli matrices acting on the space of the state $|\psi_{1,2}\rangle$. 
The corresponding evolution operators $U_{r \pm}$ can be constructed using Eq.\ \ref{diunit} and are given by
\begin{eqnarray} 
U_{r \pm} &=& e^{-i H_r(\pm 1,-1) T/(4\hbar)} e^{-i H_r(\pm 1,1) T/(4\hbar)}, \nonumber\\
U_{r 1} &=& U_{r+} U_{r -}, \quad U_{r 2}= U_{r-} U_{r +}.  \label{ueff1}
\end{eqnarray}
A straightforward expansion of $U'(2T)= U_{r1} U_{r2}$ in $\delta_{\lambda}/w$ and $w T/\hbar$ yields $H_{\rm eff} = A' \tau_z + B'\tau_x + C'\tau_y$, where 
\begin{eqnarray}
A' &=& \delta_{\lambda}(-1 + w^2 T^2/(2\hbar^2) + \cdots) 
+ {\rm O}(\delta_{\lambda}^2), \nonumber\\
B',C' &=&  {\rm O}(\delta_{\lambda}^2), \label{qptstate}
\end{eqnarray}
where the ellipsis indicates higher order terms in $wT/\hbar$. Note that all $O(wT/\hbar)$ terms in $H'_{\rm eff}$ are absent to the leading order in $\delta_{\lambda}$. In contrast, an exactly similar expansion of $U'_r(2T,0)= U_{r 1}(w) U_{r 1}(w)$ yields an effective Hamiltonian $H'_{\rm eff}= A" \tau_z + B" \tau_x + C" \tau_y$ where
\begin{eqnarray}
A" &=& \delta_{\lambda}(-1 + w^2 T^2/(2\hbar^2) + \cdots) 
+ {\rm O}(\delta_{\lambda}^2), \label{ptstate}\\
C" &=& -\sqrt{3} w T \delta_{\lambda}/(2 \hbar) + \cdots, \non \\
B" &=& \delta_{\lambda} \sqrt{3} w \lambda T^2/(8 \hbar^2) + \cdots, \nonumber
\end{eqnarray}
showing the presence of $O(w T/\hbar)$ terms to leading order in $\delta_{\lambda}$. This clearly implies a much faster thermalization for the case of $U_{r 1} U_{r 1}$ compared
to $U_{r 1} U_{r 2}$. Further, for small $T$, the absence of a linear term in $\delta_{\lambda}$ is expected to lead to ${\mathcal F}$ staying close to unity for a long time $T_0$, where $T_0 \sim 1/(\delta_{\lambda}^2 T^2)$. This is to be contrasted with the case for the periodic protocol where  $U"(2T)= U_{r 1}(w) U_{r 1}(w)$  yields a $T_0$ which is ${\rm O}(1/(\delta_{\lambda} T))$ and is hence an order of magnitude lower than its counterpart for the quasiperiodic TM sequence. 

The band of high fidelity for the periodic protocol around $wT/\hbar=2 \pi$ can also be explained using a similar analysis. For this, we first expand $U_{r1}$ in powers of $\delta_{\lambda}/w$ for arbitrary $\lambda/w$ and $wT/\hbar$ and obtain the corresponding Floquet Hamiltonian. A straightforward computation yields $H_F= A_1 \tau_z + A_2 \tau_x + A_3 \tau_y$, where 
\begin{widetext} 
\begin{eqnarray} 
A_1 &=& \frac{\delta_{\lambda}}{x_0^6} \left[(\lambda T/\hbar)^2 + 3(wT/\hbar)^2 \cos (x_0/2) \right] \left[ (\lambda T/\hbar)^2 x_0^4  + 6  (w T/\hbar)^2 x_0 \sin (x_0/2) \right], \nonumber\\
A_2 &=& \frac{2 \sqrt{3} w T \delta_{\lambda}}{\hbar x_0^6} \sin (x_0/4) \left[ (\lambda T/\hbar)^2 x_0^4 + 6(w T/\hbar)^2 x_0 \sin (x_0/2) \right] \left[ x_0 \cos (x_0/4) + 
(\lambda T/\hbar) \sin (x_0/4) \right], \nonumber \\
A_3 &=& {\rm O}(\delta_{\lambda}^2), \label{peq1} 
\end{eqnarray} 
\end{widetext}
where $x_0 = T \sqrt{\lambda^2+3 w^2}/\hbar$. This shows that
for $\lambda=w=1$, all ${\rm O}(\delta_{\lambda})$ terms in $A_2$ vanish at $wT/\hbar= 2 p \pi$ (or equivalently for $x_0=4 p \pi$ since $w=\lambda=1$); this causes a slower thermalization around $wT/\hbar=2 \pi$ as seen in Fig.\ \ref{fig4}(a). In contrast, for the TM protocol which consists entirely of sequences of $U_{r1} U_{r2}$ and $U_{r2} U_{r1}$, the off-diagonal elements of $U_{r1} U_{r2}$ and $U_{r2} U_{r1}$ are ${\rm O}(\delta_{\lambda}^2)$ for all T; this results in a slower thermalization for a much wider band of $wT/\hbar$ as see in Fig.\ \ref{fig7}(d). The random protocol consists
of sequences of $U_{r1} U_{r2}$, $U_{r2} U_{r1}$, and single $U_{r1}$'s and $U_{r2}$'s, while the
Fibonacci protocol consists of $U_{r1} U_{r2}$ and single 
$U_{r1}$ s. Hence these two protocols show a much narrower band of slow thermalization around $w T/\hbar =2 \pi$ compared to the TM protocol. A more detailed discussion of this argument comparing the four protocols is presented
in Appendix \ref{appb}.
     
\section{Discussion}
\label{diss1} 

In this work, we have studied the properties of a PXP chain driven with  two-rate random and quasiperiodic protocols. Our study points out several novel features of such drives which are absent in their single-rate counterparts.  
Our main results can be summarized as follows.

First, we recall that when the PXP model
is driven by a sequence of square-pulses with four
specific Hamiltonians with equal time durations given by
$T/4$ each, the Floquet flat bands can be made exactly flat. This is the exact dynamic freezing limit. Each of these Hamiltonian
consists of a PXP term with coefficient $\pm w$ and a magnetic field with strength $\pm \lambda$ applied in the 
$\hat z$-direction; together, they yield the protocol given by Eq.\ \ref{prot2}. We then
study what happens if the time durations of these
four Hamiltonians differ
from $T/4$ by a small amount given by $dT$ times some
randomly chosen numbers $\eta_i = \pm 1$. We find that in this case the driven system is no longer frozen. Starting from 
an initial state in which all spins point down, we 
study the magnetization as a function of the number of such unitaries applied to the PXP chain. We find that the magnetization generally
deviates rapidly from its initial value of $-1$ {\it except} at
special values of $\lambda$ and $dT$ where $\lambda dT/\hbar$ is equal to $\pi/2$ times an integer $p$.
We quantify this effect by
defining a thermalization time $T_{\rm th}$ as the
time required for the magnetization to differ from $-1$
by a given small value. We then find that $T_{\rm th}$
is large only in the neighborhoods of the special values
of $\lambda dT$ mentioned above. We note that such a slow thermalization 
does not have any analog for single-rate aperiodic drives.

Next, we study what happens if $w$ is changed by a
small amount $\delta_w$ keeping the pulse durations random. We discover that $T_{\rm th}$ has
a rather intricate dependence on the two parameters $\lambda T/\hbar$ and $\delta_w/w$.
There are lines in the parameter space where $T_{\rm th}$
is very large; these lines lie along different values
of $\lambda T/\hbar$ depending on whether $\lambda$ has a
shorter time period ($T/2$) or a longer time period
($T$), both with some randomness. In the former case
the lines appear singly while in the latter case they
appear in pairs with a splitting; the values of
$T=T_p^{\ast}$ where the lines appear are also
different in the two cases. We have provided a
semi-analytical understanding of many of the above 
mentioned features using a Floquet perturbation theory
which is valid when $\lambda, \hbar \omega_D \gg w$. Our perturbative 
analysis which is valid for $\lambda T/(4\pi \hbar) \sim 1$ shows that 
such lines occur at $T=T_p^{\ast}$, where $\lambda T_p^{\ast}/(4 \pi \hbar) 
= (p+1/2)$ in the former case and $\lambda T_p^{\ast}/(4 \pi \hbar) = p/2$ 
in the latter case. For $\lambda T/(4\pi \hbar) \ge 6$, $T_p^{\ast}$, for both cases, is given  by $\lambda T_p^{\ast}/(4\pi \hbar) \simeq 
(p+1/4)$. This regime is not captured by our perturbative analysis. 

Next, we have studied the effect of dipolar drives where
we change $\lambda$ by a small amount $\delta_\lambda$, 
and set $dT = 0$ so that there is no
randomness in time. We consider two unitaries $U_+$ and
$U_-$ in which $w$ has opposite signs, and $\lambda$ has
a shorter time period of $T/2$. These are 
combined to form two dipole unitaries $U_1 = U_+ U_-$
and $U_2 = U_- U_+$. We then consider four different
protocols in which the dipole $U_1$ is repeated 
periodically, or $U_1$ and $U_2$ are combined to form 
either a random, or a quasiperiodic Fibonacci, or a quasiperiodic TM 
sequence. We study the fidelity (defined as the square of the
overlap between the initial state with all spins down 
and the state obtained after $m$ drives as a function of 
$m$. Remarkably, we find that the deviation of the 
fidelity from unity grows most rapidly for the periodic protocol; this is an
indication that the thermalization time would be shortest
for this protocol. The deviation of the fidelity from unity grows less rapidly
for the random and Fibonacci protocols, and 
least rapidly for the TM sequence. These unusual features
become more pronounced for larger values of $wT/\hbar$.
We note that this is in sharp contrast to what happens
for a single-rate drive protocol where periodic protocols
lead to a slower rate of thermalization than random and
quasiperiodic protocols~\cite{q13,q14}. 

We provide a qualitative understanding of
the above-mentioned phenomena by carrying out a perturbative analysis. We show that for $wT/\hbar \ll 1$ the leading terms of $U_1 U_2$
and $U_2 U_1$ which are instrumental for having finite transition matrix elements to other states are much smaller than those of single 
$U_1$'s and $U_2$'s when $wT$. Since these matrix elements are responsible for thermalization, this implies that 
for the same number of drives $m$, protocols
which contain relatively fewer sequences of $U_1 U_2$
and $U_2 U_1$ compared to single $U_1$'s and $U_2$'s
will thermalize faster. We also demonstrate this feature using an analytic calculation for a $L=3$ periodic PXP chain. 
We then note that the periodic protocol has no sequences of $U_1 U_2$ and $U_2 U_1$, random and Fibonacci
sequences have a significant number of them,
and the TM protocol is entirely made up of such sequences.
These observations can explain the relative rates of thermalization
for the four protocols.

Experimental verification of our results requires a driven Rydberg chain in the PXP limit. In this limit, the van der Waals interactions between the neighboring atoms are strong enough to  preclude Rydberg excitations on neighboring sites \cite{exp1,exp2,exp3,exp4,exp5,exp6,exp7}. Such a blockade can be represented, in the spin language where the up (down) spins represent the Rydberg excited (ground) states of the atoms, as a constraint of not having two neighboring up spins. Further, the detuning of the Rydberg atoms can be represented as a  magnetic field $\sim \lambda$ in the spin language, while the coupling between the ground and Rydberg excited states on any site $j$ can be mapped to  $\tilde \sigma_j^x$  with a strength $\sim w$. 

Our proposal is to drive both the detuning $\lambda$ and coupling $w$ according to random or quasiperiodic dipolar drive protocols such 
that their drive frequencies have an integer ratio. Such a drive can be used to construct unitaries $U_1$ and $U_2$; our prediction
is that arrangement of such unitaries, acting on a starting vacuum state where all Rydberg atoms are in the ground state, 
according to the TM sequence will lead to slowest thermalization. This can be quantified by a measurement of the number of Rydberg 
atoms in the ground states, $N_0$,  after the application of $m$ such unitaries; we predict $N_0(m)$ to have the slowest decay for the TM sequence and fastest for the periodic one. We note that this result will be in contrast to the single-rate drive protocol where, for example, only $\lambda$ is driven with some frequency. In the latter case, one expects a periodic 
sequence of unitaries to produce the slowest decay of $N(m_0)$ and a random 
sequence to produce the fastest decay~\cite{q13}.       

To conclude, we have studied the PXP spin chain driven by random and quasiperiodic two-rate drive protocols. Our study identifies a dynamic freezing limit for these protocols; the thermalization of the system has been found to slow down considerably near this freezing limit. Moreover, we have shown that the dynamics of such driven systems has several properties
which are qualitatively different from their single-rate counterparts studied earlier, and we have charted out concrete experimental proposals
that can test our theory. 

\vspace{0.6cm}
\centerline{\bf Acknowledgments}
\vspace{0.4cm}

D.S. thanks SERB, India for support through the project JBR/2020/000043.
K.S. thanks DST, India for support through the SERB project JCB/2021/000030. 

\appendix

\section{Effective Hamiltonian}
\label{appa} 

In this Appendix, we provide details of the analytical results used in Sec.\ \ref{analyt}. We begin with the 
case when $\delta_w=0$ and $\lambda dT=p \pi/2$, where $p$ is an integer. In this case, for the protocol given by 
Eq.\ \ref{prot4}, we have $U_2(T,dT;\{\eta_i\}) =  U_a U_b $ where 
\begin{eqnarray}
U_a &=& e^{i H_2(-1,1)(\eta_3 -\eta_4)dT/\hbar}, \nonumber\\
U_b &=& e^{i H_2 (1,1)\,(\eta_1 -\eta_2)dT/\hbar}.\label{appeq1} 
\end{eqnarray} 
To evaluate this expression, we consider the limit $\lambda \gg w$, so that one can expand $U$ in powers of $w/\lambda$. We shall carry out this analysis for $U_b$ explicitly; the results for $U_a$ are similar. To this end, we first write 
\begin{eqnarray} 
U_b(dT) &=& U_b^{(0)}(dT) + U_b^{(1)}(dT) + U_b^{(2)}(dT) +
\cdots, \nonumber\\ 
U_b^{(0)}(t) &=& \exp\left[-i (\eta_1-\eta_2) \lambda t \sum_j \sigma_j^z\right]. \label{appeq2}  
\end{eqnarray} 
We note that $U_b^{(0)}(dT)= I$ which follows the fact $\eta_1-\eta_2 =\pm 2,0$ for any $\eta_1$ and $\eta_2$ and $\lambda dT= p \pi/2$. 
This indicates that one can write 
\begin{widetext} 
\begin{eqnarray}
U_b(dT) &=& e^{-i H_{\rm eff} dT/\hbar} = I + U_b^{(1)}(dT) + U_b^{(2)}(dT) \cdots, \nonumber\\
H_{\rm eff}^{(1)} &=& \frac{i \hbar}{dT} U_b^{(1)} = \frac{1}{dT} \int_0^{dT} U_b^{(0) \dagger}(t) H_1 U_b^{(0)}(t) dt, \nonumber\\
H_{\rm eff}^{(2)} &=& \frac{i \hbar}{dT} U_b^{(2)} = \frac{1}{dT} \left[\left(\int_0^{dT} U_b^{(0) \dagger}(t_1) H_1 U_b^{(0)}(t_1) dt_1 \int_0^{t_1} U_b^{(0)\dagger}(t_2) H_1 U_b^{(0)}(t_2) dt_2 \right)  - (U_b^{(1)}(dT))^2/2 \right], \label{appeq3} 
\end{eqnarray} 
\end{widetext} 
where $H_1= w (\eta_2-\eta_1) \sum_j \tilde \sigma_j^x$. 

To evaluate these expressions, we first write $\tilde \sigma_j^x = \tilde \sigma_j^+ +\tilde \sigma_j^-$ and use the identity 
\begin{eqnarray} 
\left[ \sum_j \sigma_j^z, \sum_{j'} \tilde \sigma_{j'}^{\pm} \right] = \pm 2 \sum_{j'} \tilde \sigma_{j'}^{\pm}. \label{appeq4} 
\end{eqnarray}
This yields 
\begin{eqnarray} 
U_b^{(0) \dagger}(t) H_1 U_b^{(0)}(t) &=& w (\eta_2-\eta_1) \sum_{m=\pm} f_m(t), \nonumber\\
f_m(t) &=& e^{2 i m (\eta_2-\eta_1)\lambda t}. \label{appeq5} 
\end{eqnarray} 
Using this identity, one can evaluate $H_F^{(1)}$. We find that for $\lambda dT= p \pi/2$,  $\int_0^{dT} f_m(t) dt =0$ so that 
$H_F^{(1)}=0$. This result has been used in the main text. A similar analysis shows that $H_F^{(1)}$ also vanishes for $U_a$.
Thus there is no contribution of ${\rm O}(w/\lambda)$ terms to the evolution operator for $\lambda dT= p \pi/2$; this is a key aspect 
for slow thermalization as discussed in the main text. 

Next, we consider the second-order term. For this we need to evaluate the double integral in the last line of Eq.\ \ref{appeq3}. In terms of 
$f_m$ defined in Eq.\ \ref{appeq5}, this can be written as 
\begin{widetext} 
\begin{eqnarray}
I_2 &=& \int_0^{dT} U_b^{(0) \dagger}(t_1) H_1 U_b^{(0)}(t_1) dt_1 \int_0^{t_1} U_b^{(0)\dagger}(t_2) H_1 U_b^{(0)}(t_2) dt_2 \nonumber\\
&=& (\eta_2-\eta_1)^2 w^2 \sum_{m,n=-1}^1 \sum_{j_1} \tilde \sigma_{j_1}^m \sum_{j_2} \tilde \sigma_{j_2}^n   \int_0^{dT} f_m(t_1) dt_1  \int_0^{t_1} f_n(t_2)dt_2 =\sum_{m,n=-1}^1 I_2(m,n). \label{appeq6} 
\end{eqnarray} 
\end{widetext} 
An explicit evaluation of this integral is straightforward and one finds that $I_2(1,1)=I_2(-1,-1)=0$ while $I_2(1,-1)=-I_2(-1,1)$. Using this, we finally obtain
\begin{eqnarray} 
H_{\rm eff}^{(2)} &=& \frac{(\eta_2-\eta_1) w^2}{2 \lambda} \sum_{j_1,j_2} \left[ \tilde \sigma_{j_1}^+, \tilde \sigma_{j_2}^- \right] = \frac{(\eta_2-\eta_1)w^2}{2\lambda} \hat C,\nonumber\\ 
\hat C &=&  \sum_{j} [\tilde \sigma_j^z + P_{j-1}(\sigma_j^+ \sigma_{j+1}^- + \sigma_j^- \sigma_{j+1}^+) P_{j+2}], \label{appeq7}
\end{eqnarray} 
where in the second line we have evaluated the commutator between $\tilde \sigma_{j_1}^+$ and $\tilde \sigma_{j_2}^-$, $P_j= (1-\sigma_j^z)/2$ 
is the projector to the spin-down state on site $j$ defined in the main text, and $\tilde \si_j^z = P_{j-1} \si_j^z
P_{j+1}$. A similar analysis of $U_a$ gives a contribution $(\eta_3-\eta_4)w^2/\lambda \hat C$ to $H_{\rm eff}^{(2)}$; this leads us to the final expression of $H_{\rm eff}^{(2)}$ in Eq.\ \ref{seceqh} 
of the main text given by 
\begin{eqnarray}
H_{\rm eff}^{(2)} &=&  \frac{(\eta_2-\eta_1 + \eta_3-\eta_4)w^2}{2\lambda} ~\hat C.  \label{appeq8}
\end{eqnarray}

Next, we consider the case where the amplitude of random variation of time period, $dT$, is fixed so that $\lambda dT/\hbar =p \pi/2$. Once again, we concentrate on the perturbative
regime where $\lambda \gg w$. First we consider the protocol given by Eq.\ \ref{prot4}, where $\lambda$ is the longer time period. To analyze it perturbatively, we first consider the limit
$w=\delta_w=0$, to obtain 
\begin{eqnarray}
U_{4}^{(0)} &=& e^{-i \lambda \sum_j \sigma_j^z t/\hbar}, \quad t\le T_{01}\label{appeq9} \\ 
&=&  e^{-i \lambda \sum_j \sigma_j^z (2T_{01}-t)/\hbar}, \quad T_{01} < t\le T_{02} \nonumber
\end{eqnarray} 
which is presented as Eq.\ \ref{u4zeroth} in the main text. Here $T_{01}= T/2 + \sum_{i=1,2} \eta_i dT$, $T_{02}= T+ \sum_{i=1}^4 \eta_i dT$, where the $\eta_{i}$ are random numbers. 
The first-order contribution to the evolution operator from  $H_1= \delta_w \sum_j \tilde \sigma_j^x = \sum_{m=-1,1} \sum_j \tilde \sigma_j^m $
can be estimated to be 
\begin{eqnarray} 
U'_1(T_{02}) &=& -~ \frac{i \delta_w}{\hbar}~ \sum_{j} (\tilde \sigma_j^+ {\mathcal A} + {\rm H.c.}), \nonumber\\
{\mathcal A} &=& \int_0^{T_{01}} g_1(t) dt + \int_{T_{01}}^{T_{02}}  g_1(2T_{01}-t) dt \nonumber\\
&=& \frac{\hbar}{2\lambda} \sin\lambda T/(2\hbar) \left( e^{i\lambda T/(2\hbar)}-1 \right), \label{appeq10} 
\end{eqnarray} 
where $g_1(t) = \exp[-2 i\lambda t/\hbar]$. Here we have used the relation $\lambda dT/\hbar=p \pi/2$ and the fact that $\eta_i \in \{-1,1\}$ to obtain the expression in the last line. 
This shows that the contribution till the first order in $\delta_w$ to the evolution operator $U_4$ after $T= T_{02}$ vanishes for $T=T_p^{\ast}$ if $\sin \lambda T_p^{\ast}/(2\hbar)=0$, {\it i.e.}, if $\lambda T_p^{\ast}/(4 \pi \hbar) = p/2$ where $p \in Z$. This reproduces the
horizontal lines of slow thermalization at integer and half-integer values of $\lambda T_p^{\ast}/(4 \pi \hbar)$ as alluded to in the main text.

Next, we consider the protocol in which $w(t)$ has the longer time period. In this case, we begin with $U_{5}^{(0)}$ given by Eq.\ \ref{u5zeroth}. To 
evaluate Eq.\ \ref{u51st}, we write 
\begin{eqnarray} 
U_5^{(1)} &=& -~ \frac{i\delta_w}{\hbar} ~\int_0^{T_4} dt U_5^{(0)\dagger} \tilde \sigma_j^x  U_5^{(0)} \nonumber\\
&=& -~ \frac{i \delta_w}{\hbar}  ~\sum_j \left(\tilde \sigma_j^+ {\mathcal B} +{\rm H.c.} \right), \label{appeq11} 
\end{eqnarray} 
where the coefficient $\mathcal B$ can be expressed in terms of $g_1(t)$ defined earlier as
\begin{eqnarray} 
{\mathcal B} &=&  \int_0^{T_1} g_1(t) dt \,\, +  \int_{T_1}^{T_2} g_1(2T_1-T) dt  \label{appeq12} \\
&& + \int_{T_2}^{T_3} g_1(t-2T_2+T_1) dt \nonumber\\
&& + \int_{T_3}^{T_4} g_1(2(T_3+T_2)-T_1 -t) dt. \nonumber
\end{eqnarray} 
The evaluation of these integrals are straightforward and for $\lambda dT/\hbar= p \pi/2$, using the fact $\eta_i\in\{-1,1\}$, one gets 
\begin{eqnarray} 
{\mathcal B} &=& \left(\frac{-i \delta_w}{\lambda \hbar}\right)  e^{i \lambda T_1/\hbar} \sin \lambda T_1/\hbar \nonumber\\ 
&=& \left(\frac{-i \delta_w}{\lambda \hbar}\right) e^{i \lambda T_1/\hbar} \cos \lambda T/(4\hbar), \label{appeq13} 
\end{eqnarray} 
where we have used $\eta_1 dT \lambda = \pm p \pi/2$. Thus we find that for $T=T_p^{\ast}$, where $\cos \lambda T_p^{\ast}/(4\hbar)= 0$ or $\lambda T_p^{\ast}/(4 \pi \hbar) = (p+1/2)$, thermalization becomes slow
due to the suppression of ${\rm O}(\delta_w)$ contributions to $U_5$, as pointed out in the main text.    

\section{Effective number of drives for different dipolar protocols}
\label{appb}

In this Appendix, we will present a qualitative argument about why
the periodic protocol thermalizes the fastest, followed
by the random and Fibonacci protocols, and the TM protocol
thermalizes the slowest as shown in Sec.~\ref{dp1} and Fig.~\ref{fig6} in particular. These
arguments will be motivated the perturbative analysis in
Sec.~\ref{dp1c} that for small values of $w$, the sequences $U_1 U_2$ and $U_2 U_1$
have much smaller off-diagonal elements than $U_1$ and $U_2$
separately, and therefore contribute much less to 
thermalization compared to single $U_1$'s and $U_2$'s. 
We will consider a very long sequence of $N$
unitaries for each of the four protocols, and estimate
how much shorter the sequence becomes if we replace 
opposite pairs of $U_1 U_2$ and $U_2 U_1$ by the identity matrix
$I$ (we will ignore the diagonal phases that appear 
along $I$). The shorter the final sequence the fewer
will be the number of single $U_1$'s and $U_2$'s and
hence the slower it will thermalize.

A periodic protocol of length $N$ only contains a sequence of $ $N$ U_1$'s.
Since the opposite $U_1 U_2$ and $U_2 U_1$ do not appear 
at all, the length of the sequence remains equal to $N$.
At the other extreme, the TM sequence only contains
the pairs $U_1 U_2$ and $U_2 U_1$ (see Eq.~\ref{tms})
and therefore reduces to just $I$. We therefore expect
a periodic protocol to lead to much faster thermalization
than the TM sequence.

Next, we turn to the random protocol. We follow the rule
given above of moving across the complete sequence from 
right to left and eliminating opposite pairs like $U_1 U_2$
and $U_2 U_1$, but keeping single $U_1$'s and $U_2$'s. Given a long
sequence $S_{i,N}$ of $N$ unitaries (where $i$ goes from
1 to $2^N$), let the length of the
final sequence obtained after applying the above rules be
$M_{i,N}$ (we will ignore appearances of $I$
when counting the length). Each of the $S_{i,N}$'s appear with probability $1/2^N$.
We will therefore be interested in the average length of
the final sequence which is given by
\beq A_N ~=~ \frac{1}{2^N} ~\sum_{i=1}^{2^N} ~M_{i,N}. 
\label{an1} \eeq
For instance, for $N=2$, there are 4 possible initial
sequences given by $U_1 U_1$, $U_1 U_2$, $U_2 U_1$ and
$U_2 U_2$, which reduce to $U_1 U_1$, $I$, $I$ and 
$U_2 U_2$ respectively, with lengths equal to 2, 0, 0 and 2.
Hence $A_2 = 1$. For $N=3$, there are 8 possible initial
sequences given by $U_1 U_1 U_1$, $U_1 U_1 U_2$, $U_1 U_2 U_1$, $U_1 U_2 U_2$, $U_2 U_1 U_1$, $U_2 U_1 U_2$, $U_2 U_2 U_1$ and $U_2 U_2 U_2$, which reduce to $U_1 U_1 U_1$,
$U_1$, $U_1$, $U_2$, $U_1$, $U_2$, $U_2$ and $U_2 U_2 U_2$.
Hence $A_3 = 3/2$.

We will now derive a recursion relation for $A_N$. 
The first two unitaries in a sequence $S_{i,N}$ can be either $U_1 U_1$, $U_1 U_2$, $U_2 U_1$ or $U_2 U_2$,
with probability $1/4$ each. If it begins with $U_1 U_1$,
they cannot be reduced; we then keep the first $U_1$ as
it is and consider the remaining unitaries as
forming a sequence $S_{i,N-1}$. The final sequence of
this will therefore have length $M_{i,N} = 1 + M_{i,N-1}$. A similar argument
applies to a sequence $S_{i,N}$ which begins with $U_2 U_2$.
However, if a $S_{i,N}$ begins with $U_1 U_2$ or $U_2 U_1$,
they get reduced to $I$ and we are then left with a sequence
$S_{i,N-2}$ with final length $M_{i,N-2}$. Averaging over
all the $M_{i,N}$'s, we conclude that the average final
length $A_N$ satisfies the recursion relation
\beq A_N ~=~ \frac{1}{2} ~(1 ~+~ A_{N-1}) ~+~ \frac{1}{2} ~A_{N-2}. \label{an2} \eeq
Using the initial values $A_2 = 1$ and $A_3 = 3/2$, we then
find the analytical expression
\beq A_N ~=~ \frac{N}{3} ~+~ \frac{4}{9} ~[ 1 ~-~ (- \frac{1}{2})^N]. \label{an3} \eeq
For $N \to \infty$, we conclude that on the average the length of the final sequence will be $1/3$ of the initial length.

Finally, we turn to the Fibonacci protocol. An inspection
of the discussion around Eq.~\ref{fs} shows that,
reading from right to left, the different sequences only contain sequences of $U_1$'s or pairs of $U_1 U_2$. Namely,
sequences with more than one $U_2$ do not appear. Hence,
eliminating the pairs $U_1 U_2$ shortens the sequence
by an amount which is exactly equal to twice the number of
$U_2$'s that are present originally, and further, the
final sequence will only have a number of $U_1$'s.
We now introduce the Fibonacci numbers $F_K$ which
satisfy the recursion relation $F_K = F_{K-1} + F_{K-2}$
and start with $F_1 = 1$ and
$F_2 = 2$. We then see from the discussion of Eq.~\ref{fs} that
a sequence $U^K$ (where $K = 1, 2, 3, \cdots$) contains
$F_K$ unitaries, of which $F_{K-1}$ are $U_1$'s and $F_{K-2}$ are $U_2$'s. Hence, eliminating the $U_2$'s 
from a 
sequence $U^K$ reduces the length from $F_K$ to 
$F_K - 2 F_{K-2} = F_{K-3}$. Since $F_K$ grows as a constant
times 
$\varphi^K$ for large $\varphi^K$, the reduction of
the length of a sequence from $F_K$ to $F_{K-3}$ means
a reduction by a factor of $\varphi^3 \simeq 4.24$.
We also observe that a Fibonacci protocol eventually
reduces to a sequence of only $U_1$'s just like a 
periodic protocol; this implies that the thermalization
properties of these two protocols should have some similar
features, except that a Fibonacci
protocol thermalizes more slowly than a periodic protocol.
This qualitatively explains the similarity between the plots in 
Figs.~\ref{fig6} (a) and (c).

To summarize, a sequence of $U_1$'s and $U_2$'s with
a large length $N$ reduces to a final sequence with
length $N$ (i.e., no reduction), $N/3$, $N/4.24$ and
zero (i.e., a reduction to I) for the periodic, random,
Fibonacci and TM sequences respectively. This provides some
understanding of why the rate of thermalization is largest for the periodic protocol, next largest for the random and 
Fibonacci protocols, and smallest for the TM protocol.

\bibliography{reference}

\begin{thebibliography}{150}%
\makeatletter
\providecommand \@ifxundefined [1]{%
 \@ifx{#1\undefined}
}%
\providecommand \@ifnum [1]{%
 \ifnum #1\expandafter \@firstoftwo
 \else \expandafter \@secondoftwo
 \fi
}%
\providecommand \@ifx [1]{%
 \ifx #1\expandafter \@firstoftwo
 \else \expandafter \@secondoftwo
 \fi
}%
\providecommand \natexlab [1]{#1}%
\providecommand \enquote  [1]{``#1''}%
\providecommand \bibnamefont  [1]{#1}%
\providecommand \bibfnamefont [1]{#1}%
\providecommand \citenamefont [1]{#1}%
\providecommand \href@noop [0]{\@secondoftwo}%
\providecommand \href [0]{\begingroup \@sanitize@url \@href}%
\providecommand \@href[1]{\@@startlink{#1}\@@href}%
\providecommand \@@href[1]{\endgroup#1\@@endlink}%
\providecommand \@sanitize@url [0]{\catcode `\\12\catcode `\$12\catcode
  `\&12\catcode `\#12\catcode `\^12\catcode `\_12\catcode `\%12\relax}%
\providecommand \@@startlink[1]{}%
\providecommand \@@endlink[0]{}%
\providecommand \url  [0]{\begingroup\@sanitize@url \@url }%
\providecommand \@url [1]{\endgroup\@href {#1}{\urlprefix }}%
\providecommand \urlprefix  [0]{URL }%
\providecommand \Eprint [0]{\href }%
\providecommand \doibase [0]{https://doi.org/}%
\providecommand \selectlanguage [0]{\@gobble}%
\providecommand \bibinfo  [0]{\@secondoftwo}%
\providecommand \bibfield  [0]{\@secondoftwo}%
\providecommand \translation [1]{[#1]}%
\providecommand \BibitemOpen [0]{}%
\providecommand \bibitemStop [0]{}%
\providecommand \bibitemNoStop [0]{.\EOS\space}%
\providecommand \EOS [0]{\spacefactor3000\relax}%
\providecommand \BibitemShut  [1]{\csname bibitem#1\endcsname}%
\let\auto@bib@innerbib\@empty
\bibitem [{\citenamefont {Dziarmaga}(2010)}]{rev1}%
  \BibitemOpen
  \bibfield  {author} {\bibinfo {author} {\bibfnamefont {J.}~\bibnamefont
  {Dziarmaga}},\ }\bibfield  {title} {\bibinfo {title} {Dynamics of a quantum
  phase transition and relaxation to a steady state},\ }\href
  {https://doi.org/10.1080/00018732.2010.514702} {\bibfield  {journal}
  {\bibinfo  {journal} {Advances in Physics}\ }\textbf {\bibinfo {volume}
  {59}},\ \bibinfo {pages} {1063–1189} (\bibinfo {year} {2010})}\BibitemShut
  {NoStop}%
\bibitem [{\citenamefont {Polkovnikov}\ \emph {et~al.}(2011)\citenamefont
  {Polkovnikov}, \citenamefont {Sengupta}, \citenamefont {Silva},\ and\
  \citenamefont {Vengalattore}}]{rev2}%
  \BibitemOpen
  \bibfield  {author} {\bibinfo {author} {\bibfnamefont {A.}~\bibnamefont
  {Polkovnikov}}, \bibinfo {author} {\bibfnamefont {K.}~\bibnamefont
  {Sengupta}}, \bibinfo {author} {\bibfnamefont {A.}~\bibnamefont {Silva}},\
  and\ \bibinfo {author} {\bibfnamefont {M.}~\bibnamefont {Vengalattore}},\
  }\bibfield  {title} {\bibinfo {title} {Colloquium: Nonequilibrium dynamics of
  closed interacting quantum systems},\ }\href
  {https://doi.org/10.1103/RevModPhys.83.863} {\bibfield  {journal} {\bibinfo
  {journal} {Rev. Mod. Phys.}\ }\textbf {\bibinfo {volume} {83}},\ \bibinfo
  {pages} {863} (\bibinfo {year} {2011})}\BibitemShut {NoStop}%
\bibitem [{\citenamefont {Dutta}\ \emph {et~al.}(2015)\citenamefont {Dutta},
  \citenamefont {Aeppli}, \citenamefont {Chakrabarti}, \citenamefont
  {Divakaran}, \citenamefont {Rosenbaum},\ and\ \citenamefont {Sen}}]{rev3}%
  \BibitemOpen
  \bibfield  {author} {\bibinfo {author} {\bibfnamefont {A.}~\bibnamefont
  {Dutta}}, \bibinfo {author} {\bibfnamefont {G.}~\bibnamefont {Aeppli}},
  \bibinfo {author} {\bibfnamefont {B.~K.}\ \bibnamefont {Chakrabarti}},
  \bibinfo {author} {\bibfnamefont {U.}~\bibnamefont {Divakaran}}, \bibinfo
  {author} {\bibfnamefont {T.~F.}\ \bibnamefont {Rosenbaum}},\ and\ \bibinfo
  {author} {\bibfnamefont {D.}~\bibnamefont {Sen}},\ }\href
  {https://doi.org/10.1017/cbo9781107706057} {\emph {\bibinfo {title} {Quantum
  Phase Transitions in Transverse Field Spin Models: From Statistical Physics
  to Quantum Information}}}\ (\bibinfo  {publisher} {Cambridge University
  Press},\ \bibinfo {year} {2015})\BibitemShut {NoStop}%
\bibitem [{\citenamefont {Chandra}\ \emph {et~al.}(2010)\citenamefont
  {Chandra}, \citenamefont {Das},\ and\ \citenamefont {Chakrabarti}}]{rev4}%
  \BibitemOpen
  \bibfield  {author} {\bibinfo {author} {\bibfnamefont {A.}~\bibnamefont
  {Chandra}}, \bibinfo {author} {\bibfnamefont {A.}~\bibnamefont {Das}},\ and\
  \bibinfo {author} {\bibfnamefont {B.}~\bibnamefont {Chakrabarti}},\ }\href
  {https://doi.org/10.1007/978-3-642-11470-0} {\emph {\bibinfo {title} {Quantum
  Quenching, Annealing and Computation}}}\ (\bibinfo  {publisher} {Springer
  Berlin Heidelberg},\ \bibinfo {year} {2010})\BibitemShut {NoStop}%
\bibitem [{\citenamefont {Bukov}\ \emph {et~al.}(2015)\citenamefont {Bukov},
  \citenamefont {D'Alessio},\ and\ \citenamefont {Polkovnikov}}]{rev5}%
  \BibitemOpen
  \bibfield  {author} {\bibinfo {author} {\bibfnamefont {M.}~\bibnamefont
  {Bukov}}, \bibinfo {author} {\bibfnamefont {L.}~\bibnamefont {D'Alessio}},\
  and\ \bibinfo {author} {\bibfnamefont {A.}~\bibnamefont {Polkovnikov}},\
  }\bibfield  {title} {\bibinfo {title} {Universal high-frequency behavior of
  periodically driven systems: from dynamical stabilization to floquet
  engineering},\ }\href {https://doi.org/10.1080/00018732.2015.1055918}
  {\bibfield  {journal} {\bibinfo  {journal} {Advances in Physics}\ }\textbf
  {\bibinfo {volume} {64}},\ \bibinfo {pages} {139} (\bibinfo {year} {2015})},\
  \Eprint {https://arxiv.org/abs/https://doi.org/10.1080/00018732.2015.1055918}
  {https://doi.org/10.1080/00018732.2015.1055918} \BibitemShut {NoStop}%
\bibitem [{\citenamefont {D’Alessio}\ and\ \citenamefont
  {Polkovnikov}(2013)}]{rev6}%
  \BibitemOpen
  \bibfield  {author} {\bibinfo {author} {\bibfnamefont {L.}~\bibnamefont
  {D’Alessio}}\ and\ \bibinfo {author} {\bibfnamefont {A.}~\bibnamefont
  {Polkovnikov}},\ }\bibfield  {title} {\bibinfo {title} {Many-body energy
  localization transition in periodically driven systems},\ }\href
  {https://doi.org/https://doi.org/10.1016/j.aop.2013.02.011} {\bibfield
  {journal} {\bibinfo  {journal} {Annals of Physics}\ }\textbf {\bibinfo
  {volume} {333}},\ \bibinfo {pages} {19} (\bibinfo {year} {2013})}\BibitemShut
  {NoStop}%
\bibitem [{\citenamefont {D'Alessio}\ \emph {et~al.}(2016)\citenamefont
  {D'Alessio}, \citenamefont {Kafri}, \citenamefont {Polkovnikov},\ and\
  \citenamefont {Rigol}}]{rev7}%
  \BibitemOpen
  \bibfield  {author} {\bibinfo {author} {\bibfnamefont {L.}~\bibnamefont
  {D'Alessio}}, \bibinfo {author} {\bibfnamefont {Y.}~\bibnamefont {Kafri}},
  \bibinfo {author} {\bibfnamefont {A.}~\bibnamefont {Polkovnikov}},\ and\
  \bibinfo {author} {\bibfnamefont {M.}~\bibnamefont {Rigol}},\ }\bibfield
  {title} {\bibinfo {title} {From quantum chaos and eigenstate thermalization
  to statistical mechanics and thermodynamics},\ }\href
  {https://doi.org/10.1080/00018732.2016.1198134} {\bibfield  {journal}
  {\bibinfo  {journal} {Advances in Physics}\ }\textbf {\bibinfo {volume}
  {65}},\ \bibinfo {pages} {239} (\bibinfo {year} {2016})},\ \Eprint
  {https://arxiv.org/abs/https://doi.org/10.1080/00018732.2016.1198134}
  {https://doi.org/10.1080/00018732.2016.1198134} \BibitemShut {NoStop}%
\bibitem [{\citenamefont {Shevchenko}\ \emph {et~al.}(2010)\citenamefont
  {Shevchenko}, \citenamefont {Ashhab},\ and\ \citenamefont {Nori}}]{rev8}%
  \BibitemOpen
  \bibfield  {author} {\bibinfo {author} {\bibfnamefont {S.}~\bibnamefont
  {Shevchenko}}, \bibinfo {author} {\bibfnamefont {S.}~\bibnamefont {Ashhab}},\
  and\ \bibinfo {author} {\bibfnamefont {F.}~\bibnamefont {Nori}},\ }\bibfield
  {title} {\bibinfo {title} {Landau–zener–st\"{u}ckelberg interferometry},\
  }\href {https://doi.org/10.1016/j.physrep.2010.03.002} {\bibfield  {journal}
  {\bibinfo  {journal} {Physics Reports}\ }\textbf {\bibinfo {volume} {492}},\
  \bibinfo {pages} {1–30} (\bibinfo {year} {2010})}\BibitemShut {NoStop}%
\bibitem [{\citenamefont {Oka}\ and\ \citenamefont {Kitamura}(2019)}]{rev9}%
  \BibitemOpen
  \bibfield  {author} {\bibinfo {author} {\bibfnamefont {T.}~\bibnamefont
  {Oka}}\ and\ \bibinfo {author} {\bibfnamefont {S.}~\bibnamefont {Kitamura}},\
  }\bibfield  {title} {\bibinfo {title} {Floquet engineering of quantum
  materials},\ }\href
  {https://doi.org/10.1146/annurev-conmatphys-031218-013423} {\bibfield
  {journal} {\bibinfo  {journal} {Annual Review of Condensed Matter Physics}\
  }\textbf {\bibinfo {volume} {10}},\ \bibinfo {pages} {387–408} (\bibinfo
  {year} {2019})}\BibitemShut {NoStop}%
\bibitem [{\citenamefont {Blanes}\ \emph {et~al.}(2009)\citenamefont {Blanes},
  \citenamefont {Casas}, \citenamefont {Oteo},\ and\ \citenamefont
  {Ros}}]{rev10}%
  \BibitemOpen
  \bibfield  {author} {\bibinfo {author} {\bibfnamefont {S.}~\bibnamefont
  {Blanes}}, \bibinfo {author} {\bibfnamefont {F.}~\bibnamefont {Casas}},
  \bibinfo {author} {\bibfnamefont {J.}~\bibnamefont {Oteo}},\ and\ \bibinfo
  {author} {\bibfnamefont {J.}~\bibnamefont {Ros}},\ }\bibfield  {title}
  {\bibinfo {title} {The magnus expansion and some of its applications},\
  }\href {https://doi.org/10.1016/j.physrep.2008.11.001} {\bibfield  {journal}
  {\bibinfo  {journal} {Physics Reports}\ }\textbf {\bibinfo {volume} {470}},\
  \bibinfo {pages} {151–238} (\bibinfo {year} {2009})}\BibitemShut {NoStop}%
\bibitem [{\citenamefont {Eckardt}(2017)}]{rev11}%
  \BibitemOpen
  \bibfield  {author} {\bibinfo {author} {\bibfnamefont {A.}~\bibnamefont
  {Eckardt}},\ }\bibfield  {title} {\bibinfo {title} {Colloquium: Atomic
  quantum gases in periodically driven optical lattices},\ }\href
  {https://doi.org/10.1103/RevModPhys.89.011004} {\bibfield  {journal}
  {\bibinfo  {journal} {Rev. Mod. Phys.}\ }\textbf {\bibinfo {volume} {89}},\
  \bibinfo {pages} {011004} (\bibinfo {year} {2017})}\BibitemShut {NoStop}%
\bibitem [{\citenamefont {Sen}\ \emph {et~al.}(2021)\citenamefont {Sen},
  \citenamefont {Sen},\ and\ \citenamefont {Sengupta}}]{rev12}%
  \BibitemOpen
  \bibfield  {author} {\bibinfo {author} {\bibfnamefont {A.}~\bibnamefont
  {Sen}}, \bibinfo {author} {\bibfnamefont {D.}~\bibnamefont {Sen}},\ and\
  \bibinfo {author} {\bibfnamefont {K.}~\bibnamefont {Sengupta}},\ }\bibfield
  {title} {\bibinfo {title} {Analytic approaches to periodically driven closed
  quantum systems: methods and applications},\ }\href
  {https://doi.org/10.1088/1361-648X/ac1b61} {\bibfield  {journal} {\bibinfo
  {journal} {Journal of Physics: Condensed Matter}\ }\textbf {\bibinfo {volume}
  {33}},\ \bibinfo {pages} {443003} (\bibinfo {year} {2021})}\BibitemShut
  {NoStop}%
\bibitem [{\citenamefont {Bloch}\ \emph {et~al.}(2008)\citenamefont {Bloch},
  \citenamefont {Dalibard},\ and\ \citenamefont {Zwerger}}]{rev13}%
  \BibitemOpen
  \bibfield  {author} {\bibinfo {author} {\bibfnamefont {I.}~\bibnamefont
  {Bloch}}, \bibinfo {author} {\bibfnamefont {J.}~\bibnamefont {Dalibard}},\
  and\ \bibinfo {author} {\bibfnamefont {W.}~\bibnamefont {Zwerger}},\
  }\bibfield  {title} {\bibinfo {title} {Many-body physics with ultracold
  gases},\ }\href {https://doi.org/10.1103/RevModPhys.80.885} {\bibfield
  {journal} {\bibinfo  {journal} {Rev. Mod. Phys.}\ }\textbf {\bibinfo {volume}
  {80}},\ \bibinfo {pages} {885} (\bibinfo {year} {2008})}\BibitemShut
  {NoStop}%
\bibitem [{\citenamefont {Tarruell}\ and\ \citenamefont
  {Sanchez-Palencia}(2018)}]{rev14}%
  \BibitemOpen
  \bibfield  {author} {\bibinfo {author} {\bibfnamefont {L.}~\bibnamefont
  {Tarruell}}\ and\ \bibinfo {author} {\bibfnamefont {L.}~\bibnamefont
  {Sanchez-Palencia}},\ }\bibfield  {title} {\bibinfo {title} {Quantum
  simulation of the hubbard model with ultracold fermions in optical
  lattices},\ }\href {https://doi.org/10.1016/j.crhy.2018.10.013} {\bibfield
  {journal} {\bibinfo  {journal} {Comptes Rendus. Physique}\ }\textbf {\bibinfo
  {volume} {19}},\ \bibinfo {pages} {365–393} (\bibinfo {year}
  {2018})}\BibitemShut {NoStop}%
\bibitem [{\citenamefont {Ho}\ \emph {et~al.}(2023)\citenamefont {Ho},
  \citenamefont {Mori}, \citenamefont {Abanin},\ and\ \citenamefont
  {Dalla~Torre}}]{rev15}%
  \BibitemOpen
  \bibfield  {author} {\bibinfo {author} {\bibfnamefont {W.~W.}\ \bibnamefont
  {Ho}}, \bibinfo {author} {\bibfnamefont {T.}~\bibnamefont {Mori}}, \bibinfo
  {author} {\bibfnamefont {D.~A.}\ \bibnamefont {Abanin}},\ and\ \bibinfo
  {author} {\bibfnamefont {E.~G.}\ \bibnamefont {Dalla~Torre}},\ }\bibfield
  {title} {\bibinfo {title} {Quantum and classical floquet prethermalization},\
  }\href {https://doi.org/10.1016/j.aop.2023.169297} {\bibfield  {journal}
  {\bibinfo  {journal} {Annals of Physics}\ }\textbf {\bibinfo {volume}
  {454}},\ \bibinfo {pages} {169297} (\bibinfo {year} {2023})}\BibitemShut
  {NoStop}%
\bibitem [{\citenamefont {Mori}\ \emph {et~al.}(2018)\citenamefont {Mori},
  \citenamefont {Ikeda}, \citenamefont {Kaminishi},\ and\ \citenamefont
  {Ueda}}]{rev16}%
  \BibitemOpen
  \bibfield  {author} {\bibinfo {author} {\bibfnamefont {T.}~\bibnamefont
  {Mori}}, \bibinfo {author} {\bibfnamefont {T.~N.}\ \bibnamefont {Ikeda}},
  \bibinfo {author} {\bibfnamefont {E.}~\bibnamefont {Kaminishi}},\ and\
  \bibinfo {author} {\bibfnamefont {M.}~\bibnamefont {Ueda}},\ }\bibfield
  {title} {\bibinfo {title} {Thermalization and prethermalization in isolated
  quantum systems: a theoretical overview},\ }\href
  {https://doi.org/10.1088/1361-6455/aabcdf} {\bibfield  {journal} {\bibinfo
  {journal} {Journal of Physics B: Atomic, Molecular and Optical Physics}\
  }\textbf {\bibinfo {volume} {51}},\ \bibinfo {pages} {112001} (\bibinfo
  {year} {2018})}\BibitemShut {NoStop}%
\bibitem [{\citenamefont {Banerjee}\ and\ \citenamefont
  {Sengupta}(2025)}]{rev17}%
  \BibitemOpen
  \bibfield  {author} {\bibinfo {author} {\bibfnamefont {T.}~\bibnamefont
  {Banerjee}}\ and\ \bibinfo {author} {\bibfnamefont {K.}~\bibnamefont
  {Sengupta}},\ }\bibfield  {title} {\bibinfo {title} {Emergent symmetries in
  prethermal phases of periodically driven quantum systems},\ }\href
  {https://doi.org/10.1088/1361-648x/ada860} {\bibfield  {journal} {\bibinfo
  {journal} {Journal of Physics: Condensed Matter}\ }\textbf {\bibinfo {volume}
  {37}},\ \bibinfo {pages} {133002} (\bibinfo {year} {2025})}\BibitemShut
  {NoStop}%
\bibitem [{\citenamefont {Sengupta}\ \emph {et~al.}(2004)\citenamefont
  {Sengupta}, \citenamefont {Powell},\ and\ \citenamefont {Sachdev}}]{subir1}%
  \BibitemOpen
  \bibfield  {author} {\bibinfo {author} {\bibfnamefont {K.}~\bibnamefont
  {Sengupta}}, \bibinfo {author} {\bibfnamefont {S.}~\bibnamefont {Powell}},\
  and\ \bibinfo {author} {\bibfnamefont {S.}~\bibnamefont {Sachdev}},\
  }\bibfield  {title} {\bibinfo {title} {Quench dynamics across quantum
  critical points},\ }\href {https://doi.org/10.1103/PhysRevA.69.053616}
  {\bibfield  {journal} {\bibinfo  {journal} {Phys. Rev. A}\ }\textbf {\bibinfo
  {volume} {69}},\ \bibinfo {pages} {053616} (\bibinfo {year}
  {2004})}\BibitemShut {NoStop}%
\bibitem [{\citenamefont {Das}\ \emph {et~al.}(2006)\citenamefont {Das},
  \citenamefont {Sengupta}, \citenamefont {Sen},\ and\ \citenamefont
  {Chakrabarti}}]{dsen}%
  \BibitemOpen
  \bibfield  {author} {\bibinfo {author} {\bibfnamefont {A.}~\bibnamefont
  {Das}}, \bibinfo {author} {\bibfnamefont {K.}~\bibnamefont {Sengupta}},
  \bibinfo {author} {\bibfnamefont {D.}~\bibnamefont {Sen}},\ and\ \bibinfo
  {author} {\bibfnamefont {B.~K.}\ \bibnamefont {Chakrabarti}},\ }\bibfield
  {title} {\bibinfo {title} {Infinite-range ising ferromagnet in a
  time-dependent transverse magnetic field: Quench and ac dynamics near the
  quantum critical point},\ }\href {https://doi.org/10.1103/PhysRevB.74.144423}
  {\bibfield  {journal} {\bibinfo  {journal} {Phys. Rev. B}\ }\textbf {\bibinfo
  {volume} {74}},\ \bibinfo {pages} {144423} (\bibinfo {year}
  {2006})}\BibitemShut {NoStop}%
\bibitem [{\citenamefont {De~Grandi}\ \emph
  {et~al.}(2010{\natexlab{a}})\citenamefont {De~Grandi}, \citenamefont
  {Gritsev},\ and\ \citenamefont {Polkovnikov}}]{anatoly0_part1}%
  \BibitemOpen
  \bibfield  {author} {\bibinfo {author} {\bibfnamefont {C.}~\bibnamefont
  {De~Grandi}}, \bibinfo {author} {\bibfnamefont {V.}~\bibnamefont {Gritsev}},\
  and\ \bibinfo {author} {\bibfnamefont {A.}~\bibnamefont {Polkovnikov}},\
  }\bibfield  {title} {\bibinfo {title} {Quench dynamics near a quantum
  critical point},\ }\href {https://doi.org/10.1103/PhysRevB.81.012303}
  {\bibfield  {journal} {\bibinfo  {journal} {Phys. Rev. B}\ }\textbf {\bibinfo
  {volume} {81}},\ \bibinfo {pages} {012303} (\bibinfo {year}
  {2010}{\natexlab{a}})}\BibitemShut {NoStop}%
\bibitem [{\citenamefont {De~Grandi}\ \emph
  {et~al.}(2010{\natexlab{b}})\citenamefont {De~Grandi}, \citenamefont
  {Gritsev},\ and\ \citenamefont {Polkovnikov}}]{anatoly0_part2}%
  \BibitemOpen
  \bibfield  {author} {\bibinfo {author} {\bibfnamefont {C.}~\bibnamefont
  {De~Grandi}}, \bibinfo {author} {\bibfnamefont {V.}~\bibnamefont {Gritsev}},\
  and\ \bibinfo {author} {\bibfnamefont {A.}~\bibnamefont {Polkovnikov}},\
  }\bibfield  {title} {\bibinfo {title} {Quench dynamics near a quantum
  critical point: Application to the sine-gordon model},\ }\bibfield  {journal}
  {\bibinfo  {journal} {Physical Review B}\ }\textbf {\bibinfo {volume} {81}},\
  \href {https://doi.org/10.1103/physrevb.81.224301}
  {10.1103/physrevb.81.224301} (\bibinfo {year}
  {2010}{\natexlab{b}})\BibitemShut {NoStop}%
\bibitem [{\citenamefont {Das}\ and\ \citenamefont
  {Sengupta}(2012)}]{sdas0_part1}%
  \BibitemOpen
  \bibfield  {author} {\bibinfo {author} {\bibfnamefont {S.~R.}\ \bibnamefont
  {Das}}\ and\ \bibinfo {author} {\bibfnamefont {K.}~\bibnamefont {Sengupta}},\
  }\bibfield  {title} {\bibinfo {title} {Non-equilibrium dynamics of o(n)
  nonlinear sigma models: a large-n approach},\ }\bibfield  {journal} {\bibinfo
   {journal} {Journal of High Energy Physics}\ }\textbf {\bibinfo {volume}
  {2012}},\ \href {https://doi.org/10.1007/jhep09(2012)072}
  {10.1007/jhep09(2012)072} (\bibinfo {year} {2012})\BibitemShut {NoStop}%
\bibitem [{\citenamefont {Basu}\ \emph {et~al.}(2013)\citenamefont {Basu},
  \citenamefont {Das}, \citenamefont {Das},\ and\ \citenamefont
  {Sengupta}}]{sdas0_part2}%
  \BibitemOpen
  \bibfield  {author} {\bibinfo {author} {\bibfnamefont {P.}~\bibnamefont
  {Basu}}, \bibinfo {author} {\bibfnamefont {D.}~\bibnamefont {Das}}, \bibinfo
  {author} {\bibfnamefont {S.~R.}\ \bibnamefont {Das}},\ and\ \bibinfo {author}
  {\bibfnamefont {K.}~\bibnamefont {Sengupta}},\ }\bibfield  {title} {\bibinfo
  {title} {Quantum quench and double trace couplings},\ }\bibfield  {journal}
  {\bibinfo  {journal} {Journal of High Energy Physics}\ }\textbf {\bibinfo
  {volume} {2013}},\ \href {https://doi.org/10.1007/jhep12(2013)070}
  {10.1007/jhep12(2013)070} (\bibinfo {year} {2013})\BibitemShut {NoStop}%
\bibitem [{\citenamefont {Das}\ \emph {et~al.}(2014)\citenamefont {Das},
  \citenamefont {Galante},\ and\ \citenamefont {Myers}}]{sdas1}%
  \BibitemOpen
  \bibfield  {author} {\bibinfo {author} {\bibfnamefont {S.~R.}\ \bibnamefont
  {Das}}, \bibinfo {author} {\bibfnamefont {D.~A.}\ \bibnamefont {Galante}},\
  and\ \bibinfo {author} {\bibfnamefont {R.~C.}\ \bibnamefont {Myers}},\
  }\bibfield  {title} {\bibinfo {title} {Universal scaling in fast quantum
  quenches in conformal field theories},\ }\href
  {https://doi.org/10.1103/PhysRevLett.112.171601} {\bibfield  {journal}
  {\bibinfo  {journal} {Phys. Rev. Lett.}\ }\textbf {\bibinfo {volume} {112}},\
  \bibinfo {pages} {171601} (\bibinfo {year} {2014})}\BibitemShut {NoStop}%
\bibitem [{\citenamefont {Das}\ \emph {et~al.}(2015{\natexlab{a}})\citenamefont
  {Das}, \citenamefont {Galante},\ and\ \citenamefont {Myers}}]{sdas2_part1}%
  \BibitemOpen
  \bibfield  {author} {\bibinfo {author} {\bibfnamefont {S.~R.}\ \bibnamefont
  {Das}}, \bibinfo {author} {\bibfnamefont {D.~A.}\ \bibnamefont {Galante}},\
  and\ \bibinfo {author} {\bibfnamefont {R.~C.}\ \bibnamefont {Myers}},\
  }\bibfield  {title} {\bibinfo {title} {Universality in fast quantum
  quenches},\ }\bibfield  {journal} {\bibinfo  {journal} {Journal of High
  Energy Physics}\ }\textbf {\bibinfo {volume} {2015}},\ \href
  {https://doi.org/10.1007/jhep02(2015)167} {10.1007/jhep02(2015)167} (\bibinfo
  {year} {2015}{\natexlab{a}})\BibitemShut {NoStop}%
\bibitem [{\citenamefont {Das}\ \emph {et~al.}(2019)\citenamefont {Das},
  \citenamefont {Hampton},\ and\ \citenamefont {Liu}}]{sdas2_part2}%
  \BibitemOpen
  \bibfield  {author} {\bibinfo {author} {\bibfnamefont {S.~R.}\ \bibnamefont
  {Das}}, \bibinfo {author} {\bibfnamefont {S.}~\bibnamefont {Hampton}},\ and\
  \bibinfo {author} {\bibfnamefont {S.}~\bibnamefont {Liu}},\ }\bibfield
  {title} {\bibinfo {title} {Quantum quench in non-relativistic fermionic field
  theory: harmonic traps and 2d string theory},\ }\bibfield  {journal}
  {\bibinfo  {journal} {Journal of High Energy Physics}\ }\textbf {\bibinfo
  {volume} {2019}},\ \href {https://doi.org/10.1007/jhep08(2019)176}
  {10.1007/jhep08(2019)176} (\bibinfo {year} {2019})\BibitemShut {NoStop}%
\bibitem [{\citenamefont {Chakraborty}\ \emph {et~al.}(2019)\citenamefont
  {Chakraborty}, \citenamefont {Gorantla},\ and\ \citenamefont
  {Sensarma}}]{rajdeep1_part1}%
  \BibitemOpen
  \bibfield  {author} {\bibinfo {author} {\bibfnamefont {A.}~\bibnamefont
  {Chakraborty}}, \bibinfo {author} {\bibfnamefont {P.}~\bibnamefont
  {Gorantla}},\ and\ \bibinfo {author} {\bibfnamefont {R.}~\bibnamefont
  {Sensarma}},\ }\bibfield  {title} {\bibinfo {title} {Nonequilibrium field
  theory for dynamics starting from arbitrary athermal initial conditions},\
  }\bibfield  {journal} {\bibinfo  {journal} {Physical Review B}\ }\textbf
  {\bibinfo {volume} {99}},\ \href {https://doi.org/10.1103/physrevb.99.054306}
  {10.1103/physrevb.99.054306} (\bibinfo {year} {2019})\BibitemShut {NoStop}%
\bibitem [{\citenamefont {Chakraborty}\ and\ \citenamefont
  {Sensarma}(2021)}]{rajdeep1_part2}%
  \BibitemOpen
  \bibfield  {author} {\bibinfo {author} {\bibfnamefont {A.}~\bibnamefont
  {Chakraborty}}\ and\ \bibinfo {author} {\bibfnamefont {R.}~\bibnamefont
  {Sensarma}},\ }\bibfield  {title} {\bibinfo {title} {Nonequilibrium dynamics
  of renyi entropy for bosonic many-particle systems},\ }\href
  {https://doi.org/10.1103/PhysRevLett.127.200603} {\bibfield  {journal}
  {\bibinfo  {journal} {Phys. Rev. Lett.}\ }\textbf {\bibinfo {volume} {127}},\
  \bibinfo {pages} {200603} (\bibinfo {year} {2021})}\BibitemShut {NoStop}%
\bibitem [{\citenamefont {Islam}\ and\ \citenamefont
  {Sensarma}(2022)}]{rajdeep2_part1}%
  \BibitemOpen
  \bibfield  {author} {\bibinfo {author} {\bibfnamefont {M.~M.}\ \bibnamefont
  {Islam}}\ and\ \bibinfo {author} {\bibfnamefont {R.}~\bibnamefont
  {Sensarma}},\ }\bibfield  {title} {\bibinfo {title} {Nonequilibrium scalar
  field dynamics starting from fock states: Absence of thermalization in
  one-dimensional phonons coupled to fermions},\ }\bibfield  {journal}
  {\bibinfo  {journal} {Physical Review B}\ }\textbf {\bibinfo {volume}
  {106}},\ \href {https://doi.org/10.1103/physrevb.106.024306}
  {10.1103/physrevb.106.024306} (\bibinfo {year} {2022})\BibitemShut {NoStop}%
\bibitem [{\citenamefont {Islam}\ \emph {et~al.}(2023)\citenamefont {Islam},
  \citenamefont {Sengupta},\ and\ \citenamefont {Sensarma}}]{rajdeep2_part2}%
  \BibitemOpen
  \bibfield  {author} {\bibinfo {author} {\bibfnamefont {M.~M.}\ \bibnamefont
  {Islam}}, \bibinfo {author} {\bibfnamefont {K.}~\bibnamefont {Sengupta}},\
  and\ \bibinfo {author} {\bibfnamefont {R.}~\bibnamefont {Sensarma}},\
  }\bibfield  {title} {\bibinfo {title} {Nonequilibrium dynamics of bosons with
  dipole symmetry: Large-$n$ keldysh approach},\ }\href
  {https://doi.org/10.1103/PhysRevB.108.214314} {\bibfield  {journal} {\bibinfo
   {journal} {Phys. Rev. B}\ }\textbf {\bibinfo {volume} {108}},\ \bibinfo
  {pages} {214314} (\bibinfo {year} {2023})}\BibitemShut {NoStop}%
\bibitem [{\citenamefont {Lankhorst}\ \emph {et~al.}(2018)\citenamefont
  {Lankhorst}, \citenamefont {Poccia}, \citenamefont {Stehno}, \citenamefont
  {Galda}, \citenamefont {Barman}, \citenamefont {Coneri}, \citenamefont
  {Hilgenkamp}, \citenamefont {Brinkman}, \citenamefont {Golubov},
  \citenamefont {Tripathi}, \citenamefont {Baturina},\ and\ \citenamefont
  {Vinokur}}]{trip1_part1}%
  \BibitemOpen
  \bibfield  {author} {\bibinfo {author} {\bibfnamefont {M.}~\bibnamefont
  {Lankhorst}}, \bibinfo {author} {\bibfnamefont {N.}~\bibnamefont {Poccia}},
  \bibinfo {author} {\bibfnamefont {M.~P.}\ \bibnamefont {Stehno}}, \bibinfo
  {author} {\bibfnamefont {A.}~\bibnamefont {Galda}}, \bibinfo {author}
  {\bibfnamefont {H.}~\bibnamefont {Barman}}, \bibinfo {author} {\bibfnamefont
  {F.}~\bibnamefont {Coneri}}, \bibinfo {author} {\bibfnamefont
  {H.}~\bibnamefont {Hilgenkamp}}, \bibinfo {author} {\bibfnamefont
  {A.}~\bibnamefont {Brinkman}}, \bibinfo {author} {\bibfnamefont {A.~A.}\
  \bibnamefont {Golubov}}, \bibinfo {author} {\bibfnamefont {V.}~\bibnamefont
  {Tripathi}}, \bibinfo {author} {\bibfnamefont {T.~I.}\ \bibnamefont
  {Baturina}},\ and\ \bibinfo {author} {\bibfnamefont {V.~M.}\ \bibnamefont
  {Vinokur}},\ }\bibfield  {title} {\bibinfo {title} {Scaling universality at
  the dynamic vortex mott transition},\ }\bibfield  {journal} {\bibinfo
  {journal} {Physical Review B}\ }\textbf {\bibinfo {volume} {97}},\ \href
  {https://doi.org/10.1103/physrevb.97.020504} {10.1103/physrevb.97.020504}
  (\bibinfo {year} {2018})\BibitemShut {NoStop}%
\bibitem [{\citenamefont {Sankar}\ and\ \citenamefont
  {Tripathi}(2019)}]{trip1_part2}%
  \BibitemOpen
  \bibfield  {author} {\bibinfo {author} {\bibfnamefont {S.}~\bibnamefont
  {Sankar}}\ and\ \bibinfo {author} {\bibfnamefont {V.}~\bibnamefont
  {Tripathi}},\ }\bibfield  {title} {\bibinfo {title} {Keldysh field theory of
  a driven dissipative mott insulator: Nonequilibrium response and phase
  transitions},\ }\bibfield  {journal} {\bibinfo  {journal} {Physical Review
  B}\ }\textbf {\bibinfo {volume} {99}},\ \href
  {https://doi.org/10.1103/physrevb.99.245113} {10.1103/physrevb.99.245113}
  (\bibinfo {year} {2019})\BibitemShut {NoStop}%
\bibitem [{\citenamefont {Kibble}(1976)}]{kibble1}%
  \BibitemOpen
  \bibfield  {author} {\bibinfo {author} {\bibfnamefont {T.~W.~B.}\
  \bibnamefont {Kibble}},\ }\bibfield  {title} {\bibinfo {title} {Topology of
  cosmic domains and strings},\ }\href
  {https://doi.org/10.1088/0305-4470/9/8/029} {\bibfield  {journal} {\bibinfo
  {journal} {Journal of Physics A: Mathematical and General}\ }\textbf
  {\bibinfo {volume} {9}},\ \bibinfo {pages} {1387–1398} (\bibinfo {year}
  {1976})}\BibitemShut {NoStop}%
\bibitem [{\citenamefont {Zurek}(1985)}]{zurek1}%
  \BibitemOpen
  \bibfield  {author} {\bibinfo {author} {\bibfnamefont {W.~H.}\ \bibnamefont
  {Zurek}},\ }\bibfield  {title} {\bibinfo {title} {Cosmological experiments in
  superfluid helium?},\ }\href {https://doi.org/10.1038/317505a0} {\bibfield
  {journal} {\bibinfo  {journal} {Nature}\ }\textbf {\bibinfo {volume} {317}},\
  \bibinfo {pages} {505–508} (\bibinfo {year} {1985})}\BibitemShut {NoStop}%
\bibitem [{\citenamefont {Polkovnikov}(2005)}]{anatoly1}%
  \BibitemOpen
  \bibfield  {author} {\bibinfo {author} {\bibfnamefont {A.}~\bibnamefont
  {Polkovnikov}},\ }\bibfield  {title} {\bibinfo {title} {Universal adiabatic
  dynamics in the vicinity of a quantum critical point},\ }\href
  {https://doi.org/10.1103/PhysRevB.72.161201} {\bibfield  {journal} {\bibinfo
  {journal} {Phys. Rev. B}\ }\textbf {\bibinfo {volume} {72}},\ \bibinfo
  {pages} {161201} (\bibinfo {year} {2005})}\BibitemShut {NoStop}%
\bibitem [{\citenamefont {Polkovnikov}\ and\ \citenamefont
  {Gritsev}(2008)}]{anatoly2}%
  \BibitemOpen
  \bibfield  {author} {\bibinfo {author} {\bibfnamefont {A.}~\bibnamefont
  {Polkovnikov}}\ and\ \bibinfo {author} {\bibfnamefont {V.}~\bibnamefont
  {Gritsev}},\ }\bibfield  {title} {\bibinfo {title} {Breakdown of the
  adiabatic limit in low-dimensional gapless systems},\ }\href
  {https://doi.org/10.1038/nphys963} {\bibfield  {journal} {\bibinfo  {journal}
  {Nature Physics}\ }\textbf {\bibinfo {volume} {4}},\ \bibinfo {pages}
  {477–481} (\bibinfo {year} {2008})}\BibitemShut {NoStop}%
\bibitem [{\citenamefont {Sengupta}\ \emph {et~al.}(2008)\citenamefont
  {Sengupta}, \citenamefont {Sen},\ and\ \citenamefont {Mondal}}]{dsen1}%
  \BibitemOpen
  \bibfield  {author} {\bibinfo {author} {\bibfnamefont {K.}~\bibnamefont
  {Sengupta}}, \bibinfo {author} {\bibfnamefont {D.}~\bibnamefont {Sen}},\ and\
  \bibinfo {author} {\bibfnamefont {S.}~\bibnamefont {Mondal}},\ }\bibfield
  {title} {\bibinfo {title} {Exact results for quench dynamics and defect
  production in a two-dimensional model},\ }\href
  {https://doi.org/10.1103/PhysRevLett.100.077204} {\bibfield  {journal}
  {\bibinfo  {journal} {Phys. Rev. Lett.}\ }\textbf {\bibinfo {volume} {100}},\
  \bibinfo {pages} {077204} (\bibinfo {year} {2008})}\BibitemShut {NoStop}%
\bibitem [{\citenamefont {Sen}\ \emph {et~al.}(2008)\citenamefont {Sen},
  \citenamefont {Sengupta},\ and\ \citenamefont {Mondal}}]{dsen2_part1}%
  \BibitemOpen
  \bibfield  {author} {\bibinfo {author} {\bibfnamefont {D.}~\bibnamefont
  {Sen}}, \bibinfo {author} {\bibfnamefont {K.}~\bibnamefont {Sengupta}},\ and\
  \bibinfo {author} {\bibfnamefont {S.}~\bibnamefont {Mondal}},\ }\bibfield
  {title} {\bibinfo {title} {Defect production in nonlinear quench across a
  quantum critical point},\ }\href
  {https://doi.org/10.1103/PhysRevLett.101.016806} {\bibfield  {journal}
  {\bibinfo  {journal} {Phys. Rev. Lett.}\ }\textbf {\bibinfo {volume} {101}},\
  \bibinfo {pages} {016806} (\bibinfo {year} {2008})}\BibitemShut {NoStop}%
\bibitem [{\citenamefont {Barankov}\ and\ \citenamefont
  {Polkovnikov}(2008)}]{polkovnikov_barankov_part2}%
  \BibitemOpen
  \bibfield  {author} {\bibinfo {author} {\bibfnamefont {R.}~\bibnamefont
  {Barankov}}\ and\ \bibinfo {author} {\bibfnamefont {A.}~\bibnamefont
  {Polkovnikov}},\ }\bibfield  {title} {\bibinfo {title} {Optimal nonlinear
  passage through a quantum critical point},\ }\href
  {https://doi.org/10.1103/PhysRevLett.101.076801} {\bibfield  {journal}
  {\bibinfo  {journal} {Phys. Rev. Lett.}\ }\textbf {\bibinfo {volume} {101}},\
  \bibinfo {pages} {076801} (\bibinfo {year} {2008})}\BibitemShut {NoStop}%
\bibitem [{\citenamefont {Das}\ \emph {et~al.}(2016)\citenamefont {Das},
  \citenamefont {Galante},\ and\ \citenamefont {Myers}}]{sdas3_part1}%
  \BibitemOpen
  \bibfield  {author} {\bibinfo {author} {\bibfnamefont {S.~R.}\ \bibnamefont
  {Das}}, \bibinfo {author} {\bibfnamefont {D.~A.}\ \bibnamefont {Galante}},\
  and\ \bibinfo {author} {\bibfnamefont {R.~C.}\ \bibnamefont {Myers}},\
  }\bibfield  {title} {\bibinfo {title} {Quantum quenches in free field theory:
  universal scaling at any rate},\ }\bibfield  {journal} {\bibinfo  {journal}
  {Journal of High Energy Physics}\ }\textbf {\bibinfo {volume} {2016}},\ \href
  {https://doi.org/10.1007/jhep05(2016)164} {10.1007/jhep05(2016)164} (\bibinfo
  {year} {2016})\BibitemShut {NoStop}%
\bibitem [{\citenamefont {Das}\ \emph {et~al.}(2015{\natexlab{b}})\citenamefont
  {Das}, \citenamefont {Galante},\ and\ \citenamefont {Myers}}]{sdas3_part2}%
  \BibitemOpen
  \bibfield  {author} {\bibinfo {author} {\bibfnamefont {S.~R.}\ \bibnamefont
  {Das}}, \bibinfo {author} {\bibfnamefont {D.~A.}\ \bibnamefont {Galante}},\
  and\ \bibinfo {author} {\bibfnamefont {R.~C.}\ \bibnamefont {Myers}},\
  }\bibfield  {title} {\bibinfo {title} {Smooth and fast versus instantaneous
  quenches in quantum field theory},\ }\bibfield  {journal} {\bibinfo
  {journal} {Journal of High Energy Physics}\ }\textbf {\bibinfo {volume}
  {2015}},\ \href {https://doi.org/10.1007/jhep08(2015)073}
  {10.1007/jhep08(2015)073} (\bibinfo {year} {2015}{\natexlab{b}})\BibitemShut
  {NoStop}%
\bibitem [{\citenamefont {Das}\ \emph {et~al.}(2017)\citenamefont {Das},
  \citenamefont {Das}, \citenamefont {Galante}, \citenamefont {Myers},\ and\
  \citenamefont {Sengupta}}]{sdas4}%
  \BibitemOpen
  \bibfield  {author} {\bibinfo {author} {\bibfnamefont {D.}~\bibnamefont
  {Das}}, \bibinfo {author} {\bibfnamefont {S.~R.}\ \bibnamefont {Das}},
  \bibinfo {author} {\bibfnamefont {D.~A.}\ \bibnamefont {Galante}}, \bibinfo
  {author} {\bibfnamefont {R.~C.}\ \bibnamefont {Myers}},\ and\ \bibinfo
  {author} {\bibfnamefont {K.}~\bibnamefont {Sengupta}},\ }\bibfield  {title}
  {\bibinfo {title} {An exactly solvable quench protocol for integrable spin
  models},\ }\bibfield  {journal} {\bibinfo  {journal} {Journal of High Energy
  Physics}\ }\textbf {\bibinfo {volume} {2017}},\ \href
  {https://doi.org/10.1007/jhep11(2017)157} {10.1007/jhep11(2017)157} (\bibinfo
  {year} {2017})\BibitemShut {NoStop}%
\bibitem [{\citenamefont {Trefzger}\ and\ \citenamefont
  {Sengupta}(2011)}]{adutta1_part1}%
  \BibitemOpen
  \bibfield  {author} {\bibinfo {author} {\bibfnamefont {C.}~\bibnamefont
  {Trefzger}}\ and\ \bibinfo {author} {\bibfnamefont {K.}~\bibnamefont
  {Sengupta}},\ }\bibfield  {title} {\bibinfo {title} {Nonequilibrium dynamics
  of the bose-hubbard model: A projection-operator approach},\ }\href
  {https://doi.org/10.1103/PhysRevLett.106.095702} {\bibfield  {journal}
  {\bibinfo  {journal} {Phys. Rev. Lett.}\ }\textbf {\bibinfo {volume} {106}},\
  \bibinfo {pages} {095702} (\bibinfo {year} {2011})}\BibitemShut {NoStop}%
\bibitem [{\citenamefont {Dutta}\ \emph {et~al.}(2012)\citenamefont {Dutta},
  \citenamefont {Trefzger},\ and\ \citenamefont {Sengupta}}]{adutta1_part2}%
  \BibitemOpen
  \bibfield  {author} {\bibinfo {author} {\bibfnamefont {A.}~\bibnamefont
  {Dutta}}, \bibinfo {author} {\bibfnamefont {C.}~\bibnamefont {Trefzger}},\
  and\ \bibinfo {author} {\bibfnamefont {K.}~\bibnamefont {Sengupta}},\
  }\bibfield  {title} {\bibinfo {title} {Projection operator approach to the
  bose-hubbard model},\ }\href {https://doi.org/10.1103/PhysRevB.86.085140}
  {\bibfield  {journal} {\bibinfo  {journal} {Phys. Rev. B}\ }\textbf {\bibinfo
  {volume} {86}},\ \bibinfo {pages} {085140} (\bibinfo {year}
  {2012})}\BibitemShut {NoStop}%
\bibitem [{\citenamefont {Lin}\ \emph {et~al.}(2012)\citenamefont {Lin},
  \citenamefont {Sensarma}, \citenamefont {Sengupta},\ and\ \citenamefont
  {Das~Sarma}}]{rajdeep3}%
  \BibitemOpen
  \bibfield  {author} {\bibinfo {author} {\bibfnamefont {C.-H.}\ \bibnamefont
  {Lin}}, \bibinfo {author} {\bibfnamefont {R.}~\bibnamefont {Sensarma}},
  \bibinfo {author} {\bibfnamefont {K.}~\bibnamefont {Sengupta}},\ and\
  \bibinfo {author} {\bibfnamefont {S.}~\bibnamefont {Das~Sarma}},\ }\bibfield
  {title} {\bibinfo {title} {Quantum dynamics of disordered bosons in an
  optical lattice},\ }\bibfield  {journal} {\bibinfo  {journal} {Physical
  Review B}\ }\textbf {\bibinfo {volume} {86}},\ \href
  {https://doi.org/10.1103/physrevb.86.214207} {10.1103/physrevb.86.214207}
  (\bibinfo {year} {2012})\BibitemShut {NoStop}%
\bibitem [{\citenamefont {Das}(2010)}]{adas1}%
  \BibitemOpen
  \bibfield  {author} {\bibinfo {author} {\bibfnamefont {A.}~\bibnamefont
  {Das}},\ }\bibfield  {title} {\bibinfo {title} {Exotic freezing of response
  in a quantum many-body system},\ }\href
  {https://doi.org/10.1103/PhysRevB.82.172402} {\bibfield  {journal} {\bibinfo
  {journal} {Phys. Rev. B}\ }\textbf {\bibinfo {volume} {82}},\ \bibinfo
  {pages} {172402} (\bibinfo {year} {2010})}\BibitemShut {NoStop}%
\bibitem [{\citenamefont {Bhattacharyya}\ \emph {et~al.}(2012)\citenamefont
  {Bhattacharyya}, \citenamefont {Das},\ and\ \citenamefont
  {Dasgupta}}]{adas2}%
  \BibitemOpen
  \bibfield  {author} {\bibinfo {author} {\bibfnamefont {S.}~\bibnamefont
  {Bhattacharyya}}, \bibinfo {author} {\bibfnamefont {A.}~\bibnamefont {Das}},\
  and\ \bibinfo {author} {\bibfnamefont {S.}~\bibnamefont {Dasgupta}},\
  }\bibfield  {title} {\bibinfo {title} {Transverse ising chain under periodic
  instantaneous quenches: Dynamical many-body freezing and emergence of slow
  solitary oscillations},\ }\href {https://doi.org/10.1103/PhysRevB.86.054410}
  {\bibfield  {journal} {\bibinfo  {journal} {Phys. Rev. B}\ }\textbf {\bibinfo
  {volume} {86}},\ \bibinfo {pages} {054410} (\bibinfo {year}
  {2012})}\BibitemShut {NoStop}%
\bibitem [{\citenamefont {Hegde}\ \emph {et~al.}(2014)\citenamefont {Hegde},
  \citenamefont {Katiyar}, \citenamefont {Mahesh},\ and\ \citenamefont
  {Das}}]{adas3}%
  \BibitemOpen
  \bibfield  {author} {\bibinfo {author} {\bibfnamefont {S.~S.}\ \bibnamefont
  {Hegde}}, \bibinfo {author} {\bibfnamefont {H.}~\bibnamefont {Katiyar}},
  \bibinfo {author} {\bibfnamefont {T.~S.}\ \bibnamefont {Mahesh}},\ and\
  \bibinfo {author} {\bibfnamefont {A.}~\bibnamefont {Das}},\ }\bibfield
  {title} {\bibinfo {title} {Freezing a quantum magnet by repeated quantum
  interference: An experimental realization},\ }\href
  {https://doi.org/10.1103/PhysRevB.90.174407} {\bibfield  {journal} {\bibinfo
  {journal} {Phys. Rev. B}\ }\textbf {\bibinfo {volume} {90}},\ \bibinfo
  {pages} {174407} (\bibinfo {year} {2014})}\BibitemShut {NoStop}%
\bibitem [{\citenamefont {Mondal}\ \emph {et~al.}(2012)\citenamefont {Mondal},
  \citenamefont {Pekker},\ and\ \citenamefont {Sengupta}}]{pekker1}%
  \BibitemOpen
  \bibfield  {author} {\bibinfo {author} {\bibfnamefont {S.}~\bibnamefont
  {Mondal}}, \bibinfo {author} {\bibfnamefont {D.}~\bibnamefont {Pekker}},\
  and\ \bibinfo {author} {\bibfnamefont {K.}~\bibnamefont {Sengupta}},\
  }\bibfield  {title} {\bibinfo {title} {Dynamics-induced freezing of strongly
  correlated ultracold bosons},\ }\href
  {https://doi.org/10.1209/0295-5075/100/60007} {\bibfield  {journal} {\bibinfo
   {journal} {EPL (Europhysics Letters)}\ }\textbf {\bibinfo {volume} {100}},\
  \bibinfo {pages} {60007} (\bibinfo {year} {2012})}\BibitemShut {NoStop}%
\bibitem [{\citenamefont {Guo}\ \emph {et~al.}(2025)\citenamefont {Guo},
  \citenamefont {Mukherjee},\ and\ \citenamefont {Chowdhury}}]{deb1}%
  \BibitemOpen
  \bibfield  {author} {\bibinfo {author} {\bibfnamefont {H.}~\bibnamefont
  {Guo}}, \bibinfo {author} {\bibfnamefont {R.}~\bibnamefont {Mukherjee}},\
  and\ \bibinfo {author} {\bibfnamefont {D.}~\bibnamefont {Chowdhury}},\
  }\bibfield  {title} {\bibinfo {title} {Dynamical freezing in exactly solvable
  models of driven chaotic quantum dots},\ }\bibfield  {journal} {\bibinfo
  {journal} {Physical Review Letters}\ }\textbf {\bibinfo {volume} {134}},\
  \href {https://doi.org/10.1103/ggk3-6cf8} {10.1103/ggk3-6cf8} (\bibinfo
  {year} {2025})\BibitemShut {NoStop}%
\bibitem [{\citenamefont {Divakaran}\ and\ \citenamefont
  {Sengupta}(2014)}]{uma1}%
  \BibitemOpen
  \bibfield  {author} {\bibinfo {author} {\bibfnamefont {U.}~\bibnamefont
  {Divakaran}}\ and\ \bibinfo {author} {\bibfnamefont {K.}~\bibnamefont
  {Sengupta}},\ }\bibfield  {title} {\bibinfo {title} {Dynamic freezing and
  defect suppression in the tilted one-dimensional bose-hubbard model},\
  }\bibfield  {journal} {\bibinfo  {journal} {Physical Review B}\ }\textbf
  {\bibinfo {volume} {90}},\ \href {https://doi.org/10.1103/physrevb.90.184303}
  {10.1103/physrevb.90.184303} (\bibinfo {year} {2014})\BibitemShut {NoStop}%
\bibitem [{\citenamefont {Camilo}\ and\ \citenamefont
  {Teixeira}(2020)}]{camilo1}%
  \BibitemOpen
  \bibfield  {author} {\bibinfo {author} {\bibfnamefont {G.}~\bibnamefont
  {Camilo}}\ and\ \bibinfo {author} {\bibfnamefont {D.}~\bibnamefont
  {Teixeira}},\ }\bibfield  {title} {\bibinfo {title} {Complexity and floquet
  dynamics: Nonequilibrium ising phase transitions},\ }\bibfield  {journal}
  {\bibinfo  {journal} {Physical Review B}\ }\textbf {\bibinfo {volume}
  {102}},\ \href {https://doi.org/10.1103/physrevb.102.174304}
  {10.1103/physrevb.102.174304} (\bibinfo {year} {2020})\BibitemShut {NoStop}%
\bibitem [{\citenamefont {Mukherjee}\ \emph {et~al.}(2024)\citenamefont
  {Mukherjee}, \citenamefont {Melendrez}, \citenamefont {Szyniszewski},
  \citenamefont {Changlani},\ and\ \citenamefont {Pal}}]{apal1}%
  \BibitemOpen
  \bibfield  {author} {\bibinfo {author} {\bibfnamefont {B.}~\bibnamefont
  {Mukherjee}}, \bibinfo {author} {\bibfnamefont {R.}~\bibnamefont
  {Melendrez}}, \bibinfo {author} {\bibfnamefont {M.}~\bibnamefont
  {Szyniszewski}}, \bibinfo {author} {\bibfnamefont {H.~J.}\ \bibnamefont
  {Changlani}},\ and\ \bibinfo {author} {\bibfnamefont {A.}~\bibnamefont
  {Pal}},\ }\bibfield  {title} {\bibinfo {title} {Emergent strong zero mode
  through local floquet engineering},\ }\bibfield  {journal} {\bibinfo
  {journal} {Physical Review B}\ }\textbf {\bibinfo {volume} {109}},\ \href
  {https://doi.org/10.1103/physrevb.109.064303} {10.1103/physrevb.109.064303}
  (\bibinfo {year} {2024})\BibitemShut {NoStop}%
\bibitem [{\citenamefont {Koch}\ \emph {et~al.}(2023)\citenamefont {Koch},
  \citenamefont {Hunanyan}, \citenamefont {Ockenfels}, \citenamefont {Rico},
  \citenamefont {Solano},\ and\ \citenamefont {Weitz}}]{koch1}%
  \BibitemOpen
  \bibfield  {author} {\bibinfo {author} {\bibfnamefont {J.}~\bibnamefont
  {Koch}}, \bibinfo {author} {\bibfnamefont {G.~R.}\ \bibnamefont {Hunanyan}},
  \bibinfo {author} {\bibfnamefont {T.}~\bibnamefont {Ockenfels}}, \bibinfo
  {author} {\bibfnamefont {E.}~\bibnamefont {Rico}}, \bibinfo {author}
  {\bibfnamefont {E.}~\bibnamefont {Solano}},\ and\ \bibinfo {author}
  {\bibfnamefont {M.}~\bibnamefont {Weitz}},\ }\bibfield  {title} {\bibinfo
  {title} {Quantum rabi dynamics of trapped atoms far in the deep strong
  coupling regime},\ }\bibfield  {journal} {\bibinfo  {journal} {Nature
  Communications}\ }\textbf {\bibinfo {volume} {14}},\ \href
  {https://doi.org/10.1038/s41467-023-36611-z} {10.1038/s41467-023-36611-z}
  (\bibinfo {year} {2023})\BibitemShut {NoStop}%
\bibitem [{\citenamefont {Haldar}\ \emph {et~al.}(2021)\citenamefont {Haldar},
  \citenamefont {Sen}, \citenamefont {Moessner},\ and\ \citenamefont
  {Das}}]{adasnew}%
  \BibitemOpen
  \bibfield  {author} {\bibinfo {author} {\bibfnamefont {A.}~\bibnamefont
  {Haldar}}, \bibinfo {author} {\bibfnamefont {D.}~\bibnamefont {Sen}},
  \bibinfo {author} {\bibfnamefont {R.}~\bibnamefont {Moessner}},\ and\
  \bibinfo {author} {\bibfnamefont {A.}~\bibnamefont {Das}},\ }\bibfield
  {title} {\bibinfo {title} {Dynamical freezing and scar points in strongly
  driven floquet matter: Resonance vs emergent conservation laws},\ }\href
  {https://doi.org/10.1103/PhysRevX.11.021008} {\bibfield  {journal} {\bibinfo
  {journal} {Phys. Rev. X}\ }\textbf {\bibinfo {volume} {11}},\ \bibinfo
  {pages} {021008} (\bibinfo {year} {2021})}\BibitemShut {NoStop}%
\bibitem [{\citenamefont {Banerjee}\ and\ \citenamefont
  {Sengupta}(2023)}]{tb1}%
  \BibitemOpen
  \bibfield  {author} {\bibinfo {author} {\bibfnamefont {T.}~\bibnamefont
  {Banerjee}}\ and\ \bibinfo {author} {\bibfnamefont {K.}~\bibnamefont
  {Sengupta}},\ }\bibfield  {title} {\bibinfo {title} {Emergent conservation in
  the floquet dynamics of integrable non-hermitian models},\ }\href
  {https://doi.org/10.1103/PhysRevB.107.155117} {\bibfield  {journal} {\bibinfo
   {journal} {Phys. Rev. B}\ }\textbf {\bibinfo {volume} {107}},\ \bibinfo
  {pages} {155117} (\bibinfo {year} {2023})}\BibitemShut {NoStop}%
\bibitem [{\citenamefont {Banerjee}\ and\ \citenamefont
  {Sengupta}(2024)}]{tb2_part1}%
  \BibitemOpen
  \bibfield  {author} {\bibinfo {author} {\bibfnamefont {T.}~\bibnamefont
  {Banerjee}}\ and\ \bibinfo {author} {\bibfnamefont {K.}~\bibnamefont
  {Sengupta}},\ }\bibfield  {title} {\bibinfo {title} {Entanglement transitions
  in a periodically driven non-hermitian ising chain},\ }\href
  {https://doi.org/10.1103/PhysRevB.109.094306} {\bibfield  {journal} {\bibinfo
   {journal} {Phys. Rev. B}\ }\textbf {\bibinfo {volume} {109}},\ \bibinfo
  {pages} {094306} (\bibinfo {year} {2024})}\BibitemShut {NoStop}%
\bibitem [{\citenamefont {Turkeshi}\ and\ \citenamefont
  {Schir\'o}(2023)}]{Turkeshi_Schiro_part2}%
  \BibitemOpen
  \bibfield  {author} {\bibinfo {author} {\bibfnamefont {X.}~\bibnamefont
  {Turkeshi}}\ and\ \bibinfo {author} {\bibfnamefont {M.}~\bibnamefont
  {Schir\'o}},\ }\bibfield  {title} {\bibinfo {title} {Entanglement and
  correlation spreading in non-hermitian spin chains},\ }\href
  {https://doi.org/10.1103/PhysRevB.107.L020403} {\bibfield  {journal}
  {\bibinfo  {journal} {Phys. Rev. B}\ }\textbf {\bibinfo {volume} {107}},\
  \bibinfo {pages} {L020403} (\bibinfo {year} {2023})}\BibitemShut {NoStop}%
\bibitem [{\citenamefont {Gangopadhay}\ and\ \citenamefont
  {Choudhury}(2025)}]{cd1}%
  \BibitemOpen
  \bibfield  {author} {\bibinfo {author} {\bibfnamefont {N.}~\bibnamefont
  {Gangopadhay}}\ and\ \bibinfo {author} {\bibfnamefont {S.}~\bibnamefont
  {Choudhury}},\ }\bibfield  {title} {\bibinfo {title} {Counterdiabatic route
  to entanglement steering and dynamical freezing in the floquet
  lipkin-meshkov-glick model},\ }\href {https://doi.org/10.1103/bzcf-gm89}
  {\bibfield  {journal} {\bibinfo  {journal} {Phys. Rev. Lett.}\ }\textbf
  {\bibinfo {volume} {135}},\ \bibinfo {pages} {020407} (\bibinfo {year}
  {2025})}\BibitemShut {NoStop}%
\bibitem [{\citenamefont {Pai}\ and\ \citenamefont {Pretko}(2019)}]{pretko1}%
  \BibitemOpen
  \bibfield  {author} {\bibinfo {author} {\bibfnamefont {S.}~\bibnamefont
  {Pai}}\ and\ \bibinfo {author} {\bibfnamefont {M.}~\bibnamefont {Pretko}},\
  }\bibfield  {title} {\bibinfo {title} {Dynamical scar states in driven
  fracton systems},\ }\href {https://doi.org/10.1103/PhysRevLett.123.136401}
  {\bibfield  {journal} {\bibinfo  {journal} {Phys. Rev. Lett.}\ }\textbf
  {\bibinfo {volume} {123}},\ \bibinfo {pages} {136401} (\bibinfo {year}
  {2019})}\BibitemShut {NoStop}%
\bibitem [{\citenamefont {Mukherjee}\ \emph
  {et~al.}(2020{\natexlab{a}})\citenamefont {Mukherjee}, \citenamefont {Nandy},
  \citenamefont {Sen}, \citenamefont {Sen},\ and\ \citenamefont
  {Sengupta}}]{bm1}%
  \BibitemOpen
  \bibfield  {author} {\bibinfo {author} {\bibfnamefont {B.}~\bibnamefont
  {Mukherjee}}, \bibinfo {author} {\bibfnamefont {S.}~\bibnamefont {Nandy}},
  \bibinfo {author} {\bibfnamefont {A.}~\bibnamefont {Sen}}, \bibinfo {author}
  {\bibfnamefont {D.}~\bibnamefont {Sen}},\ and\ \bibinfo {author}
  {\bibfnamefont {K.}~\bibnamefont {Sengupta}},\ }\bibfield  {title} {\bibinfo
  {title} {Collapse and revival of quantum many-body scars via floquet
  engineering},\ }\href {https://doi.org/10.1103/PhysRevB.101.245107}
  {\bibfield  {journal} {\bibinfo  {journal} {Phys. Rev. B}\ }\textbf {\bibinfo
  {volume} {101}},\ \bibinfo {pages} {245107} (\bibinfo {year}
  {2020}{\natexlab{a}})}\BibitemShut {NoStop}%
\bibitem [{\citenamefont {Mizuta}\ \emph {et~al.}(2020)\citenamefont {Mizuta},
  \citenamefont {Takasan},\ and\ \citenamefont {Kawakami}}]{mituza1_part1}%
  \BibitemOpen
  \bibfield  {author} {\bibinfo {author} {\bibfnamefont {K.}~\bibnamefont
  {Mizuta}}, \bibinfo {author} {\bibfnamefont {K.}~\bibnamefont {Takasan}},\
  and\ \bibinfo {author} {\bibfnamefont {N.}~\bibnamefont {Kawakami}},\
  }\bibfield  {title} {\bibinfo {title} {Exact floquet quantum many-body scars
  under rydberg blockade},\ }\href
  {https://doi.org/10.1103/PhysRevResearch.2.033284} {\bibfield  {journal}
  {\bibinfo  {journal} {Phys. Rev. Res.}\ }\textbf {\bibinfo {volume} {2}},\
  \bibinfo {pages} {033284} (\bibinfo {year} {2020})}\BibitemShut {NoStop}%
\bibitem [{\citenamefont {Sugiura}\ \emph {et~al.}(2021)\citenamefont
  {Sugiura}, \citenamefont {Kuwahara},\ and\ \citenamefont
  {Saito}}]{Sug_Kun_Saito_part2}%
  \BibitemOpen
  \bibfield  {author} {\bibinfo {author} {\bibfnamefont {S.}~\bibnamefont
  {Sugiura}}, \bibinfo {author} {\bibfnamefont {T.}~\bibnamefont {Kuwahara}},\
  and\ \bibinfo {author} {\bibfnamefont {K.}~\bibnamefont {Saito}},\ }\bibfield
   {title} {\bibinfo {title} {Many-body scar state intrinsic to periodically
  driven system},\ }\href {https://doi.org/10.1103/PhysRevResearch.3.L012010}
  {\bibfield  {journal} {\bibinfo  {journal} {Phys. Rev. Res.}\ }\textbf
  {\bibinfo {volume} {3}},\ \bibinfo {pages} {L012010} (\bibinfo {year}
  {2021})}\BibitemShut {NoStop}%
\bibitem [{\citenamefont {Mukherjee}\ \emph
  {et~al.}(2020{\natexlab{b}})\citenamefont {Mukherjee}, \citenamefont {Sen},
  \citenamefont {Sen},\ and\ \citenamefont {Sengupta}}]{bm2_part1}%
  \BibitemOpen
  \bibfield  {author} {\bibinfo {author} {\bibfnamefont {B.}~\bibnamefont
  {Mukherjee}}, \bibinfo {author} {\bibfnamefont {A.}~\bibnamefont {Sen}},
  \bibinfo {author} {\bibfnamefont {D.}~\bibnamefont {Sen}},\ and\ \bibinfo
  {author} {\bibfnamefont {K.}~\bibnamefont {Sengupta}},\ }\bibfield  {title}
  {\bibinfo {title} {Dynamics of the vacuum state in a periodically driven
  rydberg chain},\ }\href {https://doi.org/10.1103/PhysRevB.102.075123}
  {\bibfield  {journal} {\bibinfo  {journal} {Phys. Rev. B}\ }\textbf {\bibinfo
  {volume} {102}},\ \bibinfo {pages} {075123} (\bibinfo {year}
  {2020}{\natexlab{b}})}\BibitemShut {NoStop}%
\bibitem [{\citenamefont {Maskara}\ \emph {et~al.}(2021)\citenamefont
  {Maskara}, \citenamefont {Michailidis}, \citenamefont {Ho}, \citenamefont
  {Bluvstein}, \citenamefont {Choi}, \citenamefont {Lukin},\ and\ \citenamefont
  {Serbyn}}]{lukinsc1}%
  \BibitemOpen
  \bibfield  {author} {\bibinfo {author} {\bibfnamefont {N.}~\bibnamefont
  {Maskara}}, \bibinfo {author} {\bibfnamefont {A.~A.}\ \bibnamefont
  {Michailidis}}, \bibinfo {author} {\bibfnamefont {W.~W.}\ \bibnamefont {Ho}},
  \bibinfo {author} {\bibfnamefont {D.}~\bibnamefont {Bluvstein}}, \bibinfo
  {author} {\bibfnamefont {S.}~\bibnamefont {Choi}}, \bibinfo {author}
  {\bibfnamefont {M.~D.}\ \bibnamefont {Lukin}},\ and\ \bibinfo {author}
  {\bibfnamefont {M.}~\bibnamefont {Serbyn}},\ }\bibfield  {title} {\bibinfo
  {title} {Discrete time-crystalline order enabled by quantum many-body scars:
  Entanglement steering via periodic driving},\ }\href
  {https://doi.org/10.1103/PhysRevLett.127.090602} {\bibfield  {journal}
  {\bibinfo  {journal} {Phys. Rev. Lett.}\ }\textbf {\bibinfo {volume} {127}},\
  \bibinfo {pages} {090602} (\bibinfo {year} {2021})}\BibitemShut {NoStop}%
\bibitem [{\citenamefont {Hudomal}\ \emph {et~al.}(2022)\citenamefont
  {Hudomal}, \citenamefont {Desaules}, \citenamefont {Mukherjee}, \citenamefont
  {Su}, \citenamefont {Halimeh},\ and\ \citenamefont {Papić}}]{papic1}%
  \BibitemOpen
  \bibfield  {author} {\bibinfo {author} {\bibfnamefont {A.}~\bibnamefont
  {Hudomal}}, \bibinfo {author} {\bibfnamefont {J.-Y.}\ \bibnamefont
  {Desaules}}, \bibinfo {author} {\bibfnamefont {B.}~\bibnamefont {Mukherjee}},
  \bibinfo {author} {\bibfnamefont {G.-X.}\ \bibnamefont {Su}}, \bibinfo
  {author} {\bibfnamefont {J.~C.}\ \bibnamefont {Halimeh}},\ and\ \bibinfo
  {author} {\bibfnamefont {Z.}~\bibnamefont {Papić}},\ }\bibfield  {title}
  {\bibinfo {title} {Driving quantum many-body scars in the pxp model},\
  }\bibfield  {journal} {\bibinfo  {journal} {Physical Review B}\ }\textbf
  {\bibinfo {volume} {106}},\ \href
  {https://doi.org/10.1103/physrevb.106.104302} {10.1103/physrevb.106.104302}
  (\bibinfo {year} {2022})\BibitemShut {NoStop}%
\bibitem [{\citenamefont {Huang}\ \emph {et~al.}(2022)\citenamefont {Huang},
  \citenamefont {Leung}, \citenamefont {Stamper-Kurn},\ and\ \citenamefont
  {Liu}}]{liu1}%
  \BibitemOpen
  \bibfield  {author} {\bibinfo {author} {\bibfnamefont {B.}~\bibnamefont
  {Huang}}, \bibinfo {author} {\bibfnamefont {T.-H.}\ \bibnamefont {Leung}},
  \bibinfo {author} {\bibfnamefont {D.~M.}\ \bibnamefont {Stamper-Kurn}},\ and\
  \bibinfo {author} {\bibfnamefont {W.~V.}\ \bibnamefont {Liu}},\ }\bibfield
  {title} {\bibinfo {title} {Discrete time crystals enforced by floquet-bloch
  scars},\ }\href {https://doi.org/10.1103/PhysRevLett.129.133001} {\bibfield
  {journal} {\bibinfo  {journal} {Phys. Rev. Lett.}\ }\textbf {\bibinfo
  {volume} {129}},\ \bibinfo {pages} {133001} (\bibinfo {year}
  {2022})}\BibitemShut {NoStop}%
\bibitem [{\citenamefont {Ghosh}\ \emph {et~al.}(2023)\citenamefont {Ghosh},
  \citenamefont {Paul},\ and\ \citenamefont {Sengupta}}]{sg1}%
  \BibitemOpen
  \bibfield  {author} {\bibinfo {author} {\bibfnamefont {S.}~\bibnamefont
  {Ghosh}}, \bibinfo {author} {\bibfnamefont {I.}~\bibnamefont {Paul}},\ and\
  \bibinfo {author} {\bibfnamefont {K.}~\bibnamefont {Sengupta}},\ }\bibfield
  {title} {\bibinfo {title} {Prethermal fragmentation in a periodically driven
  fermionic chain},\ }\href {https://doi.org/10.1103/PhysRevLett.130.120401}
  {\bibfield  {journal} {\bibinfo  {journal} {Phys. Rev. Lett.}\ }\textbf
  {\bibinfo {volume} {130}},\ \bibinfo {pages} {120401} (\bibinfo {year}
  {2023})}\BibitemShut {NoStop}%
\bibitem [{\citenamefont {Ghosh}\ \emph {et~al.}(2024)\citenamefont {Ghosh},
  \citenamefont {Paul},\ and\ \citenamefont {Sengupta}}]{sg2}%
  \BibitemOpen
  \bibfield  {author} {\bibinfo {author} {\bibfnamefont {S.}~\bibnamefont
  {Ghosh}}, \bibinfo {author} {\bibfnamefont {I.}~\bibnamefont {Paul}},\ and\
  \bibinfo {author} {\bibfnamefont {K.}~\bibnamefont {Sengupta}},\ }\bibfield
  {title} {\bibinfo {title} {Signatures of fragmentation for periodically
  driven fermions},\ }\href {https://doi.org/10.1103/PhysRevB.109.214304}
  {\bibfield  {journal} {\bibinfo  {journal} {Phys. Rev. B}\ }\textbf {\bibinfo
  {volume} {109}},\ \bibinfo {pages} {214304} (\bibinfo {year}
  {2024})}\BibitemShut {NoStop}%
\bibitem [{\citenamefont {Langlett}\ and\ \citenamefont {Xu}(2021)}]{xu1}%
  \BibitemOpen
  \bibfield  {author} {\bibinfo {author} {\bibfnamefont {C.~M.}\ \bibnamefont
  {Langlett}}\ and\ \bibinfo {author} {\bibfnamefont {S.}~\bibnamefont {Xu}},\
  }\bibfield  {title} {\bibinfo {title} {Hilbert space fragmentation and exact
  scars of generalized fredkin spin chains},\ }\bibfield  {journal} {\bibinfo
  {journal} {Physical Review B}\ }\textbf {\bibinfo {volume} {103}},\ \href
  {https://doi.org/10.1103/physrevb.103.l220304} {10.1103/physrevb.103.l220304}
  (\bibinfo {year} {2021})\BibitemShut {NoStop}%
\bibitem [{\citenamefont {Zhang}\ \emph {et~al.}(2024)\citenamefont {Zhang},
  \citenamefont {Ke}, \citenamefont {Lin},\ and\ \citenamefont {Lee}}]{zhang1}%
  \BibitemOpen
  \bibfield  {author} {\bibinfo {author} {\bibfnamefont {L.}~\bibnamefont
  {Zhang}}, \bibinfo {author} {\bibfnamefont {Y.}~\bibnamefont {Ke}}, \bibinfo
  {author} {\bibfnamefont {L.}~\bibnamefont {Lin}},\ and\ \bibinfo {author}
  {\bibfnamefont {C.}~\bibnamefont {Lee}},\ }\bibfield  {title} {\bibinfo
  {title} {Floquet engineering of hilbert space fragmentation in stark
  lattices},\ }\bibfield  {journal} {\bibinfo  {journal} {Physical Review B}\
  }\textbf {\bibinfo {volume} {109}},\ \href
  {https://doi.org/10.1103/physrevb.109.184313} {10.1103/physrevb.109.184313}
  (\bibinfo {year} {2024})\BibitemShut {NoStop}%
\bibitem [{\citenamefont {Baum}\ \emph {et~al.}(2018)\citenamefont {Baum},
  \citenamefont {van Nieuwenburg},\ and\ \citenamefont {Refael}}]{dynloc1}%
  \BibitemOpen
  \bibfield  {author} {\bibinfo {author} {\bibfnamefont {Y.}~\bibnamefont
  {Baum}}, \bibinfo {author} {\bibfnamefont {E.}~\bibnamefont {van
  Nieuwenburg}},\ and\ \bibinfo {author} {\bibfnamefont {G.}~\bibnamefont
  {Refael}},\ }\bibfield  {title} {\bibinfo {title} {From dynamical
  localization to bunching in interacting floquet systems},\ }\bibfield
  {journal} {\bibinfo  {journal} {SciPost Physics}\ }\textbf {\bibinfo {volume}
  {5}},\ \href {https://doi.org/10.21468/scipostphys.5.2.017}
  {10.21468/scipostphys.5.2.017} (\bibinfo {year} {2018})\BibitemShut {NoStop}%
\bibitem [{\citenamefont {Luitz}\ \emph {et~al.}(2017)\citenamefont {Luitz},
  \citenamefont {Bar~Lev},\ and\ \citenamefont {Lazarides}}]{dynloc2}%
  \BibitemOpen
  \bibfield  {author} {\bibinfo {author} {\bibfnamefont {D.~J.}\ \bibnamefont
  {Luitz}}, \bibinfo {author} {\bibfnamefont {Y.}~\bibnamefont {Bar~Lev}},\
  and\ \bibinfo {author} {\bibfnamefont {A.}~\bibnamefont {Lazarides}},\
  }\bibfield  {title} {\bibinfo {title} {Absence of dynamical localization in
  interacting driven systems},\ }\bibfield  {journal} {\bibinfo  {journal}
  {SciPost Physics}\ }\textbf {\bibinfo {volume} {3}},\ \href
  {https://doi.org/10.21468/scipostphys.3.4.029} {10.21468/scipostphys.3.4.029}
  (\bibinfo {year} {2017})\BibitemShut {NoStop}%
\bibitem [{\citenamefont {Agarwala}\ and\ \citenamefont {Sen}(2017)}]{dynloc3}%
  \BibitemOpen
  \bibfield  {author} {\bibinfo {author} {\bibfnamefont {A.}~\bibnamefont
  {Agarwala}}\ and\ \bibinfo {author} {\bibfnamefont {D.}~\bibnamefont {Sen}},\
  }\bibfield  {title} {\bibinfo {title} {Effects of interactions on
  periodically driven dynamically localized systems},\ }\bibfield  {journal}
  {\bibinfo  {journal} {Physical Review B}\ }\textbf {\bibinfo {volume} {95}},\
  \href {https://doi.org/10.1103/physrevb.95.014305}
  {10.1103/physrevb.95.014305} (\bibinfo {year} {2017})\BibitemShut {NoStop}%
\bibitem [{\citenamefont {Aditya}\ and\ \citenamefont {Sen}(2023)}]{dynloc4}%
  \BibitemOpen
  \bibfield  {author} {\bibinfo {author} {\bibfnamefont {S.}~\bibnamefont
  {Aditya}}\ and\ \bibinfo {author} {\bibfnamefont {D.}~\bibnamefont {Sen}},\
  }\bibfield  {title} {\bibinfo {title} {{Dynamical localization and slow
  thermalization in a class of disorder-free periodically driven
  one-dimensional interacting systems}},\ }\href
  {https://doi.org/10.21468/SciPostPhysCore.6.4.083} {\bibfield  {journal}
  {\bibinfo  {journal} {SciPost Phys. Core}\ }\textbf {\bibinfo {volume} {6}},\
  \bibinfo {pages} {083} (\bibinfo {year} {2023})}\BibitemShut {NoStop}%
\bibitem [{\citenamefont {Nag}\ \emph {et~al.}(2014)\citenamefont {Nag},
  \citenamefont {Roy}, \citenamefont {Dutta},\ and\ \citenamefont
  {Sen}}]{dynloc5}%
  \BibitemOpen
  \bibfield  {author} {\bibinfo {author} {\bibfnamefont {T.}~\bibnamefont
  {Nag}}, \bibinfo {author} {\bibfnamefont {S.}~\bibnamefont {Roy}}, \bibinfo
  {author} {\bibfnamefont {A.}~\bibnamefont {Dutta}},\ and\ \bibinfo {author}
  {\bibfnamefont {D.}~\bibnamefont {Sen}},\ }\bibfield  {title} {\bibinfo
  {title} {Dynamical localization in a chain of hard core bosons under periodic
  driving},\ }\href {https://doi.org/10.1103/PhysRevB.89.165425} {\bibfield
  {journal} {\bibinfo  {journal} {Phys. Rev. B}\ }\textbf {\bibinfo {volume}
  {89}},\ \bibinfo {pages} {165425} (\bibinfo {year} {2014})}\BibitemShut
  {NoStop}%
\bibitem [{\citenamefont {Tamang}\ \emph {et~al.}(2021)\citenamefont {Tamang},
  \citenamefont {Nag},\ and\ \citenamefont {Biswas}}]{tanay1}%
  \BibitemOpen
  \bibfield  {author} {\bibinfo {author} {\bibfnamefont {L.}~\bibnamefont
  {Tamang}}, \bibinfo {author} {\bibfnamefont {T.}~\bibnamefont {Nag}},\ and\
  \bibinfo {author} {\bibfnamefont {T.}~\bibnamefont {Biswas}},\ }\bibfield
  {title} {\bibinfo {title} {Floquet engineering of low-energy dispersions and
  dynamical localization in a periodically kicked three-band system},\
  }\bibfield  {journal} {\bibinfo  {journal} {Physical Review B}\ }\textbf
  {\bibinfo {volume} {104}},\ \href
  {https://doi.org/10.1103/physrevb.104.174308} {10.1103/physrevb.104.174308}
  (\bibinfo {year} {2021})\BibitemShut {NoStop}%
\bibitem [{\citenamefont {Fava}\ \emph {et~al.}(2020)\citenamefont {Fava},
  \citenamefont {Fazio},\ and\ \citenamefont {Russomanno}}]{fava1_part1}%
  \BibitemOpen
  \bibfield  {author} {\bibinfo {author} {\bibfnamefont {M.}~\bibnamefont
  {Fava}}, \bibinfo {author} {\bibfnamefont {R.}~\bibnamefont {Fazio}},\ and\
  \bibinfo {author} {\bibfnamefont {A.}~\bibnamefont {Russomanno}},\ }\bibfield
   {title} {\bibinfo {title} {Many-body dynamical localization in the kicked
  bose-hubbard chain},\ }\bibfield  {journal} {\bibinfo  {journal} {Physical
  Review B}\ }\textbf {\bibinfo {volume} {101}},\ \href
  {https://doi.org/10.1103/physrevb.101.064302} {10.1103/physrevb.101.064302}
  (\bibinfo {year} {2020})\BibitemShut {NoStop}%
\bibitem [{\citenamefont {Eckardt}\ \emph {et~al.}(2005)\citenamefont
  {Eckardt}, \citenamefont {Weiss},\ and\ \citenamefont
  {Holthaus}}]{holthus_part1}%
  \BibitemOpen
  \bibfield  {author} {\bibinfo {author} {\bibfnamefont {A.}~\bibnamefont
  {Eckardt}}, \bibinfo {author} {\bibfnamefont {C.}~\bibnamefont {Weiss}},\
  and\ \bibinfo {author} {\bibfnamefont {M.}~\bibnamefont {Holthaus}},\
  }\bibfield  {title} {\bibinfo {title} {Superfluid-insulator transition in a
  periodically driven optical lattice},\ }\href
  {https://doi.org/10.1103/PhysRevLett.95.260404} {\bibfield  {journal}
  {\bibinfo  {journal} {Phys. Rev. Lett.}\ }\textbf {\bibinfo {volume} {95}},\
  \bibinfo {pages} {260404} (\bibinfo {year} {2005})}\BibitemShut {NoStop}%
\bibitem [{\citenamefont {Ghosh}\ \emph {et~al.}(2020)\citenamefont {Ghosh},
  \citenamefont {Mukherjee},\ and\ \citenamefont {Sengupta}}]{rg1}%
  \BibitemOpen
  \bibfield  {author} {\bibinfo {author} {\bibfnamefont {R.}~\bibnamefont
  {Ghosh}}, \bibinfo {author} {\bibfnamefont {B.}~\bibnamefont {Mukherjee}},\
  and\ \bibinfo {author} {\bibfnamefont {K.}~\bibnamefont {Sengupta}},\
  }\bibfield  {title} {\bibinfo {title} {Floquet perturbation theory for
  periodically driven weakly interacting fermions},\ }\href
  {https://doi.org/10.1103/PhysRevB.102.235114} {\bibfield  {journal} {\bibinfo
   {journal} {Phys. Rev. B}\ }\textbf {\bibinfo {volume} {102}},\ \bibinfo
  {pages} {235114} (\bibinfo {year} {2020})}\BibitemShut {NoStop}%
\bibitem [{\citenamefont {Keser}\ \emph {et~al.}(2016)\citenamefont {Keser},
  \citenamefont {Ganeshan}, \citenamefont {Refael},\ and\ \citenamefont
  {Galitski}}]{galit1}%
  \BibitemOpen
  \bibfield  {author} {\bibinfo {author} {\bibfnamefont {A.~C.}\ \bibnamefont
  {Keser}}, \bibinfo {author} {\bibfnamefont {S.}~\bibnamefont {Ganeshan}},
  \bibinfo {author} {\bibfnamefont {G.}~\bibnamefont {Refael}},\ and\ \bibinfo
  {author} {\bibfnamefont {V.}~\bibnamefont {Galitski}},\ }\bibfield  {title}
  {\bibinfo {title} {Dynamical many-body localization in an integrable model},\
  }\bibfield  {journal} {\bibinfo  {journal} {Physical Review B}\ }\textbf
  {\bibinfo {volume} {94}},\ \href {https://doi.org/10.1103/physrevb.94.085120}
  {10.1103/physrevb.94.085120} (\bibinfo {year} {2016})\BibitemShut {NoStop}%
\bibitem [{\citenamefont {Martinez}(2005)}]{martinez1}%
  \BibitemOpen
  \bibfield  {author} {\bibinfo {author} {\bibfnamefont {D.~F.}\ \bibnamefont
  {Martinez}},\ }\bibfield  {title} {\bibinfo {title} {High-order harmonic
  generation and dynamic localization in a driven two-level system, a
  non-perturbative solution using the floquet–green formalism},\ }\href
  {https://doi.org/10.1088/0305-4470/38/46/006} {\bibfield  {journal} {\bibinfo
   {journal} {Journal of Physics A: Mathematical and General}\ }\textbf
  {\bibinfo {volume} {38}},\ \bibinfo {pages} {9979–10005} (\bibinfo {year}
  {2005})}\BibitemShut {NoStop}%
\bibitem [{\citenamefont {Creffield}(2007)}]{tun1}%
  \BibitemOpen
  \bibfield  {author} {\bibinfo {author} {\bibfnamefont {C.~E.}\ \bibnamefont
  {Creffield}},\ }\bibfield  {title} {\bibinfo {title} {Quantum control and
  entanglement using periodic driving fields},\ }\bibfield  {journal} {\bibinfo
   {journal} {Physical Review Letters}\ }\textbf {\bibinfo {volume} {99}},\
  \href {https://doi.org/10.1103/physrevlett.99.110501}
  {10.1103/physrevlett.99.110501} (\bibinfo {year} {2007})\BibitemShut
  {NoStop}%
\bibitem [{\citenamefont {Guo}\ \emph {et~al.}(2023)\citenamefont {Guo},
  \citenamefont {Dhar}, \citenamefont {Yang}, \citenamefont {Chen},
  \citenamefont {Yao}, \citenamefont {Horvath}, \citenamefont {Ying},
  \citenamefont {Landini},\ and\ \citenamefont {Nägerl}}]{guoexp}%
  \BibitemOpen
  \bibfield  {author} {\bibinfo {author} {\bibfnamefont {Y.}~\bibnamefont
  {Guo}}, \bibinfo {author} {\bibfnamefont {S.}~\bibnamefont {Dhar}}, \bibinfo
  {author} {\bibfnamefont {A.}~\bibnamefont {Yang}}, \bibinfo {author}
  {\bibfnamefont {Z.}~\bibnamefont {Chen}}, \bibinfo {author} {\bibfnamefont
  {H.}~\bibnamefont {Yao}}, \bibinfo {author} {\bibfnamefont {M.}~\bibnamefont
  {Horvath}}, \bibinfo {author} {\bibfnamefont {L.}~\bibnamefont {Ying}},
  \bibinfo {author} {\bibfnamefont {M.}~\bibnamefont {Landini}},\ and\ \bibinfo
  {author} {\bibfnamefont {H.-C.}\ \bibnamefont {Nägerl}},\ }\href
  {https://arxiv.org/abs/2312.13880} {\bibinfo {title} {Observation of
  many-body dynamical localization}} (\bibinfo {year} {2023}),\ \Eprint
  {https://arxiv.org/abs/2312.13880} {arXiv:2312.13880 [quant-ph]} \BibitemShut
  {NoStop}%
\bibitem [{\citenamefont {Heyl}\ \emph {et~al.}(2013)\citenamefont {Heyl},
  \citenamefont {Polkovnikov},\ and\ \citenamefont {Kehrein}}]{heyl1}%
  \BibitemOpen
  \bibfield  {author} {\bibinfo {author} {\bibfnamefont {M.}~\bibnamefont
  {Heyl}}, \bibinfo {author} {\bibfnamefont {A.}~\bibnamefont {Polkovnikov}},\
  and\ \bibinfo {author} {\bibfnamefont {S.}~\bibnamefont {Kehrein}},\
  }\bibfield  {title} {\bibinfo {title} {Dynamical quantum phase transitions in
  the transverse-field ising model},\ }\href
  {https://doi.org/10.1103/PhysRevLett.110.135704} {\bibfield  {journal}
  {\bibinfo  {journal} {Phys. Rev. Lett.}\ }\textbf {\bibinfo {volume} {110}},\
  \bibinfo {pages} {135704} (\bibinfo {year} {2013})}\BibitemShut {NoStop}%
\bibitem [{\citenamefont {Heyl}(2018)}]{heyl2}%
  \BibitemOpen
  \bibfield  {author} {\bibinfo {author} {\bibfnamefont {M.}~\bibnamefont
  {Heyl}},\ }\bibfield  {title} {\bibinfo {title} {Dynamical quantum phase
  transitions: a review},\ }\href {https://doi.org/10.1088/1361-6633/aaaf9a}
  {\bibfield  {journal} {\bibinfo  {journal} {Reports on Progress in Physics}\
  }\textbf {\bibinfo {volume} {81}},\ \bibinfo {pages} {054001} (\bibinfo
  {year} {2018})}\BibitemShut {NoStop}%
\bibitem [{\citenamefont {Sen}\ \emph {et~al.}(2016)\citenamefont {Sen},
  \citenamefont {Nandy},\ and\ \citenamefont {Sengupta}}]{ks1}%
  \BibitemOpen
  \bibfield  {author} {\bibinfo {author} {\bibfnamefont {A.}~\bibnamefont
  {Sen}}, \bibinfo {author} {\bibfnamefont {S.}~\bibnamefont {Nandy}},\ and\
  \bibinfo {author} {\bibfnamefont {K.}~\bibnamefont {Sengupta}},\ }\bibfield
  {title} {\bibinfo {title} {Entanglement generation in periodically driven
  integrable systems: Dynamical phase transitions and steady state},\ }\href
  {https://doi.org/10.1103/PhysRevB.94.214301} {\bibfield  {journal} {\bibinfo
  {journal} {Phys. Rev. B}\ }\textbf {\bibinfo {volume} {94}},\ \bibinfo
  {pages} {214301} (\bibinfo {year} {2016})}\BibitemShut {NoStop}%
\bibitem [{\citenamefont {Nandy}\ \emph
  {et~al.}(2018{\natexlab{a}})\citenamefont {Nandy}, \citenamefont {Sengupta},\
  and\ \citenamefont {Sen}}]{ks2}%
  \BibitemOpen
  \bibfield  {author} {\bibinfo {author} {\bibfnamefont {S.}~\bibnamefont
  {Nandy}}, \bibinfo {author} {\bibfnamefont {K.}~\bibnamefont {Sengupta}},\
  and\ \bibinfo {author} {\bibfnamefont {A.}~\bibnamefont {Sen}},\ }\bibfield
  {title} {\bibinfo {title} {Periodically driven integrable systems with
  long-range pair potentials},\ }\href
  {https://doi.org/10.1088/1751-8121/aaced6} {\bibfield  {journal} {\bibinfo
  {journal} {Journal of Physics A: Mathematical and Theoretical}\ }\textbf
  {\bibinfo {volume} {51}},\ \bibinfo {pages} {334002} (\bibinfo {year}
  {2018}{\natexlab{a}})}\BibitemShut {NoStop}%
\bibitem [{\citenamefont {Makki}\ \emph {et~al.}(2022)\citenamefont {Makki},
  \citenamefont {Bandyopadhyay}, \citenamefont {Maity},\ and\ \citenamefont
  {Dutta}}]{amit1}%
  \BibitemOpen
  \bibfield  {author} {\bibinfo {author} {\bibfnamefont {A.~A.}\ \bibnamefont
  {Makki}}, \bibinfo {author} {\bibfnamefont {S.}~\bibnamefont
  {Bandyopadhyay}}, \bibinfo {author} {\bibfnamefont {S.}~\bibnamefont
  {Maity}},\ and\ \bibinfo {author} {\bibfnamefont {A.}~\bibnamefont {Dutta}},\
  }\bibfield  {title} {\bibinfo {title} {Dynamical crossover behavior in the
  relaxation of quenched quantum many-body systems},\ }\href
  {https://doi.org/10.1103/PhysRevB.105.054301} {\bibfield  {journal} {\bibinfo
   {journal} {Phys. Rev. B}\ }\textbf {\bibinfo {volume} {105}},\ \bibinfo
  {pages} {054301} (\bibinfo {year} {2022})}\BibitemShut {NoStop}%
\bibitem [{\citenamefont {Aditya}\ \emph {et~al.}(2022)\citenamefont {Aditya},
  \citenamefont {Samanta}, \citenamefont {Sen}, \citenamefont {Sengupta},\ and\
  \citenamefont {Sen}}]{ks3}%
  \BibitemOpen
  \bibfield  {author} {\bibinfo {author} {\bibfnamefont {S.}~\bibnamefont
  {Aditya}}, \bibinfo {author} {\bibfnamefont {S.}~\bibnamefont {Samanta}},
  \bibinfo {author} {\bibfnamefont {A.}~\bibnamefont {Sen}}, \bibinfo {author}
  {\bibfnamefont {K.}~\bibnamefont {Sengupta}},\ and\ \bibinfo {author}
  {\bibfnamefont {D.}~\bibnamefont {Sen}},\ }\bibfield  {title} {\bibinfo
  {title} {Dynamical relaxation of correlators in periodically driven
  integrable quantum systems},\ }\href
  {https://doi.org/10.1103/PhysRevB.105.104303} {\bibfield  {journal} {\bibinfo
   {journal} {Phys. Rev. B}\ }\textbf {\bibinfo {volume} {105}},\ \bibinfo
  {pages} {104303} (\bibinfo {year} {2022})}\BibitemShut {NoStop}%
\bibitem [{\citenamefont {Kitagawa}\ \emph {et~al.}(2010)\citenamefont
  {Kitagawa}, \citenamefont {Berg}, \citenamefont {Rudner},\ and\ \citenamefont
  {Demler}}]{topo1}%
  \BibitemOpen
  \bibfield  {author} {\bibinfo {author} {\bibfnamefont {T.}~\bibnamefont
  {Kitagawa}}, \bibinfo {author} {\bibfnamefont {E.}~\bibnamefont {Berg}},
  \bibinfo {author} {\bibfnamefont {M.}~\bibnamefont {Rudner}},\ and\ \bibinfo
  {author} {\bibfnamefont {E.}~\bibnamefont {Demler}},\ }\bibfield  {title}
  {\bibinfo {title} {Topological characterization of periodically driven
  quantum systems},\ }\href {https://doi.org/10.1103/PhysRevB.82.235114}
  {\bibfield  {journal} {\bibinfo  {journal} {Phys. Rev. B}\ }\textbf {\bibinfo
  {volume} {82}},\ \bibinfo {pages} {235114} (\bibinfo {year}
  {2010})}\BibitemShut {NoStop}%
\bibitem [{\citenamefont {Lindner}\ \emph {et~al.}(2011)\citenamefont
  {Lindner}, \citenamefont {Refael},\ and\ \citenamefont {Galitski}}]{topo2}%
  \BibitemOpen
  \bibfield  {author} {\bibinfo {author} {\bibfnamefont {N.~H.}\ \bibnamefont
  {Lindner}}, \bibinfo {author} {\bibfnamefont {G.}~\bibnamefont {Refael}},\
  and\ \bibinfo {author} {\bibfnamefont {V.}~\bibnamefont {Galitski}},\
  }\bibfield  {title} {\bibinfo {title} {Floquet topological insulator in
  semiconductor quantum wells},\ }\href {https://doi.org/10.1038/nphys1926}
  {\bibfield  {journal} {\bibinfo  {journal} {Nature Physics}\ }\textbf
  {\bibinfo {volume} {7}},\ \bibinfo {pages} {490–495} (\bibinfo {year}
  {2011})}\BibitemShut {NoStop}%
\bibitem [{\citenamefont {Kitagawa}\ \emph {et~al.}(2011)\citenamefont
  {Kitagawa}, \citenamefont {Oka}, \citenamefont {Brataas}, \citenamefont
  {Fu},\ and\ \citenamefont {Demler}}]{topo3}%
  \BibitemOpen
  \bibfield  {author} {\bibinfo {author} {\bibfnamefont {T.}~\bibnamefont
  {Kitagawa}}, \bibinfo {author} {\bibfnamefont {T.}~\bibnamefont {Oka}},
  \bibinfo {author} {\bibfnamefont {A.}~\bibnamefont {Brataas}}, \bibinfo
  {author} {\bibfnamefont {L.}~\bibnamefont {Fu}},\ and\ \bibinfo {author}
  {\bibfnamefont {E.}~\bibnamefont {Demler}},\ }\bibfield  {title} {\bibinfo
  {title} {Transport properties of nonequilibrium systems under the application
  of light: Photoinduced quantum hall insulators without landau levels},\
  }\href {https://doi.org/10.1103/PhysRevB.84.235108} {\bibfield  {journal}
  {\bibinfo  {journal} {Phys. Rev. B}\ }\textbf {\bibinfo {volume} {84}},\
  \bibinfo {pages} {235108} (\bibinfo {year} {2011})}\BibitemShut {NoStop}%
\bibitem [{\citenamefont {Thakurathi}\ \emph {et~al.}(2013)\citenamefont
  {Thakurathi}, \citenamefont {Patel}, \citenamefont {Sen},\ and\ \citenamefont
  {Dutta}}]{topo4}%
  \BibitemOpen
  \bibfield  {author} {\bibinfo {author} {\bibfnamefont {M.}~\bibnamefont
  {Thakurathi}}, \bibinfo {author} {\bibfnamefont {A.~A.}\ \bibnamefont
  {Patel}}, \bibinfo {author} {\bibfnamefont {D.}~\bibnamefont {Sen}},\ and\
  \bibinfo {author} {\bibfnamefont {A.}~\bibnamefont {Dutta}},\ }\bibfield
  {title} {\bibinfo {title} {Floquet generation of majorana end modes and
  topological invariants},\ }\href {https://doi.org/10.1103/PhysRevB.88.155133}
  {\bibfield  {journal} {\bibinfo  {journal} {Phys. Rev. B}\ }\textbf {\bibinfo
  {volume} {88}},\ \bibinfo {pages} {155133} (\bibinfo {year}
  {2013})}\BibitemShut {NoStop}%
\bibitem [{\citenamefont {Kundu}\ \emph {et~al.}(2014)\citenamefont {Kundu},
  \citenamefont {Fertig},\ and\ \citenamefont {Seradjeh}}]{topo5}%
  \BibitemOpen
  \bibfield  {author} {\bibinfo {author} {\bibfnamefont {A.}~\bibnamefont
  {Kundu}}, \bibinfo {author} {\bibfnamefont {H.~A.}\ \bibnamefont {Fertig}},\
  and\ \bibinfo {author} {\bibfnamefont {B.}~\bibnamefont {Seradjeh}},\
  }\bibfield  {title} {\bibinfo {title} {Effective theory of floquet
  topological transitions},\ }\href
  {https://doi.org/10.1103/PhysRevLett.113.236803} {\bibfield  {journal}
  {\bibinfo  {journal} {Phys. Rev. Lett.}\ }\textbf {\bibinfo {volume} {113}},\
  \bibinfo {pages} {236803} (\bibinfo {year} {2014})}\BibitemShut {NoStop}%
\bibitem [{\citenamefont {Nathan}\ and\ \citenamefont {Rudner}(2015)}]{topo6}%
  \BibitemOpen
  \bibfield  {author} {\bibinfo {author} {\bibfnamefont {F.}~\bibnamefont
  {Nathan}}\ and\ \bibinfo {author} {\bibfnamefont {M.~S.}\ \bibnamefont
  {Rudner}},\ }\bibfield  {title} {\bibinfo {title} {Topological singularities
  and the general classification of floquet–bloch systems},\ }\href
  {https://doi.org/10.1088/1367-2630/17/12/125014} {\bibfield  {journal}
  {\bibinfo  {journal} {New Journal of Physics}\ }\textbf {\bibinfo {volume}
  {17}},\ \bibinfo {pages} {125014} (\bibinfo {year} {2015})}\BibitemShut
  {NoStop}%
\bibitem [{\citenamefont {Mukherjee}\ \emph {et~al.}(2016)\citenamefont
  {Mukherjee}, \citenamefont {Sen}, \citenamefont {Sen},\ and\ \citenamefont
  {Sengupta}}]{topo7}%
  \BibitemOpen
  \bibfield  {author} {\bibinfo {author} {\bibfnamefont {B.}~\bibnamefont
  {Mukherjee}}, \bibinfo {author} {\bibfnamefont {A.}~\bibnamefont {Sen}},
  \bibinfo {author} {\bibfnamefont {D.}~\bibnamefont {Sen}},\ and\ \bibinfo
  {author} {\bibfnamefont {K.}~\bibnamefont {Sengupta}},\ }\bibfield  {title}
  {\bibinfo {title} {Signatures and conditions for phase band crossings in
  periodically driven integrable systems},\ }\href
  {https://doi.org/10.1103/PhysRevB.94.155122} {\bibfield  {journal} {\bibinfo
  {journal} {Phys. Rev. B}\ }\textbf {\bibinfo {volume} {94}},\ \bibinfo
  {pages} {155122} (\bibinfo {year} {2016})}\BibitemShut {NoStop}%
\bibitem [{\citenamefont {Mukherjee}\ \emph {et~al.}(2018)\citenamefont
  {Mukherjee}, \citenamefont {Mohan}, \citenamefont {Sen},\ and\ \citenamefont
  {Sengupta}}]{topo8}%
  \BibitemOpen
  \bibfield  {author} {\bibinfo {author} {\bibfnamefont {B.}~\bibnamefont
  {Mukherjee}}, \bibinfo {author} {\bibfnamefont {P.}~\bibnamefont {Mohan}},
  \bibinfo {author} {\bibfnamefont {D.}~\bibnamefont {Sen}},\ and\ \bibinfo
  {author} {\bibfnamefont {K.}~\bibnamefont {Sengupta}},\ }\bibfield  {title}
  {\bibinfo {title} {Low-frequency phase diagram of irradiated graphene and a
  periodically driven spin-$\frac{1}{2} xy$ chain},\ }\href
  {https://doi.org/10.1103/PhysRevB.97.205415} {\bibfield  {journal} {\bibinfo
  {journal} {Phys. Rev. B}\ }\textbf {\bibinfo {volume} {97}},\ \bibinfo
  {pages} {205415} (\bibinfo {year} {2018})}\BibitemShut {NoStop}%
\bibitem [{\citenamefont {Mukherjee}(2018)}]{topo9}%
  \BibitemOpen
  \bibfield  {author} {\bibinfo {author} {\bibfnamefont {B.}~\bibnamefont
  {Mukherjee}},\ }\bibfield  {title} {\bibinfo {title} {Floquet topological
  transition by unpolarized light},\ }\bibfield  {journal} {\bibinfo  {journal}
  {Physical Review B}\ }\textbf {\bibinfo {volume} {98}},\ \href
  {https://doi.org/10.1103/physrevb.98.235112} {10.1103/physrevb.98.235112}
  (\bibinfo {year} {2018})\BibitemShut {NoStop}%
\bibitem [{\citenamefont {Yao}\ and\ \citenamefont {Nayak}(2018)}]{tcrev1}%
  \BibitemOpen
  \bibfield  {author} {\bibinfo {author} {\bibfnamefont {N.~Y.}\ \bibnamefont
  {Yao}}\ and\ \bibinfo {author} {\bibfnamefont {C.}~\bibnamefont {Nayak}},\
  }\bibfield  {title} {\bibinfo {title} {Time crystals in periodically driven
  systems},\ }\href {https://doi.org/10.1063/pt.3.4020} {\bibfield  {journal}
  {\bibinfo  {journal} {Physics Today}\ }\textbf {\bibinfo {volume} {71}},\
  \bibinfo {pages} {40–47} (\bibinfo {year} {2018})}\BibitemShut {NoStop}%
\bibitem [{\citenamefont {Khemani}\ \emph {et~al.}(2019)\citenamefont
  {Khemani}, \citenamefont {Moessner},\ and\ \citenamefont {Sondhi}}]{tcrev2}%
  \BibitemOpen
  \bibfield  {author} {\bibinfo {author} {\bibfnamefont {V.}~\bibnamefont
  {Khemani}}, \bibinfo {author} {\bibfnamefont {R.}~\bibnamefont {Moessner}},\
  and\ \bibinfo {author} {\bibfnamefont {S.~L.}\ \bibnamefont {Sondhi}},\
  }\href {https://arxiv.org/abs/1910.10745} {\bibinfo {title} {A brief history
  of time crystals}} (\bibinfo {year} {2019}),\ \Eprint
  {https://arxiv.org/abs/1910.10745} {arXiv:1910.10745 [cond-mat.str-el]}
  \BibitemShut {NoStop}%
\bibitem [{\citenamefont {Else}\ \emph
  {et~al.}(2020{\natexlab{a}})\citenamefont {Else}, \citenamefont {Monroe},
  \citenamefont {Nayak},\ and\ \citenamefont {Yao}}]{tcrev3}%
  \BibitemOpen
  \bibfield  {author} {\bibinfo {author} {\bibfnamefont {D.~V.}\ \bibnamefont
  {Else}}, \bibinfo {author} {\bibfnamefont {C.}~\bibnamefont {Monroe}},
  \bibinfo {author} {\bibfnamefont {C.}~\bibnamefont {Nayak}},\ and\ \bibinfo
  {author} {\bibfnamefont {N.~Y.}\ \bibnamefont {Yao}},\ }\bibfield  {title}
  {\bibinfo {title} {Discrete time crystals},\ }\href
  {https://doi.org/10.1146/annurev-conmatphys-031119-050658} {\bibfield
  {journal} {\bibinfo  {journal} {Annual Review of Condensed Matter Physics}\
  }\textbf {\bibinfo {volume} {11}},\ \bibinfo {pages} {467–499} (\bibinfo
  {year} {2020}{\natexlab{a}})}\BibitemShut {NoStop}%
\bibitem [{\citenamefont {Zaletel}\ \emph {et~al.}(2023)\citenamefont
  {Zaletel}, \citenamefont {Lukin}, \citenamefont {Monroe}, \citenamefont
  {Nayak}, \citenamefont {Wilczek},\ and\ \citenamefont {Yao}}]{tcrev4}%
  \BibitemOpen
  \bibfield  {author} {\bibinfo {author} {\bibfnamefont {M.~P.}\ \bibnamefont
  {Zaletel}}, \bibinfo {author} {\bibfnamefont {M.}~\bibnamefont {Lukin}},
  \bibinfo {author} {\bibfnamefont {C.}~\bibnamefont {Monroe}}, \bibinfo
  {author} {\bibfnamefont {C.}~\bibnamefont {Nayak}}, \bibinfo {author}
  {\bibfnamefont {F.}~\bibnamefont {Wilczek}},\ and\ \bibinfo {author}
  {\bibfnamefont {N.~Y.}\ \bibnamefont {Yao}},\ }\bibfield  {title} {\bibinfo
  {title} {Colloquium : Quantum and classical discrete time crystals},\
  }\bibfield  {journal} {\bibinfo  {journal} {Reviews of Modern Physics}\
  }\textbf {\bibinfo {volume} {95}},\ \href
  {https://doi.org/10.1103/revmodphys.95.031001} {10.1103/revmodphys.95.031001}
  (\bibinfo {year} {2023})\BibitemShut {NoStop}%
\bibitem [{\citenamefont {Sacha}\ and\ \citenamefont
  {Zakrzewski}(2017)}]{tcrev5}%
  \BibitemOpen
  \bibfield  {author} {\bibinfo {author} {\bibfnamefont {K.}~\bibnamefont
  {Sacha}}\ and\ \bibinfo {author} {\bibfnamefont {J.}~\bibnamefont
  {Zakrzewski}},\ }\bibfield  {title} {\bibinfo {title} {Time crystals: a
  review},\ }\href {https://doi.org/10.1088/1361-6633/aa8b38} {\bibfield
  {journal} {\bibinfo  {journal} {Reports on Progress in Physics}\ }\textbf
  {\bibinfo {volume} {81}},\ \bibinfo {pages} {016401} (\bibinfo {year}
  {2017})}\BibitemShut {NoStop}%
\bibitem [{\citenamefont {von Keyserlingk}\ \emph {et~al.}(2016)\citenamefont
  {von Keyserlingk}, \citenamefont {Khemani},\ and\ \citenamefont
  {Sondhi}}]{tcpap1_part1}%
  \BibitemOpen
  \bibfield  {author} {\bibinfo {author} {\bibfnamefont {C.~W.}\ \bibnamefont
  {von Keyserlingk}}, \bibinfo {author} {\bibfnamefont {V.}~\bibnamefont
  {Khemani}},\ and\ \bibinfo {author} {\bibfnamefont {S.~L.}\ \bibnamefont
  {Sondhi}},\ }\bibfield  {title} {\bibinfo {title} {Absolute stability and
  spatiotemporal long-range order in floquet systems},\ }\href
  {https://doi.org/10.1103/PhysRevB.94.085112} {\bibfield  {journal} {\bibinfo
  {journal} {Phys. Rev. B}\ }\textbf {\bibinfo {volume} {94}},\ \bibinfo
  {pages} {085112} (\bibinfo {year} {2016})}\BibitemShut {NoStop}%
\bibitem [{\citenamefont {Moessner}\ and\ \citenamefont
  {Sondhi}(2017)}]{tcpap1_part2}%
  \BibitemOpen
  \bibfield  {author} {\bibinfo {author} {\bibfnamefont {R.}~\bibnamefont
  {Moessner}}\ and\ \bibinfo {author} {\bibfnamefont {S.~L.}\ \bibnamefont
  {Sondhi}},\ }\bibfield  {title} {\bibinfo {title} {Equilibration and order in
  quantum floquet matter},\ }\href {https://doi.org/10.1038/nphys4106}
  {\bibfield  {journal} {\bibinfo  {journal} {Nature Physics}\ }\textbf
  {\bibinfo {volume} {13}},\ \bibinfo {pages} {424–428} (\bibinfo {year}
  {2017})}\BibitemShut {NoStop}%
\bibitem [{\citenamefont {Else}\ \emph {et~al.}(2016)\citenamefont {Else},
  \citenamefont {Bauer},\ and\ \citenamefont {Nayak}}]{tcpap2_part1}%
  \BibitemOpen
  \bibfield  {author} {\bibinfo {author} {\bibfnamefont {D.~V.}\ \bibnamefont
  {Else}}, \bibinfo {author} {\bibfnamefont {B.}~\bibnamefont {Bauer}},\ and\
  \bibinfo {author} {\bibfnamefont {C.}~\bibnamefont {Nayak}},\ }\bibfield
  {title} {\bibinfo {title} {Floquet time crystals},\ }\href
  {https://doi.org/10.1103/PhysRevLett.117.090402} {\bibfield  {journal}
  {\bibinfo  {journal} {Phys. Rev. Lett.}\ }\textbf {\bibinfo {volume} {117}},\
  \bibinfo {pages} {090402} (\bibinfo {year} {2016})}\BibitemShut {NoStop}%
\bibitem [{\citenamefont {Else}\ \emph {et~al.}(2017)\citenamefont {Else},
  \citenamefont {Bauer},\ and\ \citenamefont {Nayak}}]{tcpap2_part2}%
  \BibitemOpen
  \bibfield  {author} {\bibinfo {author} {\bibfnamefont {D.~V.}\ \bibnamefont
  {Else}}, \bibinfo {author} {\bibfnamefont {B.}~\bibnamefont {Bauer}},\ and\
  \bibinfo {author} {\bibfnamefont {C.}~\bibnamefont {Nayak}},\ }\bibfield
  {title} {\bibinfo {title} {Prethermal phases of matter protected by
  time-translation symmetry},\ }\href
  {https://doi.org/10.1103/PhysRevX.7.011026} {\bibfield  {journal} {\bibinfo
  {journal} {Phys. Rev. X}\ }\textbf {\bibinfo {volume} {7}},\ \bibinfo {pages}
  {011026} (\bibinfo {year} {2017})}\BibitemShut {NoStop}%
\bibitem [{\citenamefont {Yao}\ \emph {et~al.}(2017)\citenamefont {Yao},
  \citenamefont {Potter}, \citenamefont {Potirniche},\ and\ \citenamefont
  {Vishwanath}}]{tcpap3}%
  \BibitemOpen
  \bibfield  {author} {\bibinfo {author} {\bibfnamefont {N.~Y.}\ \bibnamefont
  {Yao}}, \bibinfo {author} {\bibfnamefont {A.~C.}\ \bibnamefont {Potter}},
  \bibinfo {author} {\bibfnamefont {I.-D.}\ \bibnamefont {Potirniche}},\ and\
  \bibinfo {author} {\bibfnamefont {A.}~\bibnamefont {Vishwanath}},\ }\bibfield
   {title} {\bibinfo {title} {Discrete time crystals: Rigidity, criticality,
  and realizations},\ }\href {https://doi.org/10.1103/PhysRevLett.118.030401}
  {\bibfield  {journal} {\bibinfo  {journal} {Phys. Rev. Lett.}\ }\textbf
  {\bibinfo {volume} {118}},\ \bibinfo {pages} {030401} (\bibinfo {year}
  {2017})}\BibitemShut {NoStop}%
\bibitem [{\citenamefont {Abal}\ \emph {et~al.}(2002)\citenamefont {Abal},
  \citenamefont {Donangelo}, \citenamefont {Romanelli}, \citenamefont
  {Sicardi~Schifino},\ and\ \citenamefont {Siri}}]{q1}%
  \BibitemOpen
  \bibfield  {author} {\bibinfo {author} {\bibfnamefont {G.}~\bibnamefont
  {Abal}}, \bibinfo {author} {\bibfnamefont {R.}~\bibnamefont {Donangelo}},
  \bibinfo {author} {\bibfnamefont {A.}~\bibnamefont {Romanelli}}, \bibinfo
  {author} {\bibfnamefont {A.~C.}\ \bibnamefont {Sicardi~Schifino}},\ and\
  \bibinfo {author} {\bibfnamefont {R.}~\bibnamefont {Siri}},\ }\bibfield
  {title} {\bibinfo {title} {Dynamical localization in quasiperiodic driven
  systems},\ }\bibfield  {journal} {\bibinfo  {journal} {Physical Review E}\
  }\textbf {\bibinfo {volume} {65}},\ \href
  {https://doi.org/10.1103/physreve.65.046236} {10.1103/physreve.65.046236}
  (\bibinfo {year} {2002})\BibitemShut {NoStop}%
\bibitem [{\citenamefont {Verdeny}\ \emph {et~al.}(2016)\citenamefont
  {Verdeny}, \citenamefont {Puig},\ and\ \citenamefont {Mintert}}]{q2}%
  \BibitemOpen
  \bibfield  {author} {\bibinfo {author} {\bibfnamefont {A.}~\bibnamefont
  {Verdeny}}, \bibinfo {author} {\bibfnamefont {J.}~\bibnamefont {Puig}},\ and\
  \bibinfo {author} {\bibfnamefont {F.}~\bibnamefont {Mintert}},\ }\bibfield
  {title} {\bibinfo {title} {Quasi-periodically driven quantum systems},\
  }\href {https://doi.org/10.1515/zna-2016-0079} {\bibfield  {journal}
  {\bibinfo  {journal} {Zeitschrift f\"{u}r Naturforschung A}\ }\textbf
  {\bibinfo {volume} {71}},\ \bibinfo {pages} {897–907} (\bibinfo {year}
  {2016})}\BibitemShut {NoStop}%
\bibitem [{\citenamefont {Nandy}\ \emph {et~al.}(2017)\citenamefont {Nandy},
  \citenamefont {Sen},\ and\ \citenamefont {Sen}}]{q3}%
  \BibitemOpen
  \bibfield  {author} {\bibinfo {author} {\bibfnamefont {S.}~\bibnamefont
  {Nandy}}, \bibinfo {author} {\bibfnamefont {A.}~\bibnamefont {Sen}},\ and\
  \bibinfo {author} {\bibfnamefont {D.}~\bibnamefont {Sen}},\ }\bibfield
  {title} {\bibinfo {title} {Aperiodically driven integrable systems and their
  emergent steady states},\ }\href {https://doi.org/10.1103/PhysRevX.7.031034}
  {\bibfield  {journal} {\bibinfo  {journal} {Phys. Rev. X}\ }\textbf {\bibinfo
  {volume} {7}},\ \bibinfo {pages} {031034} (\bibinfo {year}
  {2017})}\BibitemShut {NoStop}%
\bibitem [{\citenamefont {Dumitrescu}\ \emph {et~al.}(2018)\citenamefont
  {Dumitrescu}, \citenamefont {Vasseur},\ and\ \citenamefont {Potter}}]{q4}%
  \BibitemOpen
  \bibfield  {author} {\bibinfo {author} {\bibfnamefont {P.~T.}\ \bibnamefont
  {Dumitrescu}}, \bibinfo {author} {\bibfnamefont {R.}~\bibnamefont
  {Vasseur}},\ and\ \bibinfo {author} {\bibfnamefont {A.~C.}\ \bibnamefont
  {Potter}},\ }\bibfield  {title} {\bibinfo {title} {Logarithmically slow
  relaxation in quasiperiodically driven random spin chains},\ }\href
  {https://doi.org/10.1103/PhysRevLett.120.070602} {\bibfield  {journal}
  {\bibinfo  {journal} {Phys. Rev. Lett.}\ }\textbf {\bibinfo {volume} {120}},\
  \bibinfo {pages} {070602} (\bibinfo {year} {2018})}\BibitemShut {NoStop}%
\bibitem [{\citenamefont {Nandy}\ \emph
  {et~al.}(2018{\natexlab{b}})\citenamefont {Nandy}, \citenamefont {Sen},\ and\
  \citenamefont {Sen}}]{q5}%
  \BibitemOpen
  \bibfield  {author} {\bibinfo {author} {\bibfnamefont {S.}~\bibnamefont
  {Nandy}}, \bibinfo {author} {\bibfnamefont {A.}~\bibnamefont {Sen}},\ and\
  \bibinfo {author} {\bibfnamefont {D.}~\bibnamefont {Sen}},\ }\bibfield
  {title} {\bibinfo {title} {Steady states of a quasiperiodically driven
  integrable system},\ }\bibfield  {journal} {\bibinfo  {journal} {Physical
  Review B}\ }\textbf {\bibinfo {volume} {98}},\ \href
  {https://doi.org/10.1103/physrevb.98.245144} {10.1103/physrevb.98.245144}
  (\bibinfo {year} {2018}{\natexlab{b}})\BibitemShut {NoStop}%
\bibitem [{\citenamefont {Giergiel}\ \emph {et~al.}(2019)\citenamefont
  {Giergiel}, \citenamefont {Kuro\ifmmode~\acute{s}\else \'{s}\fi{}},\ and\
  \citenamefont {Sacha}}]{q6}%
  \BibitemOpen
  \bibfield  {author} {\bibinfo {author} {\bibfnamefont {K.}~\bibnamefont
  {Giergiel}}, \bibinfo {author} {\bibfnamefont {A.}~\bibnamefont
  {Kuro\ifmmode~\acute{s}\else \'{s}\fi{}}},\ and\ \bibinfo {author}
  {\bibfnamefont {K.}~\bibnamefont {Sacha}},\ }\bibfield  {title} {\bibinfo
  {title} {Discrete time quasicrystals},\ }\href
  {https://doi.org/10.1103/PhysRevB.99.220303} {\bibfield  {journal} {\bibinfo
  {journal} {Phys. Rev. B}\ }\textbf {\bibinfo {volume} {99}},\ \bibinfo
  {pages} {220303} (\bibinfo {year} {2019})}\BibitemShut {NoStop}%
\bibitem [{\citenamefont {Ray}\ \emph {et~al.}(2019)\citenamefont {Ray},
  \citenamefont {Sinha},\ and\ \citenamefont {Sen}}]{q7}%
  \BibitemOpen
  \bibfield  {author} {\bibinfo {author} {\bibfnamefont {S.}~\bibnamefont
  {Ray}}, \bibinfo {author} {\bibfnamefont {S.}~\bibnamefont {Sinha}},\ and\
  \bibinfo {author} {\bibfnamefont {D.}~\bibnamefont {Sen}},\ }\bibfield
  {title} {\bibinfo {title} {Dynamics of quasiperiodically driven spin
  systems},\ }\bibfield  {journal} {\bibinfo  {journal} {Physical Review E}\
  }\textbf {\bibinfo {volume} {100}},\ \href
  {https://doi.org/10.1103/physreve.100.052129} {10.1103/physreve.100.052129}
  (\bibinfo {year} {2019})\BibitemShut {NoStop}%
\bibitem [{\citenamefont {Zhao}\ \emph {et~al.}(2019)\citenamefont {Zhao},
  \citenamefont {Mintert},\ and\ \citenamefont {Knolle}}]{q8}%
  \BibitemOpen
  \bibfield  {author} {\bibinfo {author} {\bibfnamefont {H.}~\bibnamefont
  {Zhao}}, \bibinfo {author} {\bibfnamefont {F.}~\bibnamefont {Mintert}},\ and\
  \bibinfo {author} {\bibfnamefont {J.}~\bibnamefont {Knolle}},\ }\bibfield
  {title} {\bibinfo {title} {Floquet time spirals and stable discrete-time
  quasicrystals in quasiperiodically driven quantum many-body systems},\
  }\bibfield  {journal} {\bibinfo  {journal} {Physical Review B}\ }\textbf
  {\bibinfo {volume} {100}},\ \href
  {https://doi.org/10.1103/physrevb.100.134302} {10.1103/physrevb.100.134302}
  (\bibinfo {year} {2019})\BibitemShut {NoStop}%
\bibitem [{\citenamefont {Maity}\ \emph {et~al.}(2019)\citenamefont {Maity},
  \citenamefont {Bhattacharya}, \citenamefont {Dutta},\ and\ \citenamefont
  {Sen}}]{q9}%
  \BibitemOpen
  \bibfield  {author} {\bibinfo {author} {\bibfnamefont {S.}~\bibnamefont
  {Maity}}, \bibinfo {author} {\bibfnamefont {U.}~\bibnamefont {Bhattacharya}},
  \bibinfo {author} {\bibfnamefont {A.}~\bibnamefont {Dutta}},\ and\ \bibinfo
  {author} {\bibfnamefont {D.}~\bibnamefont {Sen}},\ }\bibfield  {title}
  {\bibinfo {title} {Fibonacci steady states in a driven integrable quantum
  system},\ }\bibfield  {journal} {\bibinfo  {journal} {Physical Review B}\
  }\textbf {\bibinfo {volume} {99}},\ \href
  {https://doi.org/10.1103/physrevb.99.020306} {10.1103/physrevb.99.020306}
  (\bibinfo {year} {2019})\BibitemShut {NoStop}%
\bibitem [{\citenamefont {Else}\ \emph
  {et~al.}(2020{\natexlab{b}})\citenamefont {Else}, \citenamefont {Ho},\ and\
  \citenamefont {Dumitrescu}}]{q10}%
  \BibitemOpen
  \bibfield  {author} {\bibinfo {author} {\bibfnamefont {D.~V.}\ \bibnamefont
  {Else}}, \bibinfo {author} {\bibfnamefont {W.~W.}\ \bibnamefont {Ho}},\ and\
  \bibinfo {author} {\bibfnamefont {P.~T.}\ \bibnamefont {Dumitrescu}},\
  }\bibfield  {title} {\bibinfo {title} {Long-lived interacting phases of
  matter protected by multiple time-translation symmetries in quasiperiodically
  driven systems},\ }\bibfield  {journal} {\bibinfo  {journal} {Physical Review
  X}\ }\textbf {\bibinfo {volume} {10}},\ \href
  {https://doi.org/10.1103/physrevx.10.021032} {10.1103/physrevx.10.021032}
  (\bibinfo {year} {2020}{\natexlab{b}})\BibitemShut {NoStop}%
\bibitem [{\citenamefont {Mukherjee}\ \emph
  {et~al.}(2020{\natexlab{c}})\citenamefont {Mukherjee}, \citenamefont {Sen},
  \citenamefont {Sen},\ and\ \citenamefont {Sengupta}}]{q11}%
  \BibitemOpen
  \bibfield  {author} {\bibinfo {author} {\bibfnamefont {B.}~\bibnamefont
  {Mukherjee}}, \bibinfo {author} {\bibfnamefont {A.}~\bibnamefont {Sen}},
  \bibinfo {author} {\bibfnamefont {D.}~\bibnamefont {Sen}},\ and\ \bibinfo
  {author} {\bibfnamefont {K.}~\bibnamefont {Sengupta}},\ }\bibfield  {title}
  {\bibinfo {title} {Restoring coherence via aperiodic drives in a many-body
  quantum system},\ }\bibfield  {journal} {\bibinfo  {journal} {Physical Review
  B}\ }\textbf {\bibinfo {volume} {102}},\ \href
  {https://doi.org/10.1103/physrevb.102.014301} {10.1103/physrevb.102.014301}
  (\bibinfo {year} {2020}{\natexlab{c}})\BibitemShut {NoStop}%
\bibitem [{\citenamefont {Zhao}\ \emph {et~al.}(2021)\citenamefont {Zhao},
  \citenamefont {Mintert}, \citenamefont {Moessner},\ and\ \citenamefont
  {Knolle}}]{q12}%
  \BibitemOpen
  \bibfield  {author} {\bibinfo {author} {\bibfnamefont {H.}~\bibnamefont
  {Zhao}}, \bibinfo {author} {\bibfnamefont {F.}~\bibnamefont {Mintert}},
  \bibinfo {author} {\bibfnamefont {R.}~\bibnamefont {Moessner}},\ and\
  \bibinfo {author} {\bibfnamefont {J.}~\bibnamefont {Knolle}},\ }\bibfield
  {title} {\bibinfo {title} {Random multipolar driving: Tunably slow heating
  through spectral engineering},\ }\bibfield  {journal} {\bibinfo  {journal}
  {Physical Review Letters}\ }\textbf {\bibinfo {volume} {126}},\ \href
  {https://doi.org/10.1103/physrevlett.126.040601}
  {10.1103/physrevlett.126.040601} (\bibinfo {year} {2021})\BibitemShut
  {NoStop}%
\bibitem [{\citenamefont {Mori}\ \emph {et~al.}(2021)\citenamefont {Mori},
  \citenamefont {Zhao}, \citenamefont {Mintert}, \citenamefont {Knolle},\ and\
  \citenamefont {Moessner}}]{q13}%
  \BibitemOpen
  \bibfield  {author} {\bibinfo {author} {\bibfnamefont {T.}~\bibnamefont
  {Mori}}, \bibinfo {author} {\bibfnamefont {H.}~\bibnamefont {Zhao}}, \bibinfo
  {author} {\bibfnamefont {F.}~\bibnamefont {Mintert}}, \bibinfo {author}
  {\bibfnamefont {J.}~\bibnamefont {Knolle}},\ and\ \bibinfo {author}
  {\bibfnamefont {R.}~\bibnamefont {Moessner}},\ }\bibfield  {title} {\bibinfo
  {title} {Rigorous bounds on the heating rate in thue-morse quasiperiodically
  and randomly driven quantum many-body systems},\ }\bibfield  {journal}
  {\bibinfo  {journal} {Physical Review Letters}\ }\textbf {\bibinfo {volume}
  {127}},\ \href {https://doi.org/10.1103/physrevlett.127.050602}
  {10.1103/physrevlett.127.050602} (\bibinfo {year} {2021})\BibitemShut
  {NoStop}%
\bibitem [{\citenamefont {Zhao}\ \emph
  {et~al.}(2022{\natexlab{a}})\citenamefont {Zhao}, \citenamefont {Mintert},
  \citenamefont {Knolle},\ and\ \citenamefont {Moessner}}]{q14}%
  \BibitemOpen
  \bibfield  {author} {\bibinfo {author} {\bibfnamefont {H.}~\bibnamefont
  {Zhao}}, \bibinfo {author} {\bibfnamefont {F.}~\bibnamefont {Mintert}},
  \bibinfo {author} {\bibfnamefont {J.}~\bibnamefont {Knolle}},\ and\ \bibinfo
  {author} {\bibfnamefont {R.}~\bibnamefont {Moessner}},\ }\bibfield  {title}
  {\bibinfo {title} {Localization persisting under aperiodic driving},\
  }\bibfield  {journal} {\bibinfo  {journal} {Physical Review B}\ }\textbf
  {\bibinfo {volume} {105}},\ \href
  {https://doi.org/10.1103/physrevb.105.l220202} {10.1103/physrevb.105.l220202}
  (\bibinfo {year} {2022}{\natexlab{a}})\BibitemShut {NoStop}%
\bibitem [{\citenamefont {Long}\ \emph {et~al.}(2022)\citenamefont {Long},
  \citenamefont {Crowley},\ and\ \citenamefont {Chandran}}]{q15}%
  \BibitemOpen
  \bibfield  {author} {\bibinfo {author} {\bibfnamefont {D.~M.}\ \bibnamefont
  {Long}}, \bibinfo {author} {\bibfnamefont {P.~J.~D.}\ \bibnamefont
  {Crowley}},\ and\ \bibinfo {author} {\bibfnamefont {A.}~\bibnamefont
  {Chandran}},\ }\bibfield  {title} {\bibinfo {title} {Many-body localization
  with quasiperiodic driving},\ }\bibfield  {journal} {\bibinfo  {journal}
  {Physical Review B}\ }\textbf {\bibinfo {volume} {105}},\ \href
  {https://doi.org/10.1103/physrevb.105.144204} {10.1103/physrevb.105.144204}
  (\bibinfo {year} {2022})\BibitemShut {NoStop}%
\bibitem [{\citenamefont {Martin}\ \emph {et~al.}(2022)\citenamefont {Martin},
  \citenamefont {Martin},\ and\ \citenamefont {Agarwal}}]{q16}%
  \BibitemOpen
  \bibfield  {author} {\bibinfo {author} {\bibfnamefont {T.}~\bibnamefont
  {Martin}}, \bibinfo {author} {\bibfnamefont {I.}~\bibnamefont {Martin}},\
  and\ \bibinfo {author} {\bibfnamefont {K.}~\bibnamefont {Agarwal}},\
  }\bibfield  {title} {\bibinfo {title} {Effect of quasiperiodic and random
  noise on many-body dynamical decoupling protocols},\ }\bibfield  {journal}
  {\bibinfo  {journal} {Physical Review B}\ }\textbf {\bibinfo {volume}
  {106}},\ \href {https://doi.org/10.1103/physrevb.106.134306}
  {10.1103/physrevb.106.134306} (\bibinfo {year} {2022})\BibitemShut {NoStop}%
\bibitem [{\citenamefont {Zhao}\ \emph
  {et~al.}(2022{\natexlab{b}})\citenamefont {Zhao}, \citenamefont {Knolle},
  \citenamefont {Moessner},\ and\ \citenamefont {Mintert}}]{q17}%
  \BibitemOpen
  \bibfield  {author} {\bibinfo {author} {\bibfnamefont {H.}~\bibnamefont
  {Zhao}}, \bibinfo {author} {\bibfnamefont {J.}~\bibnamefont {Knolle}},
  \bibinfo {author} {\bibfnamefont {R.}~\bibnamefont {Moessner}},\ and\
  \bibinfo {author} {\bibfnamefont {F.}~\bibnamefont {Mintert}},\ }\bibfield
  {title} {\bibinfo {title} {Suppression of interband heating for random
  driving},\ }\bibfield  {journal} {\bibinfo  {journal} {Physical Review
  Letters}\ }\textbf {\bibinfo {volume} {129}},\ \href
  {https://doi.org/10.1103/physrevlett.129.120605}
  {10.1103/physrevlett.129.120605} (\bibinfo {year}
  {2022}{\natexlab{b}})\BibitemShut {NoStop}%
\bibitem [{\citenamefont {Zhao}\ \emph
  {et~al.}(2022{\natexlab{c}})\citenamefont {Zhao}, \citenamefont {Rudner},
  \citenamefont {Moessner},\ and\ \citenamefont {Knolle}}]{q18}%
  \BibitemOpen
  \bibfield  {author} {\bibinfo {author} {\bibfnamefont {H.}~\bibnamefont
  {Zhao}}, \bibinfo {author} {\bibfnamefont {M.~S.}\ \bibnamefont {Rudner}},
  \bibinfo {author} {\bibfnamefont {R.}~\bibnamefont {Moessner}},\ and\
  \bibinfo {author} {\bibfnamefont {J.}~\bibnamefont {Knolle}},\ }\bibfield
  {title} {\bibinfo {title} {Anomalous random multipolar driven insulators},\
  }\bibfield  {journal} {\bibinfo  {journal} {Physical Review B}\ }\textbf
  {\bibinfo {volume} {105}},\ \href
  {https://doi.org/10.1103/physrevb.105.245119} {10.1103/physrevb.105.245119}
  (\bibinfo {year} {2022}{\natexlab{c}})\BibitemShut {NoStop}%
\bibitem [{\citenamefont {Das}\ \emph {et~al.}(2023)\citenamefont {Das},
  \citenamefont {Bhakuni}, \citenamefont {Santos},\ and\ \citenamefont
  {Sharma}}]{q19}%
  \BibitemOpen
  \bibfield  {author} {\bibinfo {author} {\bibfnamefont {P.}~\bibnamefont
  {Das}}, \bibinfo {author} {\bibfnamefont {D.~S.}\ \bibnamefont {Bhakuni}},
  \bibinfo {author} {\bibfnamefont {L.~F.}\ \bibnamefont {Santos}},\ and\
  \bibinfo {author} {\bibfnamefont {A.}~\bibnamefont {Sharma}},\ }\bibfield
  {title} {\bibinfo {title} {Periodically and quasiperiodically driven
  anisotropic dicke model},\ }\bibfield  {journal} {\bibinfo  {journal}
  {Physical Review A}\ }\textbf {\bibinfo {volume} {108}},\ \href
  {https://doi.org/10.1103/physreva.108.063716} {10.1103/physreva.108.063716}
  (\bibinfo {year} {2023})\BibitemShut {NoStop}%
\bibitem [{\citenamefont {Dutta}\ \emph {et~al.}(2025)\citenamefont {Dutta},
  \citenamefont {Choudhury},\ and\ \citenamefont {Shukla}}]{q20}%
  \BibitemOpen
  \bibfield  {author} {\bibinfo {author} {\bibfnamefont {P.}~\bibnamefont
  {Dutta}}, \bibinfo {author} {\bibfnamefont {S.}~\bibnamefont {Choudhury}},\
  and\ \bibinfo {author} {\bibfnamefont {V.}~\bibnamefont {Shukla}},\
  }\bibfield  {title} {\bibinfo {title} {Prethermalization in the pxp model
  under continuous quasiperiodic driving},\ }\bibfield  {journal} {\bibinfo
  {journal} {Physical Review B}\ }\textbf {\bibinfo {volume} {111}},\ \href
  {https://doi.org/10.1103/physrevb.111.064303} {10.1103/physrevb.111.064303}
  (\bibinfo {year} {2025})\BibitemShut {NoStop}%
\bibitem [{\citenamefont {Liu}\ \emph {et~al.}(2025)\citenamefont {Liu},
  \citenamefont {Liu}, \citenamefont {Liang}, \citenamefont {Deng},
  \citenamefont {Chen}, \citenamefont {Shi}, \citenamefont {Li}, \citenamefont
  {Zhang}, \citenamefont {Chen}, \citenamefont {Fang}, \citenamefont {Feng},
  \citenamefont {Gu}, \citenamefont {He}, \citenamefont {Huang}, \citenamefont
  {Li}, \citenamefont {Liu}, \citenamefont {Li}, \citenamefont {Mei},
  \citenamefont {Peng}, \citenamefont {Song}, \citenamefont {Wang},
  \citenamefont {Wang}, \citenamefont {Wang}, \citenamefont {Xiao},
  \citenamefont {Xu}, \citenamefont {Xu}, \citenamefont {Yan}, \citenamefont
  {Yu}, \citenamefont {Yuan}, \citenamefont {Zhang}, \citenamefont {Zhao},
  \citenamefont {Zhao}, \citenamefont {Zhou}, \citenamefont {Wang},
  \citenamefont {Song}, \citenamefont {Tian}, \citenamefont {Mintert},
  \citenamefont {Knolle}, \citenamefont {Moessner}, \citenamefont {Zhang},
  \citenamefont {Zhang}, \citenamefont {Xiang}, \citenamefont {Zheng},
  \citenamefont {Xu}, \citenamefont {Zhao},\ and\ \citenamefont {Fan}}]{q21}%
  \BibitemOpen
  \bibfield  {author} {\bibinfo {author} {\bibfnamefont {Z.-H.}\ \bibnamefont
  {Liu}}, \bibinfo {author} {\bibfnamefont {Y.}~\bibnamefont {Liu}}, \bibinfo
  {author} {\bibfnamefont {G.-H.}\ \bibnamefont {Liang}}, \bibinfo {author}
  {\bibfnamefont {C.-L.}\ \bibnamefont {Deng}}, \bibinfo {author}
  {\bibfnamefont {K.}~\bibnamefont {Chen}}, \bibinfo {author} {\bibfnamefont
  {Y.-H.}\ \bibnamefont {Shi}}, \bibinfo {author} {\bibfnamefont {T.-M.}\
  \bibnamefont {Li}}, \bibinfo {author} {\bibfnamefont {L.}~\bibnamefont
  {Zhang}}, \bibinfo {author} {\bibfnamefont {B.-J.}\ \bibnamefont {Chen}},
  \bibinfo {author} {\bibfnamefont {C.-P.}\ \bibnamefont {Fang}}, \bibinfo
  {author} {\bibfnamefont {D.}~\bibnamefont {Feng}}, \bibinfo {author}
  {\bibfnamefont {X.-Y.}\ \bibnamefont {Gu}}, \bibinfo {author} {\bibfnamefont
  {Y.}~\bibnamefont {He}}, \bibinfo {author} {\bibfnamefont {K.}~\bibnamefont
  {Huang}}, \bibinfo {author} {\bibfnamefont {H.}~\bibnamefont {Li}}, \bibinfo
  {author} {\bibfnamefont {H.-T.}\ \bibnamefont {Liu}}, \bibinfo {author}
  {\bibfnamefont {L.}~\bibnamefont {Li}}, \bibinfo {author} {\bibfnamefont
  {Z.-Y.}\ \bibnamefont {Mei}}, \bibinfo {author} {\bibfnamefont {Z.-Y.}\
  \bibnamefont {Peng}}, \bibinfo {author} {\bibfnamefont {J.-C.}\ \bibnamefont
  {Song}}, \bibinfo {author} {\bibfnamefont {M.-C.}\ \bibnamefont {Wang}},
  \bibinfo {author} {\bibfnamefont {S.-L.}\ \bibnamefont {Wang}}, \bibinfo
  {author} {\bibfnamefont {Z.}~\bibnamefont {Wang}}, \bibinfo {author}
  {\bibfnamefont {Y.}~\bibnamefont {Xiao}}, \bibinfo {author} {\bibfnamefont
  {M.}~\bibnamefont {Xu}}, \bibinfo {author} {\bibfnamefont {Y.-S.}\
  \bibnamefont {Xu}}, \bibinfo {author} {\bibfnamefont {Y.}~\bibnamefont
  {Yan}}, \bibinfo {author} {\bibfnamefont {Y.-H.}\ \bibnamefont {Yu}},
  \bibinfo {author} {\bibfnamefont {W.-P.}\ \bibnamefont {Yuan}}, \bibinfo
  {author} {\bibfnamefont {J.-C.}\ \bibnamefont {Zhang}}, \bibinfo {author}
  {\bibfnamefont {J.-J.}\ \bibnamefont {Zhao}}, \bibinfo {author}
  {\bibfnamefont {K.}~\bibnamefont {Zhao}}, \bibinfo {author} {\bibfnamefont
  {S.-Y.}\ \bibnamefont {Zhou}}, \bibinfo {author} {\bibfnamefont {Z.-A.}\
  \bibnamefont {Wang}}, \bibinfo {author} {\bibfnamefont {X.}~\bibnamefont
  {Song}}, \bibinfo {author} {\bibfnamefont {Y.}~\bibnamefont {Tian}}, \bibinfo
  {author} {\bibfnamefont {F.}~\bibnamefont {Mintert}}, \bibinfo {author}
  {\bibfnamefont {J.}~\bibnamefont {Knolle}}, \bibinfo {author} {\bibfnamefont
  {R.}~\bibnamefont {Moessner}}, \bibinfo {author} {\bibfnamefont {Y.-R.}\
  \bibnamefont {Zhang}}, \bibinfo {author} {\bibfnamefont {P.}~\bibnamefont
  {Zhang}}, \bibinfo {author} {\bibfnamefont {Z.}~\bibnamefont {Xiang}},
  \bibinfo {author} {\bibfnamefont {D.}~\bibnamefont {Zheng}}, \bibinfo
  {author} {\bibfnamefont {K.}~\bibnamefont {Xu}}, \bibinfo {author}
  {\bibfnamefont {H.}~\bibnamefont {Zhao}},\ and\ \bibinfo {author}
  {\bibfnamefont {H.}~\bibnamefont {Fan}},\ }\href
  {https://arxiv.org/abs/2503.21553} {\bibinfo {title} {Prethermalization by
  random multipolar driving on a 78-qubit superconducting processor}} (\bibinfo
  {year} {2025}),\ \Eprint {https://arxiv.org/abs/2503.21553} {arXiv:2503.21553
  [quant-ph]} \BibitemShut {NoStop}%
\bibitem [{\citenamefont {Kumar}\ and\ \citenamefont {Choudhury}(2024)}]{q22}%
  \BibitemOpen
  \bibfield  {author} {\bibinfo {author} {\bibfnamefont {S.}~\bibnamefont
  {Kumar}}\ and\ \bibinfo {author} {\bibfnamefont {S.}~\bibnamefont
  {Choudhury}},\ }\bibfield  {title} {\bibinfo {title} {Prethermalization in
  aperiodically driven classical spin systems},\ }\href
  {https://doi.org/10.1103/PhysRevE.110.064150} {\bibfield  {journal} {\bibinfo
   {journal} {Phys. Rev. E}\ }\textbf {\bibinfo {volume} {110}},\ \bibinfo
  {pages} {064150} (\bibinfo {year} {2024})}\BibitemShut {NoStop}%
\bibitem [{\citenamefont {Moon}\ \emph {et~al.}(2024)\citenamefont {Moon},
  \citenamefont {Schindler}, \citenamefont {Sun}, \citenamefont {Druga},
  \citenamefont {Knolle}, \citenamefont {Moessner}, \citenamefont {Zhao},
  \citenamefont {Bukov},\ and\ \citenamefont {Ajoy}}]{q23}%
  \BibitemOpen
  \bibfield  {author} {\bibinfo {author} {\bibfnamefont {L.~J.~I.}\
  \bibnamefont {Moon}}, \bibinfo {author} {\bibfnamefont {P.~M.}\ \bibnamefont
  {Schindler}}, \bibinfo {author} {\bibfnamefont {Y.}~\bibnamefont {Sun}},
  \bibinfo {author} {\bibfnamefont {E.}~\bibnamefont {Druga}}, \bibinfo
  {author} {\bibfnamefont {J.}~\bibnamefont {Knolle}}, \bibinfo {author}
  {\bibfnamefont {R.}~\bibnamefont {Moessner}}, \bibinfo {author}
  {\bibfnamefont {H.}~\bibnamefont {Zhao}}, \bibinfo {author} {\bibfnamefont
  {M.}~\bibnamefont {Bukov}},\ and\ \bibinfo {author} {\bibfnamefont
  {A.}~\bibnamefont {Ajoy}},\ }\href {https://arxiv.org/abs/2404.05620}
  {\bibinfo {title} {Experimental observation of a time rondeau crystal:
  Temporal disorder in spatiotemporal order}} (\bibinfo {year} {2024}),\
  \Eprint {https://arxiv.org/abs/2404.05620} {arXiv:2404.05620 [quant-ph]}
  \BibitemShut {NoStop}%
\bibitem [{\citenamefont {He}\ \emph {et~al.}(2025)\citenamefont {He},
  \citenamefont {Ye}, \citenamefont {Gong}, \citenamefont {Yao}, \citenamefont
  {Liu}, \citenamefont {Murch}, \citenamefont {Yao},\ and\ \citenamefont
  {Zu}}]{q24}%
  \BibitemOpen
  \bibfield  {author} {\bibinfo {author} {\bibfnamefont {G.}~\bibnamefont
  {He}}, \bibinfo {author} {\bibfnamefont {B.}~\bibnamefont {Ye}}, \bibinfo
  {author} {\bibfnamefont {R.}~\bibnamefont {Gong}}, \bibinfo {author}
  {\bibfnamefont {C.}~\bibnamefont {Yao}}, \bibinfo {author} {\bibfnamefont
  {Z.}~\bibnamefont {Liu}}, \bibinfo {author} {\bibfnamefont {K.~W.}\
  \bibnamefont {Murch}}, \bibinfo {author} {\bibfnamefont {N.~Y.}\ \bibnamefont
  {Yao}},\ and\ \bibinfo {author} {\bibfnamefont {C.}~\bibnamefont {Zu}},\
  }\bibfield  {title} {\bibinfo {title} {Experimental realization of discrete
  time quasicrystals},\ }\href {https://doi.org/10.1103/PhysRevX.15.011055}
  {\bibfield  {journal} {\bibinfo  {journal} {Phys. Rev. X}\ }\textbf {\bibinfo
  {volume} {15}},\ \bibinfo {pages} {011055} (\bibinfo {year}
  {2025})}\BibitemShut {NoStop}%
\bibitem [{\citenamefont {Banerjee}\ \emph {et~al.}(2024)\citenamefont
  {Banerjee}, \citenamefont {Choudhury},\ and\ \citenamefont
  {Sengupta}}]{flat1}%
  \BibitemOpen
  \bibfield  {author} {\bibinfo {author} {\bibfnamefont {T.}~\bibnamefont
  {Banerjee}}, \bibinfo {author} {\bibfnamefont {S.}~\bibnamefont
  {Choudhury}},\ and\ \bibinfo {author} {\bibfnamefont {K.}~\bibnamefont
  {Sengupta}},\ }\href {https://arxiv.org/abs/2404.06536} {\bibinfo {title}
  {Exact floquet flat band and heating suppression via two-rate drive
  protocols}} (\bibinfo {year} {2024}),\ \Eprint
  {https://arxiv.org/abs/2404.06536} {arXiv:2404.06536 [cond-mat.stat-mech]}
  \BibitemShut {NoStop}%
\bibitem [{\citenamefont {Deutsch}(1991)}]{eth1}%
  \BibitemOpen
  \bibfield  {author} {\bibinfo {author} {\bibfnamefont {J.~M.}\ \bibnamefont
  {Deutsch}},\ }\bibfield  {title} {\bibinfo {title} {Quantum statistical
  mechanics in a closed system},\ }\href
  {https://doi.org/10.1103/physreva.43.2046} {\bibfield  {journal} {\bibinfo
  {journal} {Physical Review A}\ }\textbf {\bibinfo {volume} {43}},\ \bibinfo
  {pages} {2046–2049} (\bibinfo {year} {1991})}\BibitemShut {NoStop}%
\bibitem [{\citenamefont {Srednicki}(1994)}]{eth2a}%
  \BibitemOpen
  \bibfield  {author} {\bibinfo {author} {\bibfnamefont {M.}~\bibnamefont
  {Srednicki}},\ }\bibfield  {title} {\bibinfo {title} {Chaos and quantum
  thermalization},\ }\href {https://doi.org/10.1103/physreve.50.888} {\bibfield
   {journal} {\bibinfo  {journal} {Physical Review E}\ }\textbf {\bibinfo
  {volume} {50}},\ \bibinfo {pages} {888–901} (\bibinfo {year}
  {1994})}\BibitemShut {NoStop}%
\bibitem [{\citenamefont {Srednicki}(1999)}]{eth2b}%
  \BibitemOpen
  \bibfield  {author} {\bibinfo {author} {\bibfnamefont {M.}~\bibnamefont
  {Srednicki}},\ }\bibfield  {title} {\bibinfo {title} {The approach to thermal
  equilibrium in quantized chaotic systems},\ }\href
  {https://doi.org/10.1088/0305-4470/32/7/007} {\bibfield  {journal} {\bibinfo
  {journal} {Journal of Physics A: Mathematical and General}\ }\textbf
  {\bibinfo {volume} {32}},\ \bibinfo {pages} {1163–1175} (\bibinfo {year}
  {1999})}\BibitemShut {NoStop}%
\bibitem [{\citenamefont {Rigol}\ \emph {et~al.}(2008)\citenamefont {Rigol},
  \citenamefont {Dunjko},\ and\ \citenamefont {Olshanii}}]{eth3}%
  \BibitemOpen
  \bibfield  {author} {\bibinfo {author} {\bibfnamefont {M.}~\bibnamefont
  {Rigol}}, \bibinfo {author} {\bibfnamefont {V.}~\bibnamefont {Dunjko}},\ and\
  \bibinfo {author} {\bibfnamefont {M.}~\bibnamefont {Olshanii}},\ }\bibfield
  {title} {\bibinfo {title} {Thermalization and its mechanism for generic
  isolated quantum systems},\ }\href {https://doi.org/10.1038/nature06838}
  {\bibfield  {journal} {\bibinfo  {journal} {Nature}\ }\textbf {\bibinfo
  {volume} {452}},\ \bibinfo {pages} {854–858} (\bibinfo {year}
  {2008})}\BibitemShut {NoStop}%
\bibitem [{\citenamefont {D’Alessio}\ and\ \citenamefont
  {Rigol}(2014)}]{eth4}%
  \BibitemOpen
  \bibfield  {author} {\bibinfo {author} {\bibfnamefont {L.}~\bibnamefont
  {D’Alessio}}\ and\ \bibinfo {author} {\bibfnamefont {M.}~\bibnamefont
  {Rigol}},\ }\bibfield  {title} {\bibinfo {title} {Long-time behavior of
  isolated periodically driven interacting lattice systems},\ }\bibfield
  {journal} {\bibinfo  {journal} {Physical Review X}\ }\textbf {\bibinfo
  {volume} {4}},\ \href {https://doi.org/10.1103/physrevx.4.041048}
  {10.1103/physrevx.4.041048} (\bibinfo {year} {2014})\BibitemShut {NoStop}%
\bibitem [{\citenamefont {Sachdev}\ \emph {et~al.}(2002)\citenamefont
  {Sachdev}, \citenamefont {Sengupta},\ and\ \citenamefont {Girvin}}]{subir2}%
  \BibitemOpen
  \bibfield  {author} {\bibinfo {author} {\bibfnamefont {S.}~\bibnamefont
  {Sachdev}}, \bibinfo {author} {\bibfnamefont {K.}~\bibnamefont {Sengupta}},\
  and\ \bibinfo {author} {\bibfnamefont {S.~M.}\ \bibnamefont {Girvin}},\
  }\bibfield  {title} {\bibinfo {title} {Mott insulators in strong electric
  fields},\ }\href {https://doi.org/10.1103/PhysRevB.66.075128} {\bibfield
  {journal} {\bibinfo  {journal} {Phys. Rev. B}\ }\textbf {\bibinfo {volume}
  {66}},\ \bibinfo {pages} {075128} (\bibinfo {year} {2002})}\BibitemShut
  {NoStop}%
\bibitem [{\citenamefont {Fendley}\ \emph {et~al.}(2004)\citenamefont
  {Fendley}, \citenamefont {Sengupta},\ and\ \citenamefont {Sachdev}}]{subir3}%
  \BibitemOpen
  \bibfield  {author} {\bibinfo {author} {\bibfnamefont {P.}~\bibnamefont
  {Fendley}}, \bibinfo {author} {\bibfnamefont {K.}~\bibnamefont {Sengupta}},\
  and\ \bibinfo {author} {\bibfnamefont {S.}~\bibnamefont {Sachdev}},\
  }\bibfield  {title} {\bibinfo {title} {Competing density-wave orders in a
  one-dimensional hard-boson model},\ }\href
  {https://doi.org/10.1103/PhysRevB.69.075106} {\bibfield  {journal} {\bibinfo
  {journal} {Phys. Rev. B}\ }\textbf {\bibinfo {volume} {69}},\ \bibinfo
  {pages} {075106} (\bibinfo {year} {2004})}\BibitemShut {NoStop}%
\bibitem [{\citenamefont {Turner}\ \emph
  {et~al.}(2018{\natexlab{a}})\citenamefont {Turner}, \citenamefont
  {Michailidis}, \citenamefont {Abanin}, \citenamefont {Serbyn},\ and\
  \citenamefont {Papić}}]{abanin1}%
  \BibitemOpen
  \bibfield  {author} {\bibinfo {author} {\bibfnamefont {C.~J.}\ \bibnamefont
  {Turner}}, \bibinfo {author} {\bibfnamefont {A.~A.}\ \bibnamefont
  {Michailidis}}, \bibinfo {author} {\bibfnamefont {D.~A.}\ \bibnamefont
  {Abanin}}, \bibinfo {author} {\bibfnamefont {M.}~\bibnamefont {Serbyn}},\
  and\ \bibinfo {author} {\bibfnamefont {Z.}~\bibnamefont {Papić}},\
  }\bibfield  {title} {\bibinfo {title} {Weak ergodicity breaking from quantum
  many-body scars},\ }\href {https://doi.org/10.1038/s41567-018-0137-5}
  {\bibfield  {journal} {\bibinfo  {journal} {Nature Physics}\ }\textbf
  {\bibinfo {volume} {14}},\ \bibinfo {pages} {745–749} (\bibinfo {year}
  {2018}{\natexlab{a}})}\BibitemShut {NoStop}%
\bibitem [{\citenamefont {Turner}\ \emph
  {et~al.}(2018{\natexlab{b}})\citenamefont {Turner}, \citenamefont
  {Michailidis}, \citenamefont {Abanin}, \citenamefont {Serbyn},\ and\
  \citenamefont {Papi\ifmmode~\acute{c}\else \'{c}\fi{}}}]{abanin2}%
  \BibitemOpen
  \bibfield  {author} {\bibinfo {author} {\bibfnamefont {C.~J.}\ \bibnamefont
  {Turner}}, \bibinfo {author} {\bibfnamefont {A.~A.}\ \bibnamefont
  {Michailidis}}, \bibinfo {author} {\bibfnamefont {D.~A.}\ \bibnamefont
  {Abanin}}, \bibinfo {author} {\bibfnamefont {M.}~\bibnamefont {Serbyn}},\
  and\ \bibinfo {author} {\bibfnamefont {Z.}~\bibnamefont
  {Papi\ifmmode~\acute{c}\else \'{c}\fi{}}},\ }\bibfield  {title} {\bibinfo
  {title} {Quantum scarred eigenstates in a rydberg atom chain: Entanglement,
  breakdown of thermalization, and stability to perturbations},\ }\href
  {https://doi.org/10.1103/PhysRevB.98.155134} {\bibfield  {journal} {\bibinfo
  {journal} {Phys. Rev. B}\ }\textbf {\bibinfo {volume} {98}},\ \bibinfo
  {pages} {155134} (\bibinfo {year} {2018}{\natexlab{b}})}\BibitemShut
  {NoStop}%
\bibitem [{\citenamefont {Bakr}\ \emph {et~al.}(2009)\citenamefont {Bakr},
  \citenamefont {Gillen}, \citenamefont {Peng}, \citenamefont {F\"{o}lling},\
  and\ \citenamefont {Greiner}}]{exp1}%
  \BibitemOpen
  \bibfield  {author} {\bibinfo {author} {\bibfnamefont {W.~S.}\ \bibnamefont
  {Bakr}}, \bibinfo {author} {\bibfnamefont {J.~I.}\ \bibnamefont {Gillen}},
  \bibinfo {author} {\bibfnamefont {A.}~\bibnamefont {Peng}}, \bibinfo {author}
  {\bibfnamefont {S.}~\bibnamefont {F\"{o}lling}},\ and\ \bibinfo {author}
  {\bibfnamefont {M.}~\bibnamefont {Greiner}},\ }\bibfield  {title} {\bibinfo
  {title} {A quantum gas microscope for detecting single atoms in a
  hubbard-regime optical lattice},\ }\href
  {https://doi.org/10.1038/nature08482} {\bibfield  {journal} {\bibinfo
  {journal} {Nature}\ }\textbf {\bibinfo {volume} {462}},\ \bibinfo {pages}
  {74–77} (\bibinfo {year} {2009})}\BibitemShut {NoStop}%
\bibitem [{\citenamefont {Bakr}\ \emph {et~al.}(2010)\citenamefont {Bakr},
  \citenamefont {Peng}, \citenamefont {Tai}, \citenamefont {Ma}, \citenamefont
  {Simon}, \citenamefont {Gillen}, \citenamefont {F\"{o}lling}, \citenamefont
  {Pollet},\ and\ \citenamefont {Greiner}}]{exp2}%
  \BibitemOpen
  \bibfield  {author} {\bibinfo {author} {\bibfnamefont {W.~S.}\ \bibnamefont
  {Bakr}}, \bibinfo {author} {\bibfnamefont {A.}~\bibnamefont {Peng}}, \bibinfo
  {author} {\bibfnamefont {M.~E.}\ \bibnamefont {Tai}}, \bibinfo {author}
  {\bibfnamefont {R.}~\bibnamefont {Ma}}, \bibinfo {author} {\bibfnamefont
  {J.}~\bibnamefont {Simon}}, \bibinfo {author} {\bibfnamefont {J.~I.}\
  \bibnamefont {Gillen}}, \bibinfo {author} {\bibfnamefont {S.}~\bibnamefont
  {F\"{o}lling}}, \bibinfo {author} {\bibfnamefont {L.}~\bibnamefont
  {Pollet}},\ and\ \bibinfo {author} {\bibfnamefont {M.}~\bibnamefont
  {Greiner}},\ }\bibfield  {title} {\bibinfo {title} {Probing the
  superfluid–to–mott insulator transition at the single-atom level},\
  }\href {https://doi.org/10.1126/science.1192368} {\bibfield  {journal}
  {\bibinfo  {journal} {Science}\ }\textbf {\bibinfo {volume} {329}},\ \bibinfo
  {pages} {547–550} (\bibinfo {year} {2010})}\BibitemShut {NoStop}%
\bibitem [{\citenamefont {Bernien}\ \emph {et~al.}(2017)\citenamefont
  {Bernien}, \citenamefont {Schwartz}, \citenamefont {Keesling}, \citenamefont
  {Levine}, \citenamefont {Omran}, \citenamefont {Pichler}, \citenamefont
  {Choi}, \citenamefont {Zibrov}, \citenamefont {Endres}, \citenamefont
  {Greiner}, \citenamefont {Vuletić},\ and\ \citenamefont {Lukin}}]{exp3}%
  \BibitemOpen
  \bibfield  {author} {\bibinfo {author} {\bibfnamefont {H.}~\bibnamefont
  {Bernien}}, \bibinfo {author} {\bibfnamefont {S.}~\bibnamefont {Schwartz}},
  \bibinfo {author} {\bibfnamefont {A.}~\bibnamefont {Keesling}}, \bibinfo
  {author} {\bibfnamefont {H.}~\bibnamefont {Levine}}, \bibinfo {author}
  {\bibfnamefont {A.}~\bibnamefont {Omran}}, \bibinfo {author} {\bibfnamefont
  {H.}~\bibnamefont {Pichler}}, \bibinfo {author} {\bibfnamefont
  {S.}~\bibnamefont {Choi}}, \bibinfo {author} {\bibfnamefont {A.~S.}\
  \bibnamefont {Zibrov}}, \bibinfo {author} {\bibfnamefont {M.}~\bibnamefont
  {Endres}}, \bibinfo {author} {\bibfnamefont {M.}~\bibnamefont {Greiner}},
  \bibinfo {author} {\bibfnamefont {V.}~\bibnamefont {Vuletić}},\ and\
  \bibinfo {author} {\bibfnamefont {M.~D.}\ \bibnamefont {Lukin}},\ }\bibfield
  {title} {\bibinfo {title} {Probing many-body dynamics on a 51-atom quantum
  simulator},\ }\href {https://doi.org/10.1038/nature24622} {\bibfield
  {journal} {\bibinfo  {journal} {Nature}\ }\textbf {\bibinfo {volume} {551}},\
  \bibinfo {pages} {579–584} (\bibinfo {year} {2017})}\BibitemShut {NoStop}%
\bibitem [{\citenamefont {Levine}\ \emph {et~al.}(2018)\citenamefont {Levine},
  \citenamefont {Keesling}, \citenamefont {Omran}, \citenamefont {Bernien},
  \citenamefont {Schwartz}, \citenamefont {Zibrov}, \citenamefont {Endres},
  \citenamefont {Greiner}, \citenamefont {Vuleti\ifmmode~\acute{c}\else
  \'{c}\fi{}},\ and\ \citenamefont {Lukin}}]{exp4}%
  \BibitemOpen
  \bibfield  {author} {\bibinfo {author} {\bibfnamefont {H.}~\bibnamefont
  {Levine}}, \bibinfo {author} {\bibfnamefont {A.}~\bibnamefont {Keesling}},
  \bibinfo {author} {\bibfnamefont {A.}~\bibnamefont {Omran}}, \bibinfo
  {author} {\bibfnamefont {H.}~\bibnamefont {Bernien}}, \bibinfo {author}
  {\bibfnamefont {S.}~\bibnamefont {Schwartz}}, \bibinfo {author}
  {\bibfnamefont {A.~S.}\ \bibnamefont {Zibrov}}, \bibinfo {author}
  {\bibfnamefont {M.}~\bibnamefont {Endres}}, \bibinfo {author} {\bibfnamefont
  {M.}~\bibnamefont {Greiner}}, \bibinfo {author} {\bibfnamefont
  {V.}~\bibnamefont {Vuleti\ifmmode~\acute{c}\else \'{c}\fi{}}},\ and\ \bibinfo
  {author} {\bibfnamefont {M.~D.}\ \bibnamefont {Lukin}},\ }\bibfield  {title}
  {\bibinfo {title} {High-fidelity control and entanglement of rydberg-atom
  qubits},\ }\href {https://doi.org/10.1103/PhysRevLett.121.123603} {\bibfield
  {journal} {\bibinfo  {journal} {Phys. Rev. Lett.}\ }\textbf {\bibinfo
  {volume} {121}},\ \bibinfo {pages} {123603} (\bibinfo {year}
  {2018})}\BibitemShut {NoStop}%
\bibitem [{\citenamefont {Bluvstein}\ \emph {et~al.}(2021)\citenamefont
  {Bluvstein}, \citenamefont {Omran}, \citenamefont {Levine}, \citenamefont
  {Keesling}, \citenamefont {Semeghini}, \citenamefont {Ebadi}, \citenamefont
  {Wang}, \citenamefont {Michailidis}, \citenamefont {Maskara}, \citenamefont
  {Ho}, \citenamefont {Choi}, \citenamefont {Serbyn}, \citenamefont {Greiner},
  \citenamefont {Vuletić},\ and\ \citenamefont {Lukin}}]{exp5}%
  \BibitemOpen
  \bibfield  {author} {\bibinfo {author} {\bibfnamefont {D.}~\bibnamefont
  {Bluvstein}}, \bibinfo {author} {\bibfnamefont {A.}~\bibnamefont {Omran}},
  \bibinfo {author} {\bibfnamefont {H.}~\bibnamefont {Levine}}, \bibinfo
  {author} {\bibfnamefont {A.}~\bibnamefont {Keesling}}, \bibinfo {author}
  {\bibfnamefont {G.}~\bibnamefont {Semeghini}}, \bibinfo {author}
  {\bibfnamefont {S.}~\bibnamefont {Ebadi}}, \bibinfo {author} {\bibfnamefont
  {T.~T.}\ \bibnamefont {Wang}}, \bibinfo {author} {\bibfnamefont {A.~A.}\
  \bibnamefont {Michailidis}}, \bibinfo {author} {\bibfnamefont
  {N.}~\bibnamefont {Maskara}}, \bibinfo {author} {\bibfnamefont {W.~W.}\
  \bibnamefont {Ho}}, \bibinfo {author} {\bibfnamefont {S.}~\bibnamefont
  {Choi}}, \bibinfo {author} {\bibfnamefont {M.}~\bibnamefont {Serbyn}},
  \bibinfo {author} {\bibfnamefont {M.}~\bibnamefont {Greiner}}, \bibinfo
  {author} {\bibfnamefont {V.}~\bibnamefont {Vuletić}},\ and\ \bibinfo
  {author} {\bibfnamefont {M.~D.}\ \bibnamefont {Lukin}},\ }\bibfield  {title}
  {\bibinfo {title} {Controlling quantum many-body dynamics in driven rydberg
  atom arrays},\ }\href {https://doi.org/10.1126/science.abg2530} {\bibfield
  {journal} {\bibinfo  {journal} {Science}\ }\textbf {\bibinfo {volume}
  {371}},\ \bibinfo {pages} {1355} (\bibinfo {year} {2021})}\BibitemShut
  {NoStop}%
\bibitem [{\citenamefont {Manovitz}\ \emph {et~al.}(2025)\citenamefont
  {Manovitz}, \citenamefont {Li}, \citenamefont {Ebadi}, \citenamefont
  {Samajdar}, \citenamefont {Geim}, \citenamefont {Evered}, \citenamefont
  {Bluvstein}, \citenamefont {Zhou}, \citenamefont {Koyluoglu}, \citenamefont
  {Feldmeier}, \citenamefont {Dolgirev}, \citenamefont {Maskara}, \citenamefont
  {Kalinowski}, \citenamefont {Sachdev}, \citenamefont {Huse}, \citenamefont
  {Greiner}, \citenamefont {Vuletić},\ and\ \citenamefont {Lukin}}]{exp6}%
  \BibitemOpen
  \bibfield  {author} {\bibinfo {author} {\bibfnamefont {T.}~\bibnamefont
  {Manovitz}}, \bibinfo {author} {\bibfnamefont {S.~H.}\ \bibnamefont {Li}},
  \bibinfo {author} {\bibfnamefont {S.}~\bibnamefont {Ebadi}}, \bibinfo
  {author} {\bibfnamefont {R.}~\bibnamefont {Samajdar}}, \bibinfo {author}
  {\bibfnamefont {A.~A.}\ \bibnamefont {Geim}}, \bibinfo {author}
  {\bibfnamefont {S.~J.}\ \bibnamefont {Evered}}, \bibinfo {author}
  {\bibfnamefont {D.}~\bibnamefont {Bluvstein}}, \bibinfo {author}
  {\bibfnamefont {H.}~\bibnamefont {Zhou}}, \bibinfo {author} {\bibfnamefont
  {N.~U.}\ \bibnamefont {Koyluoglu}}, \bibinfo {author} {\bibfnamefont
  {J.}~\bibnamefont {Feldmeier}}, \bibinfo {author} {\bibfnamefont {P.~E.}\
  \bibnamefont {Dolgirev}}, \bibinfo {author} {\bibfnamefont {N.}~\bibnamefont
  {Maskara}}, \bibinfo {author} {\bibfnamefont {M.}~\bibnamefont {Kalinowski}},
  \bibinfo {author} {\bibfnamefont {S.}~\bibnamefont {Sachdev}}, \bibinfo
  {author} {\bibfnamefont {D.~A.}\ \bibnamefont {Huse}}, \bibinfo {author}
  {\bibfnamefont {M.}~\bibnamefont {Greiner}}, \bibinfo {author} {\bibfnamefont
  {V.}~\bibnamefont {Vuletić}},\ and\ \bibinfo {author} {\bibfnamefont
  {M.~D.}\ \bibnamefont {Lukin}},\ }\bibfield  {title} {\bibinfo {title}
  {Quantum coarsening and collective dynamics on a programmable simulator},\
  }\href {https://doi.org/10.1103/w1cp-l5vq} {\bibfield  {journal} {\bibinfo
  {journal} {Nature}\ }\textbf {\bibinfo {volume} {638}},\ \bibinfo {pages}
  {86} (\bibinfo {year} {2025})}\BibitemShut {NoStop}%
\bibitem [{\citenamefont {Liang}\ \emph {et~al.}(2025)\citenamefont {Liang},
  \citenamefont {Yue}, \citenamefont {Chao}, \citenamefont {Hua}, \citenamefont
  {Lin}, \citenamefont {Tey},\ and\ \citenamefont {You}}]{exp7}%
  \BibitemOpen
  \bibfield  {author} {\bibinfo {author} {\bibfnamefont {X.}~\bibnamefont
  {Liang}}, \bibinfo {author} {\bibfnamefont {Z.}~\bibnamefont {Yue}}, \bibinfo
  {author} {\bibfnamefont {Y.-X.}\ \bibnamefont {Chao}}, \bibinfo {author}
  {\bibfnamefont {Z.-X.}\ \bibnamefont {Hua}}, \bibinfo {author} {\bibfnamefont
  {Y.}~\bibnamefont {Lin}}, \bibinfo {author} {\bibfnamefont {M.~K.}\
  \bibnamefont {Tey}},\ and\ \bibinfo {author} {\bibfnamefont {L.}~\bibnamefont
  {You}},\ }\bibfield  {title} {\bibinfo {title} {Observation of anomalous
  information scrambling in a rydberg atom array},\ }\href
  {https://doi.org/10.1103/w1cp-l5vq} {\bibfield  {journal} {\bibinfo
  {journal} {Phys. Rev. Lett.}\ }\textbf {\bibinfo {volume} {135}},\ \bibinfo
  {pages} {050201} (\bibinfo {year} {2025})}\BibitemShut {NoStop}%
\end{thebibliography}%

\end{document}